\documentclass[fleqn,usenatbib]{mnras}

\usepackage{newtxtext,newtxmath}

\usepackage[T1]{fontenc}

\DeclareRobustCommand{\VAN}[3]{#2}
\let\VANthebibliography\thebibliography
\def\thebibliography{\DeclareRobustCommand{\VAN}[3]{##3}\VANthebibliography}

\usepackage{graphicx}	
\usepackage{amsmath}	
\usepackage{threeparttable}
\usepackage{booktabs, makecell, multirow}
\usepackage{url}
\usepackage{caption}
\usepackage[position=bottom]{subcaption}
\usepackage{hyperref}
\usepackage{mathtools}
\usepackage{dirtytalk}
\usepackage{capt-of}
\usepackage{tabularx}
\usepackage{rotating}
\usepackage{adjustbox}
\usepackage{mwe}
\newlength{\tempheight}
\newlength{\tempwidth}
\pdfoutput=1

\newcommand{\rowname}[1]
{\rotatebox{90}{\makebox[\tempheight][c]{{#1}}}}
\newcommand{\columnname}[1]
{\makebox[\tempwidth][c]{\textbf{#1}}}

\renewcommand{\thesubfigure}{\alph{subfigure}}
\newcommand{\mycaption}[1]
{\refstepcounter{subfigure}\textbf{(\thesubfigure) }{\ignorespaces #1}}

\title[Stacked 610 MHz \textit{sSFR-M$_{\star}$} plane]{Star formation history of $\rm{0.1\leq\,\textit{z}\,\leq\,1.5}$ mass-selected galaxies in the ELAIS-N1 Field}

\author[Ocran et al.]{
E.F. Ocran$^{1,2}$\thanks{E-mail: ocran62@gmail.com}, M. Vaccari$^{2,3,5}$,
J.M. Stil$^{3,4}$, A.R. Taylor$^{2,3}$, C.H. Ishwara-Chandra$^{2,6}$, Jae-Woo Kim$^{1}$
\\
$^1$ Korea Astronomy and Space Science Institute, 776 Daedeokdae-ro, Daejeon 305-348, Korea\\
$^2$ Inter-university Institute for Data Intensive Astronomy, Department of Astronomy, University of Cape Town, 7701 Rondebosch, Cape Town, South Africa\\
$^3$ Inter-university Institute for Data Intensive Astronomy, Department of Physics and Astronomy, University of the Western Cape, 7535 Bellville, Cape Town,\\ South Africa\\ 
$^4$ Department of Physics and Astronomy, University of Calgary, 2500 University Drive NW, Calgary AB, T2N 1N4, Canada\\
$^5$ INAF - Istituto di Radioastronomia, via Gobetti 101, 40129 Bologna, Italy \\
$^6$ National Centre for Radio Astrophysics, Tata Institute of Fundamental Research, Pune 411007, India
}

\date{Accepted 2023 July 07. Received 2023 July 05; in original form 2023 January 08}

\pubyear{2023}

\begin{document}
\label{firstpage}
\pagerange{\pageref{firstpage}--\pageref{lastpage}}
\maketitle

\begin{abstract}
We measure the specific star formation rates of \textit{K}-band selected galaxies from the ELAIS-N1 by stacking GMRT data at 610 MHz. 
We identify a sample of SFGs, spanning $\rm{0.1\leq\,\textit{z}\,\leq\,1.5}$ and $\rm{10^{8.5}<\,{\textit{M}_{\star}}/{\textit{M}_{\odot}}<10^{12.4}}$, using a  
combination of multi-wavelength diagnostics obtained from the deep LoTSS multi-wavelength catalogue.
We measure the flux densities in the radio map and estimate the radio SFR
in order to probe the nature of the galaxies below the noise and confusion limits. 
The massive galaxies in our sample have the lowest sSFRs which is in agreement with previous studies. 
For the different populations, we show that the sSFR-mass relation steepens with redshift, with an average slope of  $\rm{\langle \beta_{All} \rangle\,=\, -0.49\pm0.01}$ for the whole sample, and $\rm{\langle \beta_{SFG} \rangle\,=\, -0.42\pm0.02}$ for the SFGs. 
Our results indicate that galaxy populations undergo 'downsizing', whereby most massive galaxies form their stars earlier and more rapidly than low mass galaxies.
Both populations show a strong decrease in their sSFR toward the present epoch.
The sSFR evolution with redshift is best described by a power law $\rm{(1\,+\,\textit{z})^\textit{n}}$, where $\rm{\langle \textit{n}_{ALL}\rangle\sim4.94\pm0.53}$ for all galaxies, and $\rm{\langle \textit{n}_{SFG}\rangle \sim3.51\pm0.52}$ for SFGs. 
Comparing our measured sSFRs to results from literature, we find a general agreement in the \textit{sSFR-M$_{\star}$} plane.

\end{abstract}

\begin{keywords}
Galaxy: evolution — radio continuum: galaxies.
\end{keywords}



\section{Introduction}

Radio surveys have now reached sufficient areal coverage that they are now dominated by the same galaxies detected by infrared (IR), optical and X-ray surveys and have become increasingly important in studies of galaxy evolution. The galaxy populations that lie beneath the sensitivity limits of the current deepest surveys has become an important area of study in recent years (e.g. see \citealt{2007ApJ...654...99W,2009MNRAS.394..105G,2009MNRAS.394....3D,2014ApJ...787...99S,2014MNRAS.439.1459Z}, and references therein). Deep radio surveys are able to probe the galaxy star formation rate (SFR) due to cosmic ray and synchrotron emission that originates from the accelerated electrons in the magnetic fields of supernova remnants which are the result of massive star formation \citep{1985ApJ...298L...7H}. 

The relation between SFR and 1.4 GHz luminosity is calibrated to the far-infrared-radio correlation (e.g. \citealt{1992ARA&A..30..575C,2000ApJ...544..641H,2001ApJ...554..803Y,2002AJ....124..675C}). At radio-wavelengths, observations are not obscured by dust and their higher angular resolution, as compared to infrared surveys, significantly reduces source confusion. However, emission from active galactic nuclei (AGN) represent a significant source of contamination (see \citealt{2014MNRAS.439.1459Z}). \citet{2021MNRAS.500.4685O}
compared the SFR derived from the IR luminosity and the radio power to show that the two are equivalently good tracers of star formation in non-active star-forming galaxies (SFGs) and also for the host galaxies of radio quiet (RQ AGN). They studied the correlation between galaxy SFR and stellar mass at different redshifts for  SFGs, RQ, and  radio loud (RL) AGN and found that the vast majority of our sources lie on the star formation main sequence (hereafter, MS) when using infrared star formation rates.

The MS of star forming galaxies is a fundamental relation in galaxy evolution which relates galaxy star formation to their stellar mass
\citep[see,][]{2007ApJ...660L..43N,2007A&A...468...33E,2009ApJ...698L.116P,2010MNRAS.405.2279O,2012ApJ...754...25R,2012ApJ...754L..29W,2014ApJ...795..104W,2019MNRAS.490.5285P,2019MNRAS.483.3213P,2020ApJ...899...58L}. However, in the literature on there is no common agreement on the  form of the MS. There is contention whether the MS is linear across all redshifts \citep[see,][]{2011ApJ...738..106W,2011ApJ...742...96W,2014ApJS..214...15S,2015A&A...575A..74S,2018A&A...615A.146P}, or has a flattening or turn-over at
stellar masses $\rm{\log_{10}(\textit{M}_{\star}/\textit{M}_{\odot})\,>\,10.5}$ \cite[see,][]{2014ApJ...795..104W,2015ApJ...801...80L,2020ApJ...899...58L,2021MNRAS.505..540T}. The specific SFR (SFR divided by stellar mass, sSFR) provides a measure of the current star formation activity related to the past history \citep{2022MNRAS.515.2951S}.
Studies have shown that the galaxy stellar mass function at high masses evolves fairly slowly up to $\rm{\textit{z}\sim0.9}$, and then more rapidly up to at least $\rm{\textit{z}\sim2.5}$, suggesting that the majority of stellar mass assembly took place at $\rm{\textit{z}\,\gtrsim\,1}$ \citep[see,][]{2007MNRAS.378..429F,2007A&A...474..443P}. At low masses, \cite{2007ApJS..172..270C} showed that the mass of a galaxy plays an important role in star formation, at $\rm{\textit{z}\,\lesssim\,1}$. However, at $\rm{\textit{z}\,\lesssim\,0.2}$, ongoing star formation in massive galaxies is almost entirely absent \citep[see,][]{2021MNRAS.505..540T}. The evolution of the slope  $\rm{\textit{sSFR}-\textit{M}_{\star}}$ plane as a function of redshift, on mass-dependent timescales, has been found to decline significantly but smoothly 
\citep[see,][]{2014ApJS..214...15S}. Moreover, the sSFR plateaus at higher redshifts, $\rm{\textit{z} \gtrsim 3}$, and  continues to increase with a relatively shallow slope as noted in \citet{2013ApJ...762L..31B}.
Studies like \cite{2018ApJ...852..107D} have used the differential evolution of the galaxy stellar mass function to infer the sSFR evolution of galaxies.

The sensitivities achieved by SKA  pathfinders and eventually the SKA itself will have a huge impact on our understanding of star formation in galaxies and its co-evolution with supermassive black holes \citep{2011MNRAS.411.1547P}. Improvements in both depth and sky coverage is being  made with these surveys, with narrow, but very deep surveys such as the MeerKAT MIGHTEE \citep{MIGHTEE2016} and wide-area radio data such as Evolutionary Map of the Universe (EMU) \citep{2011JApA...32..599N}. These new surveys are probing unexplored volume of the Universe.
Studies have shown that at the faintest radio flux densities ($\rm{S_{1.4}\,<\,10\,mJy}$), conflicting results emerge regarding whether there is a flattening of the average spectral index between a low radio frequency (325 or 610MHz) (see, \citealt{2012MNRAS.421.1644R}).
More comprehensive observations of the shape
of the radio spectrum, extending to lower frequencies, will ensure a maximum scientific return by combining the deep radio continuum data at GHz frequencies.

Stacking is a common tool which has been used to investigate the star formation properties of galaxies at far greater sensitivity by combining many observations of individual galaxies at the expense of any specific knowledge of the individual galaxies that make up the stack. For example,
\cite{2009MNRAS.394....3D} used stacking of 610 MHz and 1.4 GHz data from the VLA and the Giant Metrewave Radio Telescope (GMRT) to investigate the star formation history of \textit{BzK}-selected galaxies from the UKIRT Infrared Deep Sky Survey (UKIDSS–UDS) and computed stellar masses using the absolute \textit{K}-band magnitude.
\citet{2011ApJ...730...61K} calculated stellar
masses using SED fitting from their photometric-redshift fitting by selecting galaxies at $\rm{3.6\,\mu m}$, and stacked 1.4 GHz VLA data (A and C arrays), with a noise of $\rm{8\,\mu Jy}$ at the centre of their $\rm{1.72\, deg^{2}}$ map. They found a good  agreement in their radio-derived sSFR–redshift  evolution between their studies and that of \cite{2009MNRAS.394....3D}, however the dependence of sSFR on stellar mass was found to be much shallower for the UKIDSS data than for COSMOS. \citet{2014MNRAS.439.1459Z} stacked deep (17.5 $\mu$Jy) VLA radio observations at the positions of $\rm{K_{s}}$-selected sources in the VIDEO field for $\rm{\textit{K}_{s}\,<\,23.5}$ and sensitive to $\rm{0\,<\,\textit{z}\,\lesssim\,5}$.
They found that sSFR falls with stellar mass, in agreement with the 'downsizing' paradigm. \citet{2020ApJ...899...58L} measured the MS using mean stacks of 3 GHz radio continuum images to derive average SFRs for $\sim$200,000 mass-selected galaxies at $\rm{\textit{z}\,>\,0.3}$ in the COSMOS field.
They described the MS by adopting a new model that incorporates a linear relation at low stellar
mass $\rm{(\log(\textit{M}_{\star}/\textit{M}_{\odot})<10)}$ and a flattening at high stellar mass that becomes more prominent at low
redshift (i.e. $\rm{\textit{z}\,<\,1.5}$).

We present an independent stacking analysis of radio data from the  GMRT surveys of the ELAIS-N1 region. We stack by mass and redshift bins respectively, for sources drawn from the rich LOFAR Two-metre Sky Survey (LoTSS) \citep{2017A&A...598A.104S} deep field multi-wavelength    ancillary data available in the field. We calibrate 610 MHz rest-frame luminosity as a SFR indicator following \citet{2009MNRAS.397.1101G}, allowing us to turn radio luminosity estimates into SFR function estimates. We provide a coherent, uniform measurement of the evolution of the logarithmic specific star formation rate (sSFR)– stellar mass ($\rm{\textit{M}_{\star}}$) relation, for star forming and all galaxies out to $\rm{\textit{z}\sim 1.5}$. Using median stacked images at 610 MHz, we derive average SFRs and sSFRs for $\rm{\sim 77,047}$ mass-selected galaxies, spanning $\rm{0.1\leq\,\textit{z}\,\leq\,1.5}$ and $\rm{10^{8.5}<\,{\textit{M}_{\star}}/{\textit{M}_{\odot}}<10^{12.4}}$ in the ELAIS-N1. We aim to answer how the sSFRs  change as a function of stellar mass and redshift with regards to a deep 610 MHz low-frequency continuum survey, which are complimentary to  sSFRs derived at the high frequency observations.

The paper is arranged as follows: The datasets used in this work are described in Section~\ref{data.sec}. In Section~\ref{sample.sec} we describe the prescription we used in selecting the sample for our analyses. Section~\ref{analyses.sec} presents the analyses and results from our stacking experiment. We compare the synergies between our work and findings from the literature in Section~\ref{comp.sec}. We then provide our discussions and a summary of our work in Sections~\ref{disc.sec} and ~\ref{conc.sec} respectively. 
We assume a flat cold dark matter ($\rm{\Lambda}$CDM) cosmology with  $\rm{\Omega_{\Lambda} \ = \ 0.7}$, $\rm{\Omega_{m} \ = \ 0.3}$ and $\rm{H_{o} \ = \ 70 \ km\,s^{-1} \ Mpc^{-1}}$  and $\rm{S_{\nu}\,\propto\,\nu^{\alpha}}$ for calculation of intrinsic source properties.

\section{DATASETS}\label{data.sec}
In this section, we discuss the different datasets we use for our investigation. These data
are all taken from publicly available catalogs.

\subsection{Radio Datasets}
We employ the 610 MHz wide-area survey \citep{2020MNRAS.497.5383I} of the ELAIS-N1 (European Larg Area ISO Survey North 1) \citep{2000MNRAS.316..749O} region, using the GMRT. This data is in hexagonal configuration centred on  $\mathrm{\alpha \ = \ 16^{h} \ 10^{m} \ 30^{s}, \  \delta \ = \ 54^{\circ} \ 35 \ 00^{\prime \prime}}$
The 610 MHz wide-area survey consists of 51 pointings, mosaicked to create an image of ELAIS-N1
covering $\rm{\sim12.8~deg^{2}}$. The
FWHM of the synthesized beam varies between 4.5 and 6 arcsec. Before mosaicking, the image from each field was smoothed to a circular Gaussian beam with FWHM of 6 arcsec. The final rms in the total intensity mosaic image is $\rm{\sim40\mu Jy\,beam^{-1}}$. \citealt{2020MNRAS.497.5383I} indicated that this is
equivalent to $\rm{\sim20\,\mu Jy\,beam^{-1}}$ rms noise at 1.4 GHz for a spectral
index of -0.75, which is several times deeper than the VLA FIRST survey at similar resolution. 
The resulting mosaic is about $\rm{3.6\times3.6\,deg^{2}}$
\citep[see,][]{2020MNRAS.497.5383I}. 
The criterion \citet{2020MNRAS.497.5383I} used to distinguish point sources from resolved sources resulted in about
60 per cent of sources being are unresolved by  using $\rm{S_{total}/S_{peak}\,<\, 1}$\citep[see][]{2001A&A...365..392P}, and an extra term to fit the envelope. By considering a total to peak flux ratio <1.5, $\sim$75 per cent of sources were found to be unresolved. 

\subsection{The LOFAR science-ready multi-wavelength catalogue of the ELAIS-N1 }

The LOw Frequency ARray \citep[LOFAR;][]{2013A&A...556A...2V} Two-metre Sky Survey (LoTSS) deep field multi-wavelength data we use is only briefly described here, for much greater detail, the reader is referred to \citet{2017A&A...598A.104S,2019A&A...622A...1S,2022A&A...659A...1S}, \citet{2021A&A...648A...3K} and subsequent
data release papers.
LoTSS is an ongoing sensitive, high-resolution 120 – 168 MHz survey of the entire northern sky for which
the first full-quality public data release covers 424 square degrees with a median rms noise of $\rm{71\,\mu Jy}$ at 150 MHz \citep{2019A&A...622A..17S,2019A&A...622A...2W}. The second data release covers 27$\%$ (i.e. split into two regions spanning 4178 and 1457 square degrees) of the northern sky with a central frequency of 144 MHz down to a median rms sensitivity of $83\,\rm{\mu}$Jy beam$^{-1}$ \citep[see,][]{2022A&A...659A...1S}.  The ELAIS-N1 field is the deepest of the LoTSS deep fields to date and one of the areas that have the most extensive ancillary data \citep{2021A&A...648A...2S}. 

The LOFAR team has provided science-ready multi-wavelength data in three fields along with the full complimentary optical/IR catalogue presented by \citet{2021A&A...648A...3K}. 
The  photometric redshift estimates for all plausible counterparts were produced  from a complete, homogeneous sample of objects measured across optical to IR wavelength. This is  achieved by building a forced, matched aperture, multi-wavelength catalogue in each field spanning the UV to mid-infrared wavelengths using the latest deep datasets. 
The full details of the photo-\textit{z}
estimation, are presented in a companion release paper
\citep[see,][for more details]{refId0}.
\subsection{Stellar masses}\label{stellar_mass_estimation}

Galaxy stellar masses were obtained from science-ready multi-wavelength catalogue  \citep[see][for more details]{refId0}. This is the total stellar mass of a galaxy in units of solar mass and was estimated using the \textsc{Python}-based SED fitting code previously used by \citet{2014MNRAS.444.2960D,2019ApJ...876..110D}.
Stellar population synthesis models of \citet{2003MNRAS.344.1000B} for a \citet{2003PASP..115..763C} initial mass function (IMF) were generated for composite stellar population models with three different stellar metallicities of $\rm{Z_{\odot}\,=\,0.1,0.4,1.0}$. \cite{refId0} used a grid of star-formation histories based on the double power-law model with the priors on the range of power-law slopes and turnover ages taken from \cite{2019MNRAS.490..417C}. They argued this provides sufficient flexibility to accurately describe the star-formation histories of a wide range of possible formation and quenching mechanisms.
A simple prescription for nebular emission is included in the model SEDs. Further details of the assumed emission line ratios for Balmer and metal lines, as well as the nebular continuum prescription, can be found in \citet{2014MNRAS.444.2960D}. They also incorporate dust attenuation following the two-component dust model of \citet{2000ApJ...539..718C}. The ELAIS-N1 field is complete to significantly lower masses when using K band to select samples at \textit{z} < 1, where deeper NIR observations are provided by UKIDSS Deep Extragalactic Survey (DXS) \citep{2007MNRAS.379.1599L}. From simple estimations of the galaxy stellar mass functions (SMFs) within the ELAIS-N1 field and comparison with the literature, \citet{refId0} validated that the stellar masses provide reliable and self consistent estimates suitable for statistical studies across the whole field.

\begin{figure}
\centering
\centerline{\includegraphics[width=0.42\textwidth]{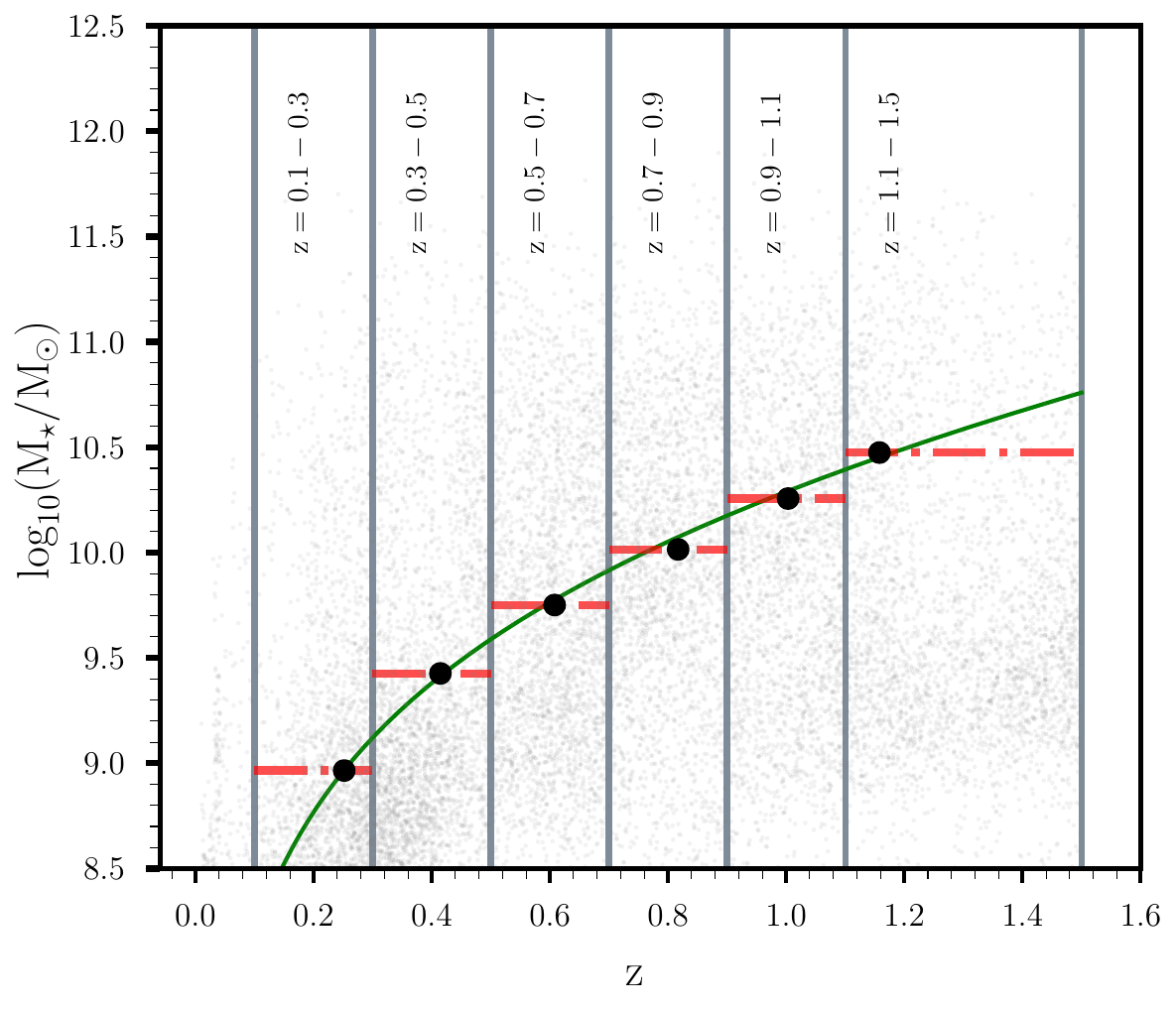}}

    \caption{Observed stellar mass distribution as a function of redshift for the ELAIS-N1 LoTSS Deep Field. The background density plot shows the mass distribution of sources with the solid black circles showing the 90 per cent mass completeness limit of the \textit{K} reference band in the ELAIS-N1 field derived empirically \citep[see,][]{2010A&A...523A..13P}  at the median redshift in each bin. Here we a plotting $\sim$ 80 per cent of the data points. The solid green curve represents the fit to $\rm{\textit{M}_{lim}}$, the mass limit}.
    \label{mass_lim.fig}
\end{figure}

Following \citet{refId0}, we empirically estimate the stellar mass completeness
\citep[][]{2010A&A...523A..13P,2013A&A...556A..55I,2013A&A...558A..23D,2016ApJS..224...24L,2017A&A...605A..70D}. This is determined by the 3$\sigma$ magnitude limit, $\rm{\textit{K}_{lim}\,=\, 22.7}$ mag.
In Figure~\ref{mass_lim.fig} we show the distribution of stellar mass $\rm{(\textit{M}_{\star})}$ with the redshift (\textit{z}) for the galaxies in the ELAIS-N1 field.  
For each redshift bin, we estimate the stellar mass
completeness $\rm{\textit{M}_{lim}}$ within which 90 per cent of the galaxies lie. The measured stellar masses for the
sample are scaled to the magnitude limit: $\rm{\log_{10}\textit{M}_{lim}\,=\,\log_{10}\textit{M}\,- \,0.4(\textit{K}_{lim}\,-\,\textit{K})}$ and the mass completeness limit derived from 95th percentile of the scaled $\rm{\textit{M}_{lim}}$ mass distribution. The black circles in Figure~\ref{mass_lim.fig} represent the mass limit in each redshift bin and the solid green curve represents the fit to $\rm{\textit{M}_{lim}}$ the mass limit. Table~\ref{masslim.tab} presents the calculated mass limits for each redshift bin in this study of ELAIS-N1 LoTSS Deep Field for the full sample. Our sources should be largely complete above the cuts. The distributions are generally consistent among different fields, supporting
the self-consistency of our results.

\begin{table}
 \centering
 \caption{Mass limits for each redshift bin in this study of ELAIS-N1 LoTSS Deep Field, for the full sample.}
 \begin{tabular}{cc}
 \hline
Bin    & $\rm{\textit{M}^{full}_{lim}}$ \\
 \hline
$\rm{0.1 - 0.3}$ &8.96  \\
$\rm{0.3 - 0.5}$ &9.42  \\
$\rm{0.5 - 0.7}$ &9.75  \\
$\rm{0.7 - 0.9}$ &10.01  \\
$\rm{0.9 - 1.1}$ &10.25\\
$\rm{1.1 - 1.5}$ &10.47 \\
\hline
\end{tabular}
\label{masslim.tab} 
\end{table}

\section{Sample selection}\label{sample.sec}

In this section, we describe the sample and, the multi-wavelength diagnostics employed to obtain a census
of galaxies showing evidence of hosting an AGN within this sample. 

Previous studies of the sSFR-$\rm{\textit{M}_{\star}}$ plane,
indicate that galaxies reside in two populations. The first is the SFG population  whose SFR is positively correlated with stellar mass out
to redshifts, $\rm{\textit{z}\sim4}$. \citep[see,][]{2011ApJ...730...61K,2015A&A...575A..74S,2016ApJ...817..118T,2020ApJ...899...58L}.  The second population consists of quiescent galaxies that are not actively forming stars and typically reside at the high-mass end (i.e.
they have lower sSFR) \citep{2015ApJ...801L..29R}. In this Section, we describe how we separate quiescent galaxies, which are systems with little or no ongoing
star-formation and large surface stellar mass density, from SFGs \citep[see,][]{2011ApJ...735...53P,2012ApJ...748L..27P}. 
This method is most efficient in excluding quiescent galaxies \citep[see][]{2019ApJ...877..140L,2019ApJ...880L...9L}. These
quiescent galaxies (QGs) have very low SFR by definition, and they are preferentially found at high $\rm{\textit{M}_{\star}}$ \citep[see,][]{2015A&A...575A..74S}. 

The sources classified as SFGs are those sources in our redshift and mass selected bins satisfy the color cuts in the diagnostics employed in this study.

\subsection{The Sample}\label{sample_.sec}
Extensive details of the photometric redshift and stellar mass (limited to \textit{z} < 1.5) estimation included in the LOFAR science-ready multi-wavelength data are outlined in  \citet{refId0}.
We followed the prescription by  \citet{2013MNRAS.428.1281J} in order  to remove sources that could be  spectroscopically and photometrically flagged as stars. This star galaxy separation criterion clearly segregates the two types of objects  in rest-frame   \textit{J - K} versus \textit{u - J} color space. We found that galaxies dominate
the catalog at \textit{K > 22.7} mag.

We then applied the prescription below to select our sample:

\begin{multline}\label{eqn1}
\rm{\left(\frac{\textit{z}_{1,max}\,-\,\textit{z}_{1,min}}{1+\textit{z}_{1,median}}\right)\times\,0.5<0.1\,\&\,\left(\frac{\textit{S}_{K}}{\textit{S}_{Kerr}}\right)>5\,\&\,(\textit{z}_{best} \leq 1.5)}
\\
{\,\&\,(10^{8.5}<\,{\textit{M}_{\star}}/{\textit{M}_{\odot}}<10^{12.4})}
\end{multline}

Where the best available estimate is $\rm{\textit{z}_{best}}$, including spectroscopic redshifts  when available and photometric redshift (photo-\textit{z}) estimates otherwise. $\rm{\textit{z}_{1,median}}$ is the primary redshift peak used when calculating the photo-\textit{z}. Whereas $\rm{\textit{z}_{1,min}}$ and $\rm{\textit{z}_{1,max}}$ are the lower and upper bounds of the primary 80\% highest probability density (HPD) credible interval (CI) peak respectively. The $\rm{\textit{S}_{K}}$ and $\rm{\textit{S}_{Kerr}}$ represent the NIR  UKIDSS Deep Extragalactic Survey (DXS) (hereafter UKIDSS-DXS) DR10  \textit{K}-band flux and flux error respectively. This NIR data covers a maximum area of around 8 deg$^2$ with a 3- and 5-$\sigma$  magnitude depths of 22.7 mag and 22.1 mag respectively, in ELAIS-N1. Following Equation~\ref{eqn1}, we select $\sim$ 77,047  sources which constitute our sample. Further selection cuts (i.e. SFG/AGN /Quiescent galaxy separation) applied to the sample are described in subsequent sections. 

\subsection{AGN removal using IRAC colour diagnostics}\label{IRAC.diag}
AGN are known to show flux variability over all observable timescales and across the entire electromagnetic spectrum hence a combination of multi-wavelength diagnostics are usually employed (see \citealt{2014NatPh..10..417V}).

We applied the original AGN flag given in the LoTSS multi-wavelength catalogue \citep[from][]{2023MNRAS.523.1729B} in order to remove galaxies showing evidence of hosting an AGN. 
This AGN flag in the catalogue incorporates:
\begin{itemize}
\item optAGN : sources included in the Million Quasar Catalog compilation or spectroscopically identified AGN. 
\item IRAGN : when a source satisfies \cite{2012ApJ...748..142D} IR AGN criterion given by:
\begin{equation}
{x \ = \ \log_{10}\left(\frac{f_{5.8\mu m}} {f_{3.6\mu m}}\right) , \    \ y = \log_{10}\left(\frac{f_{8.0\mu m}}{f_{4.5\mu m}}\right)}
\end{equation}
\begin{multline}
{x  \ge \ 0.08 \ \wedge \ y   \ge  0.15}
\\ 
{\wedge  y \ge  (1.21  \times   x)  -  0.27}
\\
{\wedge y \le  \ (1.21  \times  x)  + 0.27}
\\
{\wedge  f_{4.5\mu m}   >  f_{3.6\mu m}  >   f_{4.5\mu m}   \wedge   f_{8.0\mu m}   >   f_{5.8\mu m}} \\
\label{donley_crit}
\end{multline}
\item XrayAGN : when a source has X-ray counterpart.
\end{itemize}

\begin{figure}
\includegraphics[width=0.49\textwidth]{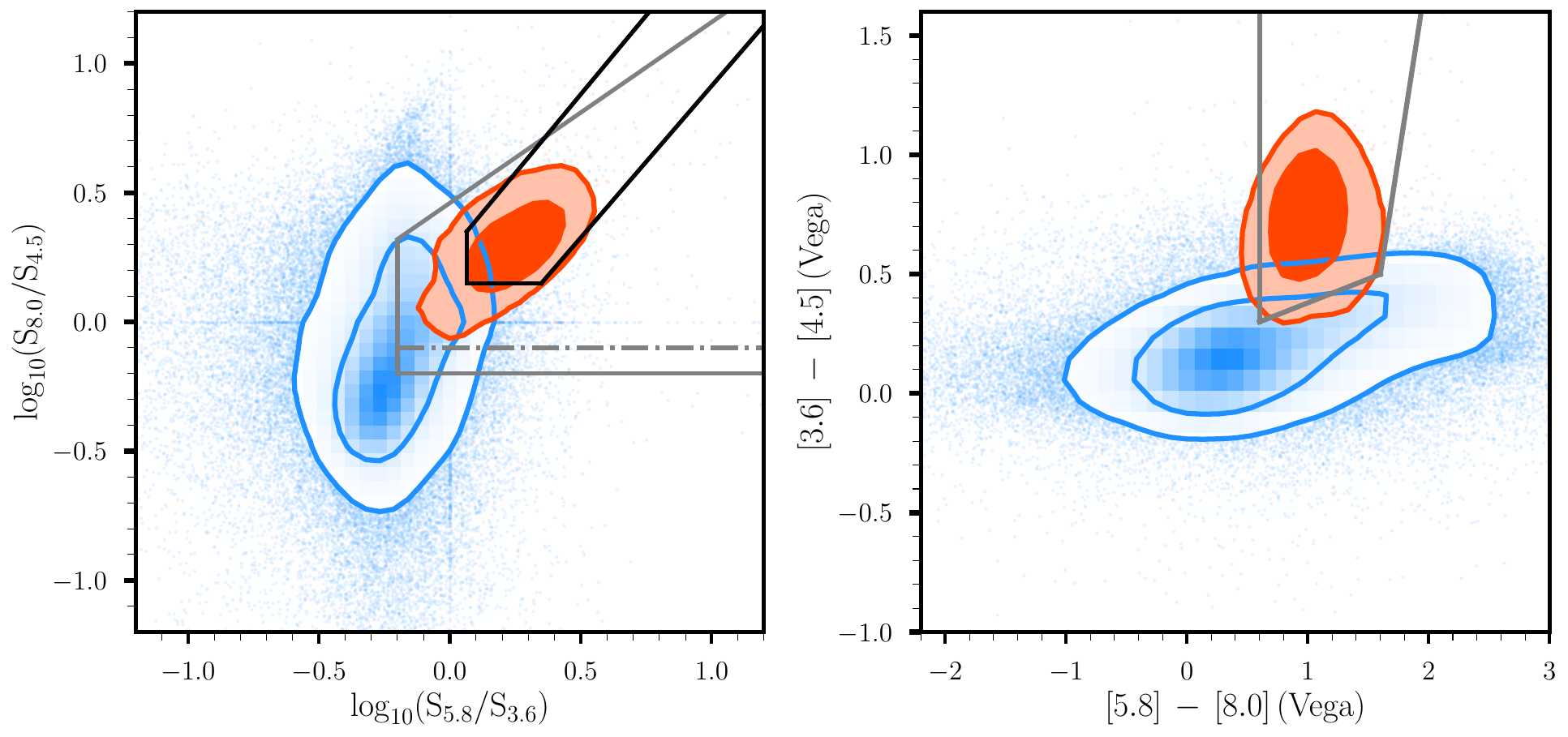}

    \caption{IRAC colour–colour diagram showing the separation between AGN (red scatter contour)
and SFG (blue scatter contour). The solid black wedge in the left panel indicates the \citet{2012ApJ...748..142D} wedge. Also shown are solid grey lines in the left and right panels indicating colour selection wedges
of the \citet{2004ApJS..154..166L} and \citet{2005ApJ...631..163S} respectively.  The dotted dashed grey line in the left panel indicate the \citet{2007ApJ...669L..61L} wedge.} 

    \label{lacy_stern.fig}
\end{figure}

Following the \textit{optAGN}, \textit{IRAGN} and \textit{XrayAGN} flags, we select 428 sources satisfying equation~\ref{eqn1}  as candidate AGN. It is important to note that every technique for selecting AGN is affected by selection biases, and these ones are no exception. The color selection means that objects whose observed mid-infrared colours are not dominated by thermal emission from AGN will be missing from the sample.
Figure~\ref{lacy_stern.fig} presents IRAC colour–colour diagram showing the separation between AGN (red scatter contour) and SFG (blue scatter contour). For comparison, we show the \citet{2012ApJ...748..142D} colour selection criterion given by Equation~\ref{donley_crit}, and  represented by the solid black wedge in the left panel. The solid grey and dotted dashed grey lines indicate the \citet{2004ApJS..154..166L} and \citet{2007ApJ...669L..61L} wedges respectively. 
We  also compare the AGN selection based to the \citet{2005ApJ...631..163S} IRAC [3.6] - [4.5] versus [5.8] -  [8.0], in Vega magnitudes shown in the right panel of Figure~\ref{lacy_stern.fig}.

\subsection{SFG/Quiescent galaxy separation}\label{SFG/QGS_galaxy separation}

\begin{figure}
\centering
\centerline{\includegraphics[width=0.5\textwidth]{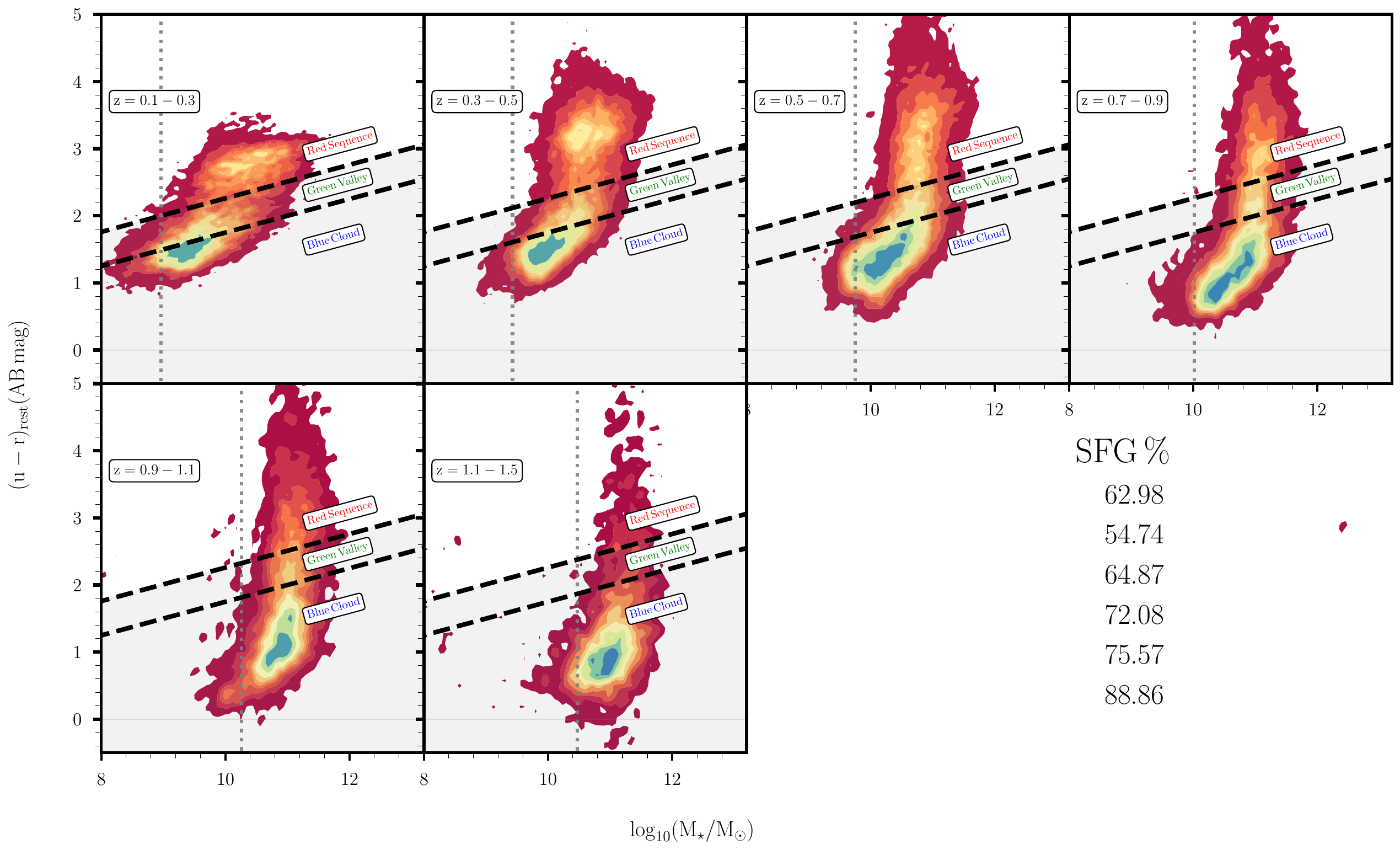}}
    \caption{The \textit{u - r} galaxy color-mass diagram in bins of redshift as density contours. The galaxy color-mass diagram showing blue, star-forming galaxies are at the bottom, in the blue cloud region. The red, quiescent/passively evolving galaxies are at the top, in the red sequence. The "green valley" is the transition zone in between. The dotted vertical lines indicate mass completeness limits for each redshift bin.
    }
\label{ur_mass.fig}
\end{figure}

We aim to create well-defined, unbiased samples of the SFG population. 
Following \citet{2014MNRAS.440..889S}, one can  separate red, green valley, and blue cloud populations using the dust-corrected colour-mass diagram for all redshift bins using:
$\rm{\textit{u\,-\,r}(\textit{M}_{\star})\,=\,-0.24\,+\,0.25\,\times\,\textit{M}_{\star}}$ and
$\rm{\textit{u\,-\,r}(\textit{M}_{\star})\,=\,-0.75\,+\,0.25\,\times\,\textit{M}_{\star}}$.
Figure~\ref{ur_mass.fig} presents the \textit{u - r} color-stellar mass dust-corrected diagram for our sample. Galaxies with "green" or intermediate colors are those galaxies in which star formation is in the process of turning off, but still have some ongoing star formation – indicating the process only shut down a short while ago, $\rm{\sim10^{8}}$ years \citep{2004ApJ...608..752B}. The "green valley" region represents the crossroads of galaxy evolution. Galaxies that constitute this population are between the blue star-forming galaxies (the "blue cloud") and the red, passively evolving galaxies (the "red sequence"). 
The colour bimodality is weakly evident in the redshift bins at 0.1 - 0.3 and 0.3 - 0.5 of the rest-frame \textit{u - r} color distribution. Subsequent redshift bins exhibit a unimodal distribution peaking in the blue (these are the main sequence star-forming galaxies). In particular, the colour-mass or colour- magnitude diagrams do not exhibit strong colour bimodality seen in  of the  \textit{UVJ} or \textit{NUVrJ} diagrams \citep[see,][]{2009ApJ...691.1879W,2013ApJS..206....8M,2014ApJ...783L..14S,2016ApJ...830...51S}.

Similarly, observational results have also been presented by \citet{2006A&A...453..869B} and \citet{2009ApJ...706L.173B}  who used the $\rm{\textit{U}\,-\,\textit{V}}$ color–mass relation to separate red galaxies from blue galaxies at $\rm{0.2\,\leq\,\textit{z}\,\leq\,1.0}$ and $\rm{0.\,\leq\,\textit{z}\,\leq\,2.5}$, respectively. \citet{2010ApJ...721..193P} used the $\rm{\textit{U}\,-\,\textit{B}}$ with redshift evolution extrapolated to $\rm{\textit{z}\,=\,1}$ to split  into red and blue galaxies. More recently, \citet{2017ApJ...835...22P} used the rest-frame $\rm{\textit{U}\,-\,\textit{R}}$ color $\rm{0.7\,\leq\,\textit{z}\,\leq\,1.3}$ to distinguish between red and blue sequences galaxies.

For simplicity, we consider only two states for galaxies, “blue star-forming” and “red passive” based on a dividing rest-frame \textit{u - r } color. Obviously, this approach is  partly simplistic, but is in accordance with our approach to identify the most basic features of the SFG population. Since the  \textit{V} and \textit{r} bands are not equivalent, we therefore adhere to using only the \citet{2014MNRAS.440..889S}, $\rm{\textit{u\,-\,r}(\textit{M}_{\star})\,=\,-0.24\,+\,0.25\,\times\,\textit{M}_{\star}}$ line to separate red, quiescent/passively evolving galaxies from blue star-forming (i.e. potentially including green valley) ones.  
Thus, the grey shaded area in each panel of  Figure~\ref{ur_mass.fig} represents the region in which we classify a source as an SFG.  We indicate the corresponding percentage of SFGs (i.e., green and blue galaxies) for each redshift bin in Figure~\ref{ur_mass.fig} from the first to the last redshift bin. Appendix~\ref{append.colourmass} provides more discussions on our colour-mass selection.

Following the AGN diagnostics in subsection~\ref{IRAC.diag}, and  the   separation of quiescent/passively evolving galaxies from candidate SFGs, we employ our final sample selected for the subsequent analyses is a follows:

\begin{enumerate}
\item[1.] All Galaxies: the original 77,047 sources that satisfying equation~\ref{eqn1}.
\item[2.] SFGs: sources from the original 77,047 sample that are classified as SFGs based on the \textit{u - r} galaxy color-mass diagnostics and sources that are not labeled as \textit{optAGN}, \textit{IRAGN} and \textit{XrayAGN} from the LoTSS multi-wavelength catalogue. Removing these flags, we obtain a subsample of 51 124 as SFGs.

\end{enumerate}


Table~\ref{tab_ppts} presents a summary of the results of subsequent analysis of the average galaxy mass in each stellar mass range. For a given stellar mass range, we show the median stellar across the entire redshift range (i.e, $\rm{0.1\leq\,\textit{z}\,\leq\,1.5}$ ) of this work, for the corresponding total and SFG populations. The lower and upper bounds represent the 16/84th percentile.

\begin{table}
 \centering
 \caption{Table showing the summary of the results of subsequent analysis of the average galaxy properties in each stellar mass range.}
 \scalebox{0.8}{
 \begin{tabular}{ccc|ccc}
 \hline
 \hline
\textit{M} range&$\rm{N_{All,galaxies}}$&  $\rm{\left(\log\frac{M_{\star}}{M_{\odot}}\right)_{ALL\,GALAXIES}}$&$\rm{N_{SFGs}}$ & $\rm{\left(\log\frac{M_{\star}}{M_{\odot}}\right)_{SFGs}}$\\
\\
 \hline
$\rm{8.5<{\textit{M}_{\star}}<9.0}$&1393 &  $\rm{8.86^{+0.12}_{-0.18}}$&1314 &$8.84^{+0.15}_{-0.18}$&\\ 
$\rm{9.0<{\textit{M}_{\star}}<9.5}$&4919 &$\rm{9.37^{+0.10}_{-0.19}}$&4522 &$9.36^{+0.11}_{-0.20}$\\
$\rm{9.5<{\textit{M}_{\star}}<10.0}$&12853 & $\rm{9.84^{+0.13}_{-0.17}}$&10812 &$9.82^{+0.13}_{-0.17}$\\
$\rm{10.0<{\textit{M}_{\star}}<10.5}$&20580 &$\rm{10.30^{+0.12}_{-0.17}}$&14343 &$10.28^{+0.12}_{-0.17}$ \\
$\rm{10.5<{\textit{M}_{\star}}<11.0}$&24766 &$\rm{10.76^{+0.12}_{-0.16}}$&13805 &$10.74^{+0.12}_{-0.17}$\\
$\rm{11.0<{\textit{M}_{\star}}<12.4}$&12536 &$\rm{11.16^{+0.24}_{-0.17}}$&6328 &$11.15^{+0.23}_{-0.17}$\\
\\
 \hline
 \end{tabular}}
 \label{tab_ppts} 
 \end{table}

\section{Analysis and Results}\label{analyses.sec}
\subsection{Stacking Methodology}\label{stackmethod.sec}
The direct detection of the radio point source population is complicated by source confusion.
Confusion is the blending of faint sources within a telescope beam. Hence statistical
techniques such as stacking, which are not strongly 
effected by confusion noise, can be 
 a powerful tool for reaching below the noise. Stacking is a tool to average together data for a given set of objects.  For an input sample of N galaxies its background noise level in a stacked image should reduce to $\rm{\sim1/\sqrt{N}}$ of the noise measured in a single radio image. Stacking is at the expense of knowledge of the individual
galaxies, but with careful application of criteria when binning the
galaxies, and with a large enough sample, it can reveal properties
of galaxies below the noise and confusion levels. The technique has
been used to great effect many times in the literature \citep[see,][for example]{2004ApJS..154..118S,2007MNRAS.380..199I,2011MNRAS.410.1155B}.

We choose six bins with a stellar mass and redshift range of $\rm{10^{8.5}<\textit{M}_{\star}/\textit{M}_{\odot}<10^{12.4}}$ and $0.1\,\leq\,z\,\leq\,1.5$ respectively. 
Out of the 77,047 sources we select as all galaxies, 51 124 sources are SFGs.
We stack the \textit{K}-band mass selected positions from the LOFAR multi-wavelength catalogue of the ELAIS-N1 on the 610 MHz  wide radio map \citep{2020MNRAS.497.5383I} of the ELAIS-N1.
Stacking was done with the Python Astronomical Stacking Tool Array (\texttt{PASTA}) \citep{2018ascl.soft09003K} program \footnote{\url{https://github.com/bwkeller/PASTA}} which measures the flux in a map from selected sources (
usually at another wavelength) and then builds  a distribution of map-extracted fluxes for the
sample \citep{2007ApJ...654...99W}. We choose fixed  bin sizes  and non-overlapping (statistically independent) bins in stellar mass and redshift space.  This allows for a statistically robust number of sources of in each bin and allows us to achieve a high signal-to-noise ratio (SNR). Notice that there is a $\sim 0.5$ dex in mass, i.e $\rm{\textit{M}_{\star}}$ bin size, up to $\rm{\textit{M}_{\star}(\textit{M}_{\odot})\,=\,11.0}$, beyond which the bin size is increased to $\sim 1.4$ dex in order to cover the full mass range. Conversely,  there is a $\sim 0.25$ dex redshift bin size over the entire redshift range. The advantage of stacking technique
is the gain in the signal-to-noise ratio, as combining many sources reduces the random noise while maintaining the average level of the signal. Median stacking analyses are less susceptible to contamination from radio AGN, which constitute a minority of the population at faint radio fluxes as compared to mean stacking \cite[][]{2017A&A...602A...6S,2020ApJ...903..139A}.

Our stacking work is summarized as follows:

\begin{itemize}
\item An input list of coordinates is created for the
number of galaxies to be stacked, taking into consideration
the selection criteria. An input image in FITS format.
\item \texttt{PASTA} reads the source list and FITS file with the number of pixels specified (i.e. $30\times30$ pixels cutouts for our 610 MHz image). The program proceeds to extract
“stamps”, square sections of the source image with a source centered within them  and generating 2-dimensional matrices of the median and mean output images. The detection threshold is improved by stacking images centered on the object coordinates.
\item The integrated and peak flux densities are computed by running \textsc{PyBDSF} source finder \citep{2015ascl.soft02007M} on the median stacked images.  \textsc{PyBDSF} fits a 2D Gaussian to
any significant emission in the center of the stack.
\end{itemize}

 
The median
estimator is more robust to outliers than the mean, and we will
demonstrate that the median is the most appropriate choice for our analysis.
The median image provides a compelling visual impression of the statistical significance of the sample median compared to nearby off positions. The premise of median stacking a survey is that the radio emission is unresolved, and that the central pixel represents the flux density of the sources in the stack. 
\cite{2007ApJ...654...99W} performed detailed calculations that show that a median stacking analysis is superior to a mean stacking, since it is robust to small numbers of bright sources, and it does not require any maximum allowed flux density cutoff prior to stacking.
It also shows patterns like the side lobes of the dirty beam that must be present around real sources of any flux density in the image.

\begin{figure*}
\centerline{\includegraphics[width=0.98\textwidth]{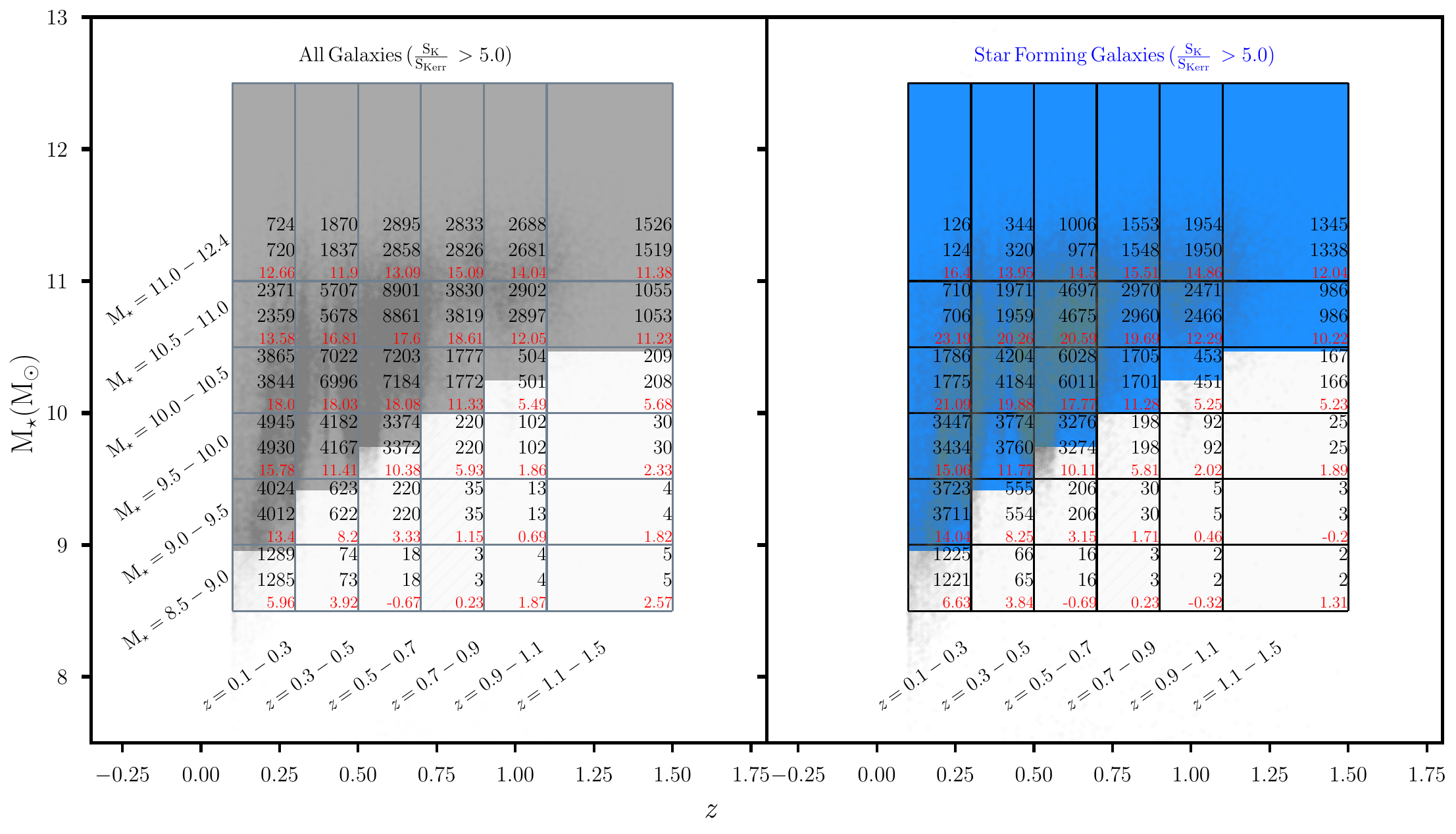}}
    \caption{Binning scheme in stellar mass and photometric redshift for the entire (left) and the SF (right) sample. The top number in each box is the total number of galaxies in each bin. The middle
number is the total number of galaxies used in the median radio stack; the bottom number shows the signal-to-noise (SNR) ratio achieved in the median radio stack. The gray and blue shading traces the mass completeness limits derived for all galaxies in subsection~\ref{stellar_mass_estimation} (see Table~\ref{masslim.tab})}.

    \label{mass_zbins.fig}
\end{figure*}

Figure~\ref{mass_zbins.fig} shows the binning scheme in stellar mass and photometric redshift for the entire (left) and the SF (right) sample. The top number in each box is the total number of galaxies in each bin. The middle number is the total number of galaxies used in the radio stack; the bottom number shows the signal-to-noise (SNR) ratio achieved in the radio stack. Ideally, one could roughly estimate stellar mass completeness limits by visual inspection of Figure~\ref{mass_zbins.fig}.
Figure~\ref{all_galaxies_image.fig} presents the stacked images of total intensity for all galaxies. The columns indicate the median stacked 610 MHz total intensity radio images for total galaxies within the range $\rm{\textit{z}\in[0.1-0.3]}$, $\rm{[0.3-0.5]}$, $\rm{[0.5-0.7]}$, $\rm{[0.7-0.9]}$, $\rm{[0.9-1.1]}$, $\rm{[1.1-1.5]}$ for the \textit{K}-band magnitude mass selected sample. The rows indicate mass range, $\rm{\textit{M}_{\star}\in[11.0-12.4]}$, $\rm{[10.5-11.0]}$, $\rm{[10.0-10.5]}$, $\rm{[9.5-10.0]}$, $\rm{[9.0-9.5]}$, $\rm{[8.5-9.0]}$ respectively, from top to bottom. All images have a size of $\rm{\sim 36\times36}$ arcsec$^2$ respectively. The image-scale ranges between 1 and 100 $\rm{\mu}$Jy beam$^{-1}$. The stacked images (GMRT, 610 MHz) of total intensity for star-forming galaxies is shown in Figure~\ref{sfg_image.fig}. In contrast to
this, Figure~\ref{mean_stack_fig} shows the mean stacked images for the same redshift and stellar mass bins for the total (top) and the SFG (bottom) populations,respectively.
All images are notably noisier than their median equivalents, and
bright sources away from the centre of the cut-out images have a
much greater effect on the stacked images. Thus, the mean images are strongly biased by a few bright radio sources and as such not a good representation of the typical sources within each flux density bin with the noise level of the
mean stacked images generally $\sim$1.5 times the noise of the median stacked images.

Figure~\ref{src_size.fig} shows the stacked median axial ratio (angular size) $\rm{\textit{B}_{maj}/\textit{B}_{min}}$ as a function of median redshift for all galaxies (left) and star forming galaxies (right). The errors represent the difference between the maximum Gaussian fit to a source and best fit Gaussian that encompasses the full source at the center of the median stacked image.
The fitted angular size is overall closely consistent with the original beam size of the 610 MHz image. At the first redshift range for SFGs (i.e. $\rm{\textit{z}\in[0.1-0.3]}$ ), the size of the gaussian fits to the median stacked image with $\rm{\textit{M}_{\star}\in[11.0-12.4]}$, seem higher than the size of the synthesized beam $\rm{\textit{B}_{maj}/\textit{B}_{min}\,=1}$ (see the horizontal solid green  line in Figure~\ref{src_size.fig}). The rest of the mass bins are consistent with the beam when compared with the horizontal solid green  line. Differences may occur for various reasons, for example, a Gaussian fit to a source convolved with a non-Gaussian point spread function can give rise to systematic errors. Errors in the positions of the input source catalog can lead to blurring of the stacked image. The same effect can occur if the radio emission is systematically offset from the IR emission, in some cases. Fitting of source sizes is a simple test one can run on the results of a stack. This can be used as
a test of both the positional accuracy, 
and in testing that the sources stacked are indeed unresolved. Since our stacked image produces a source that is almost the
same size of the beam, this confirms that the positional accuracy is sufficient, and that the stack
is dominated by unresolved sources.

\newsavebox\mybox
\savebox\mybox{%
\centering
  \begin{minipage}[t]{0.48\linewidth}
    \includegraphics[width=0.2\linewidth]{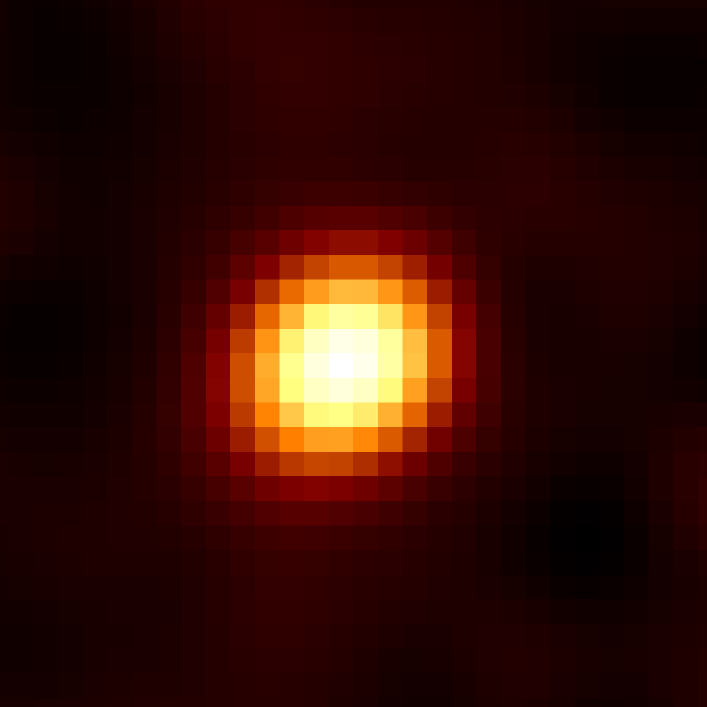}
  \end{minipage}%
}
\newlength{\ImageHt}
\setlength\ImageHt{\ht\mybox}

\begin{figure}
\begin{tabular}{@{}c@{ }c@{ }c@{ }c@{ }c@{ }c@{ }c@{}}
&\textbf{\textit{z}=0.1-0.3} & \textbf{\textit{z}=0.3-0.5} & \textbf{\textit{z}=0.5-0.7} & \textbf{\textit{z}=0.7-0.9} & \textbf{\textit{z}=0.9-1.1} & \textbf{\textit{z}=1.1-1.5}\\
\centering
\rotatebox[origin=c]{90}{\makebox[\ImageHt]{\scriptsize $\rm{11.0-12.4}$}}&
\includegraphics[width=.15\linewidth]{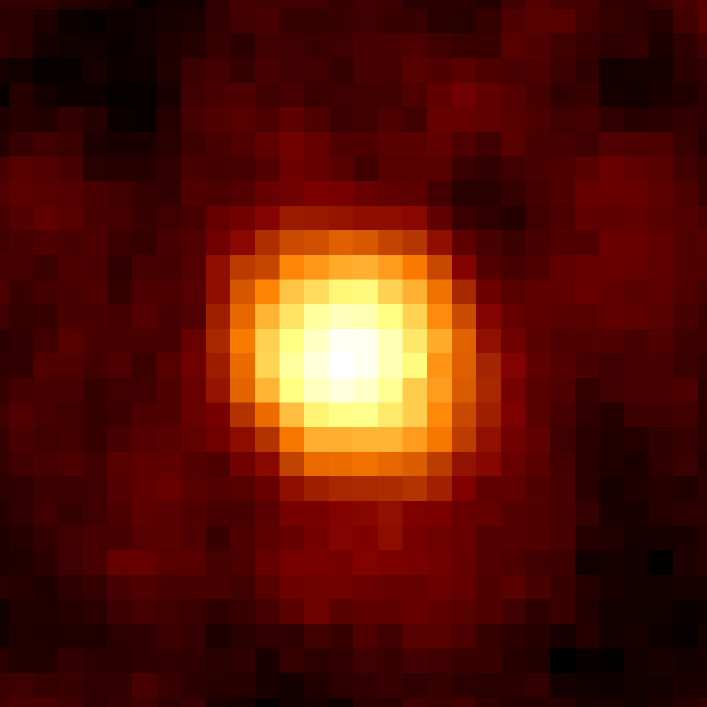}&
\includegraphics[width=.15\linewidth]{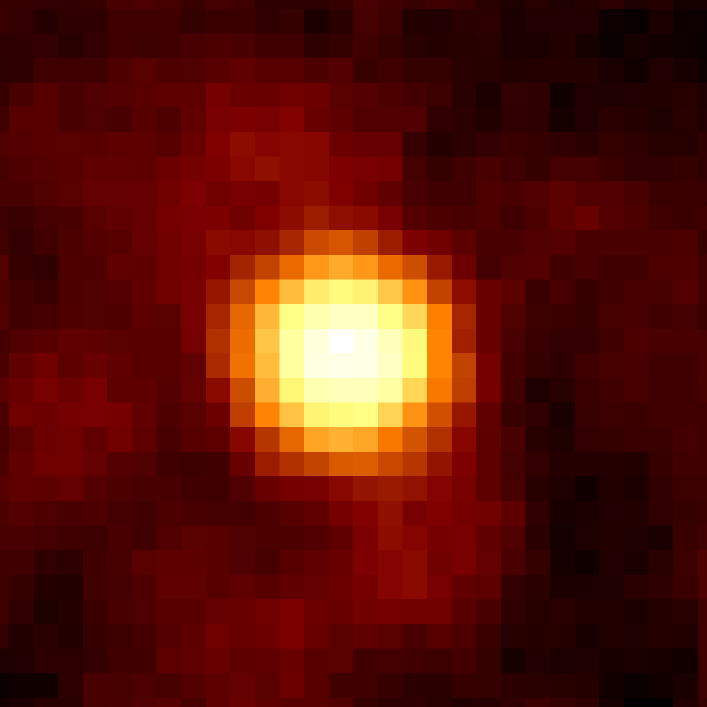}&
\includegraphics[width=.15\linewidth]{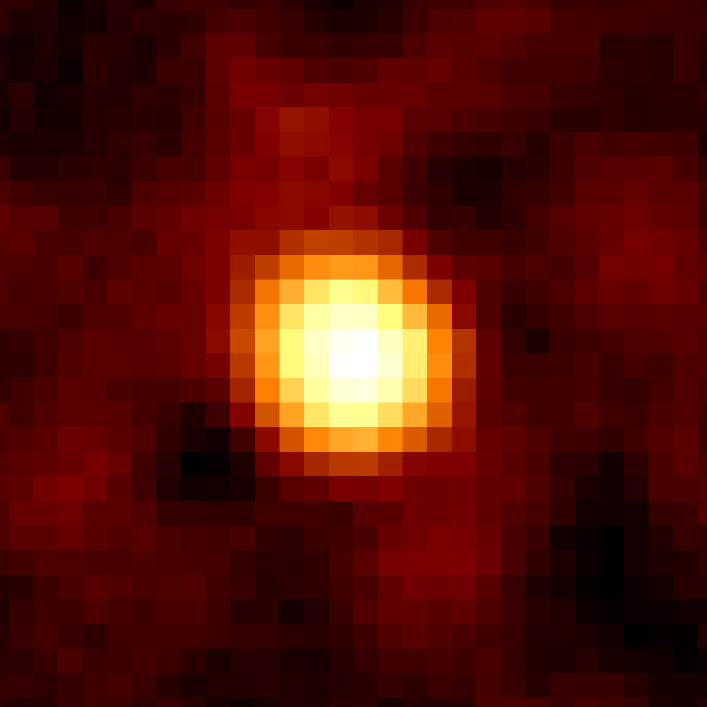}&
\includegraphics[width=.15\linewidth]{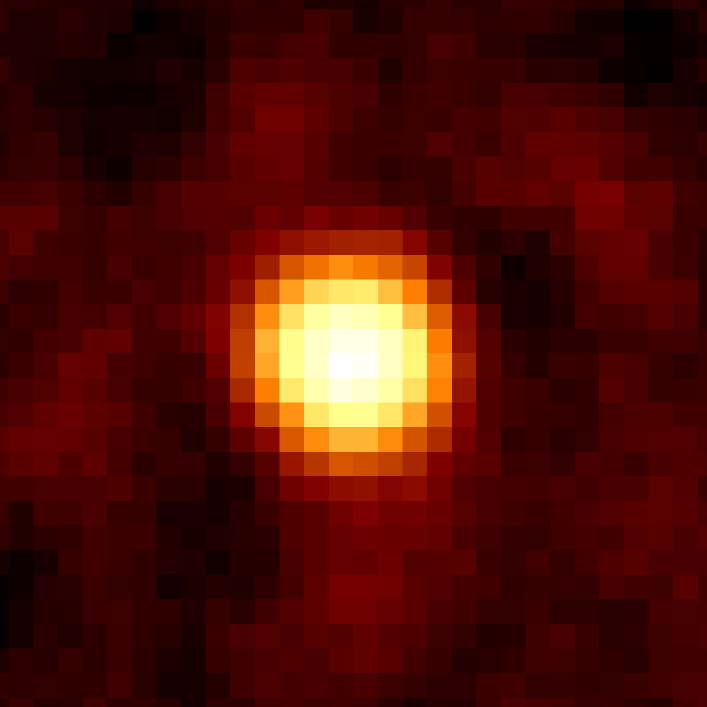}&
\includegraphics[width=.15\linewidth]{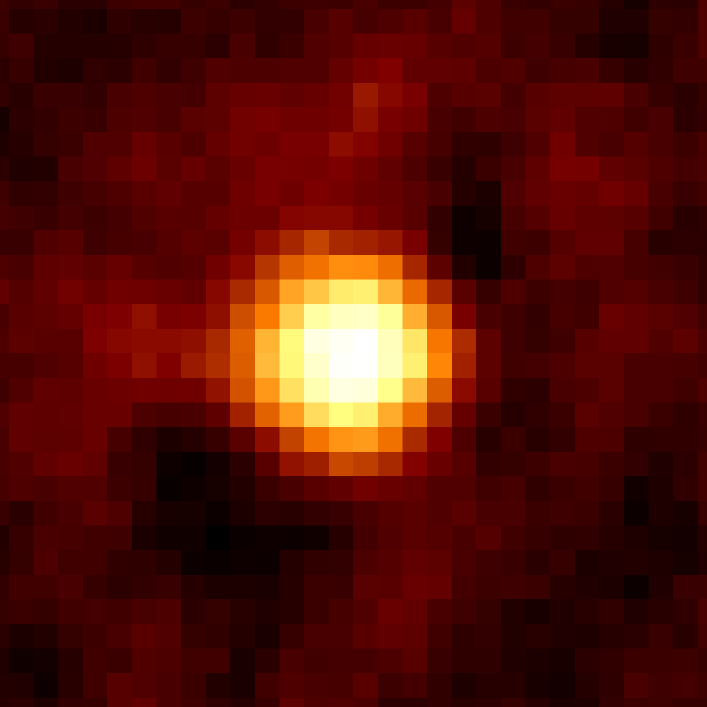}&
\includegraphics[width=.15\linewidth]{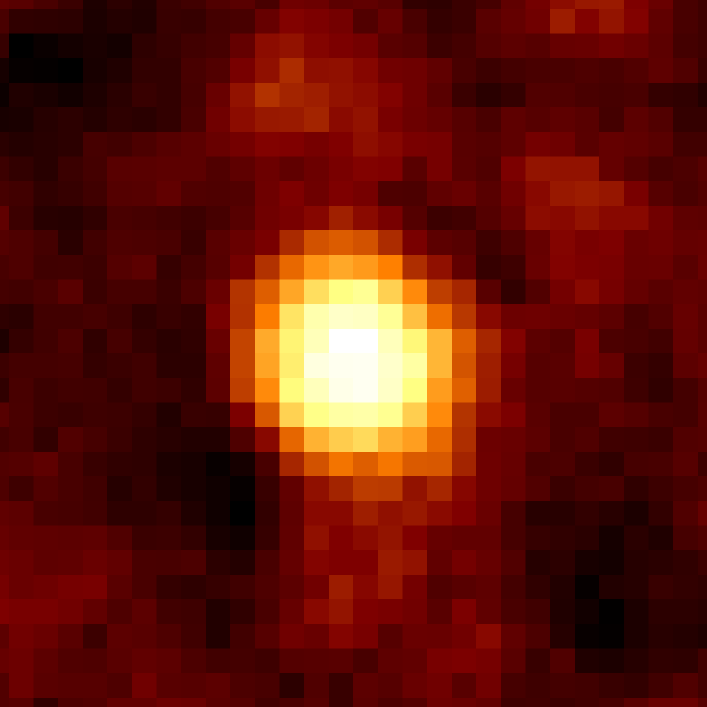}\\

\rotatebox[origin=c]{90}{\makebox[\ImageHt]{\scriptsize$\rm{10.5-11.0}$}\strut}&
\includegraphics[width=.15\linewidth]{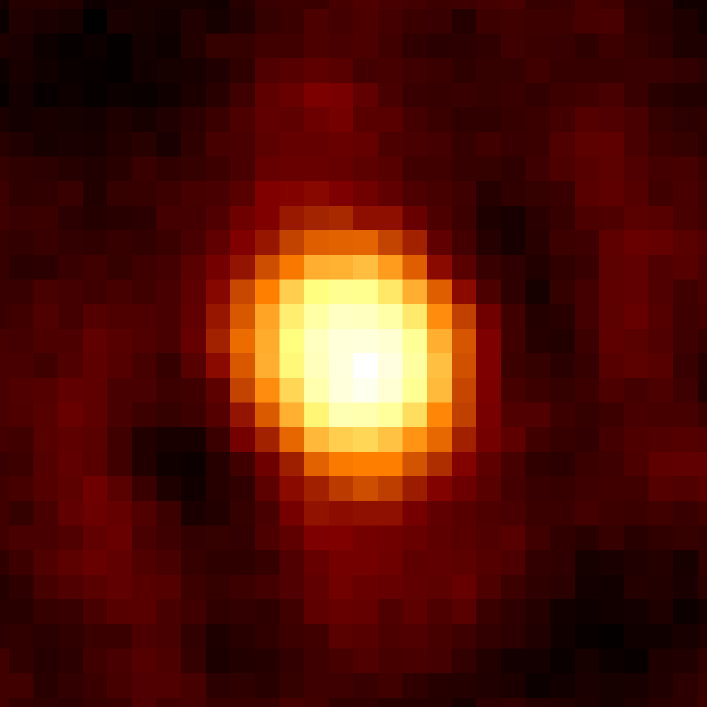}&
\includegraphics[width=.15\linewidth]{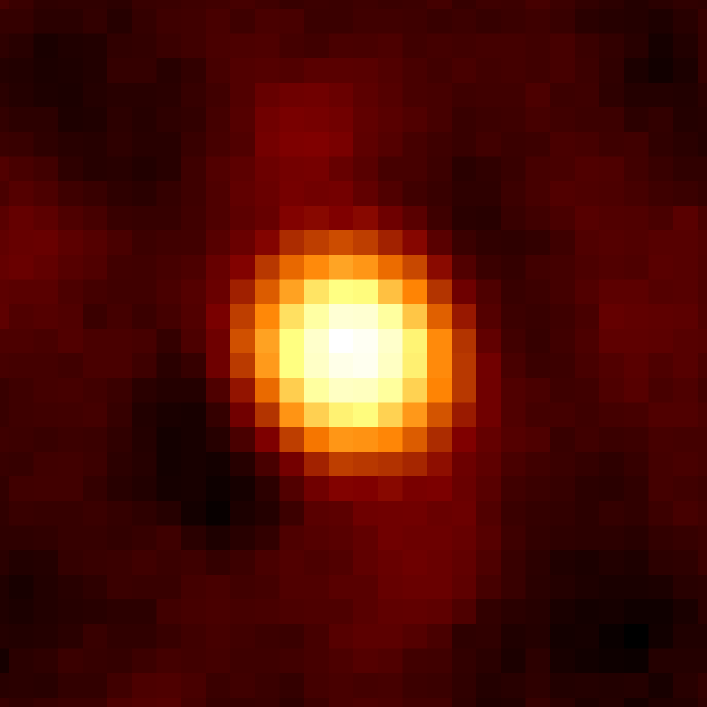}&
\includegraphics[width=.15\linewidth]{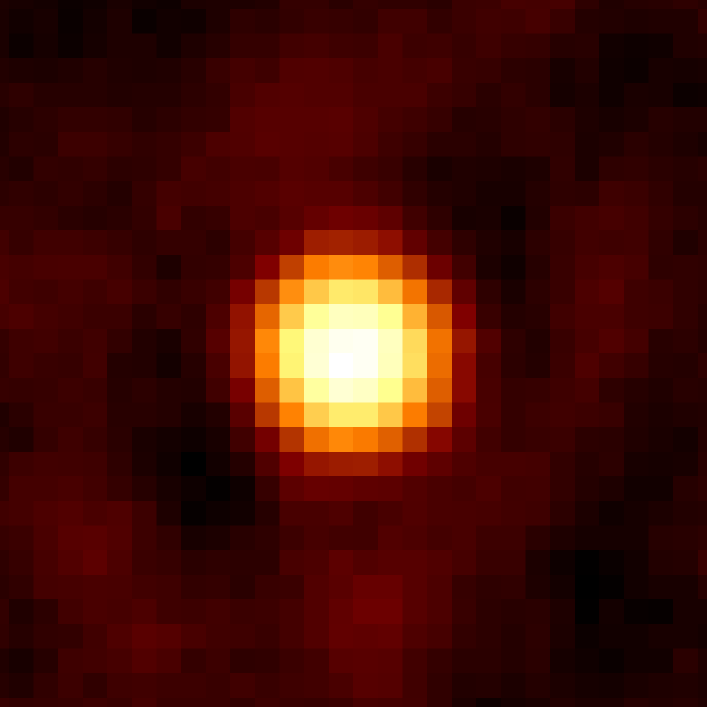}&
\includegraphics[width=.15\linewidth]{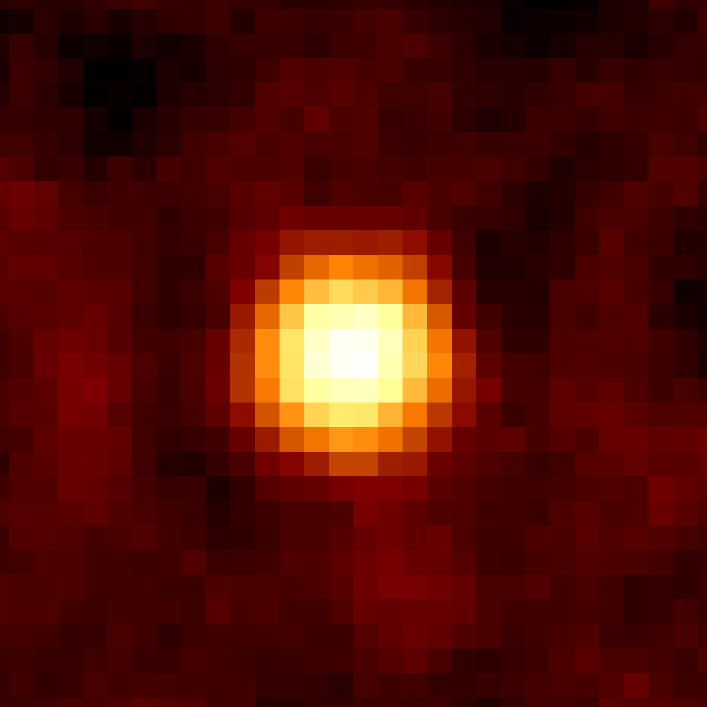}&
\includegraphics[width=.15\linewidth]{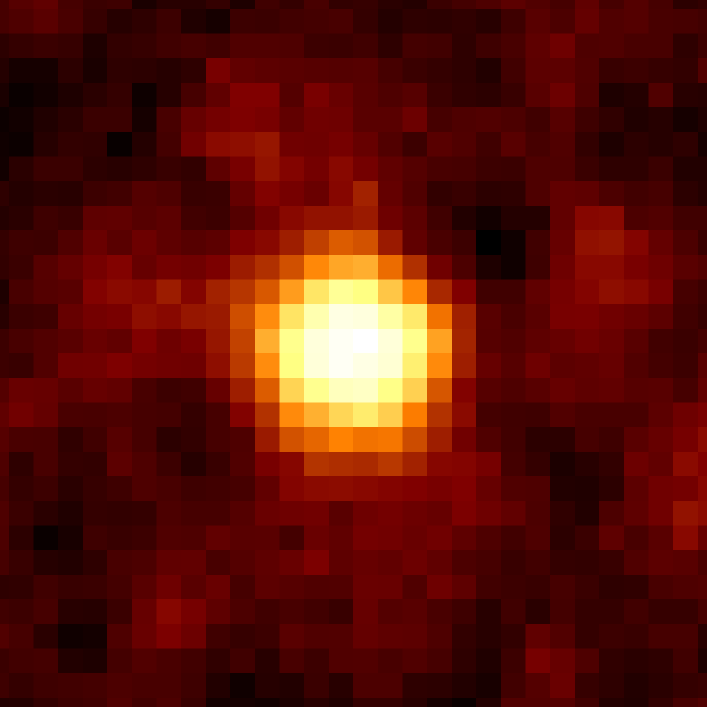}&
\includegraphics[width=.15\linewidth]{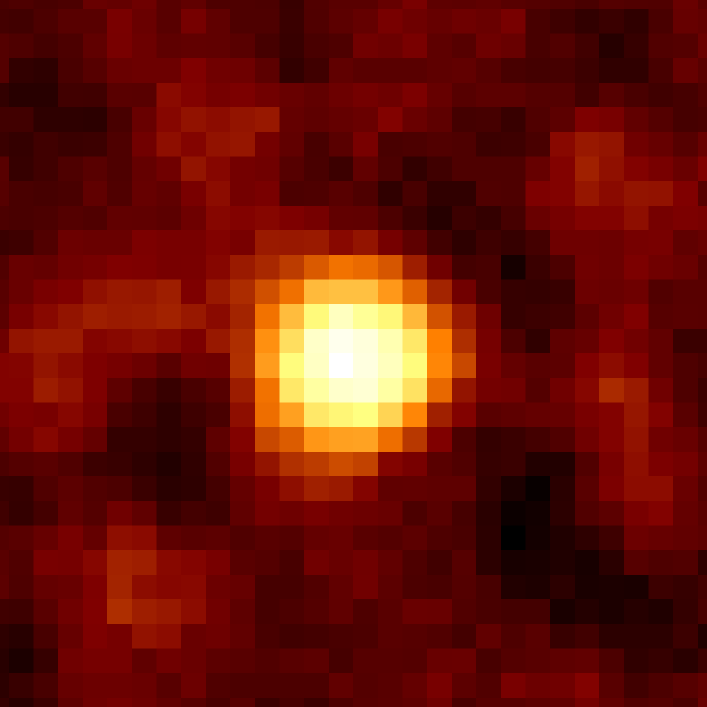}\\

\rotatebox[origin=c]{90}{\makebox[\ImageHt]{\scriptsize $\rm{10.0-10.5}$}}&
\includegraphics[width=.15\linewidth]{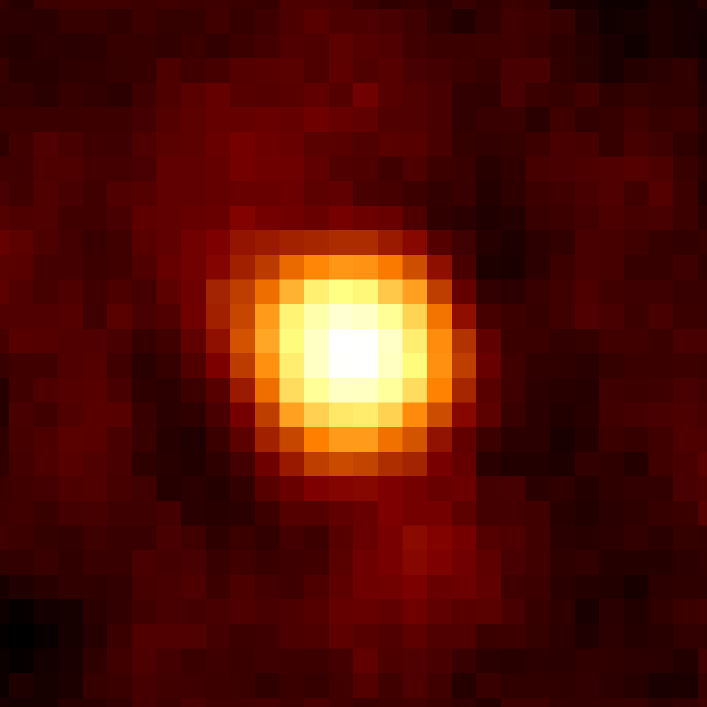}&
\includegraphics[width=.15\linewidth]{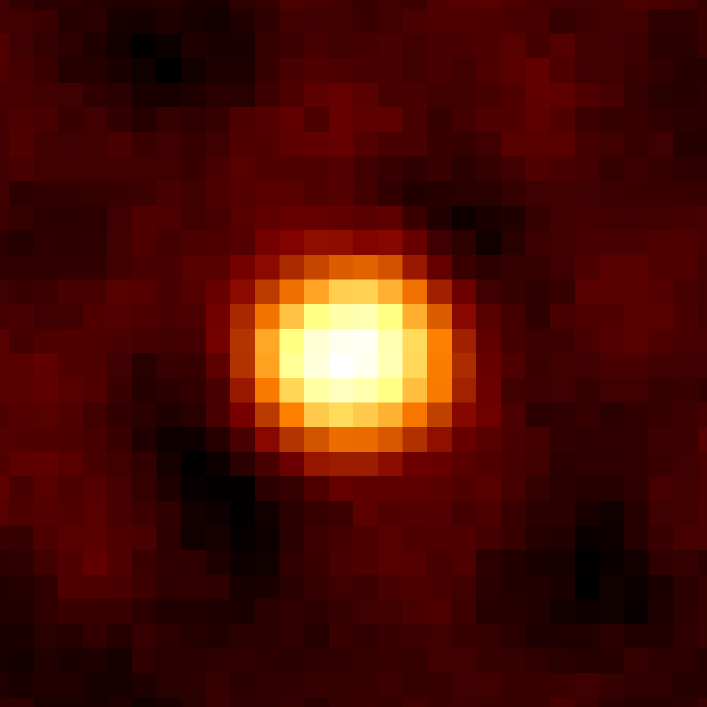}&
\includegraphics[width=.15\linewidth]{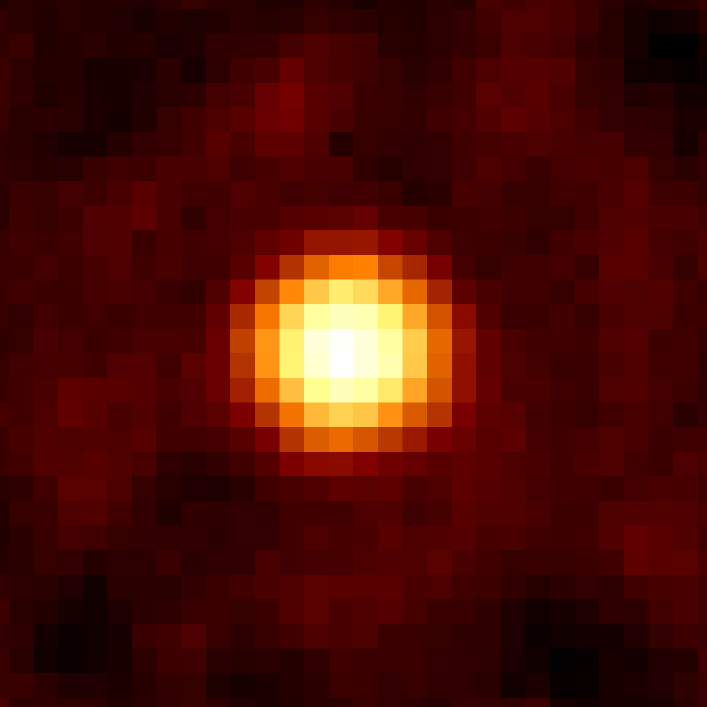}&
\includegraphics[width=.15\linewidth]{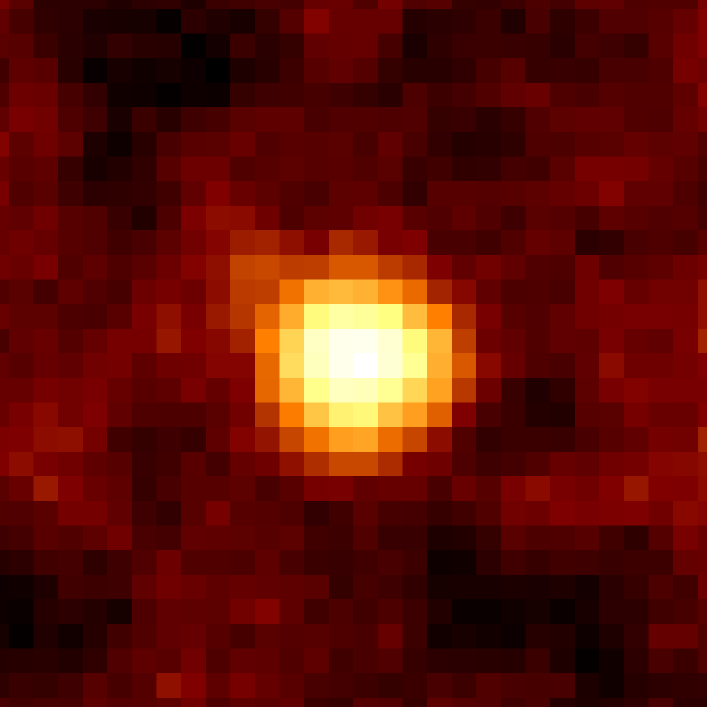}&
\includegraphics[width=.15\linewidth]{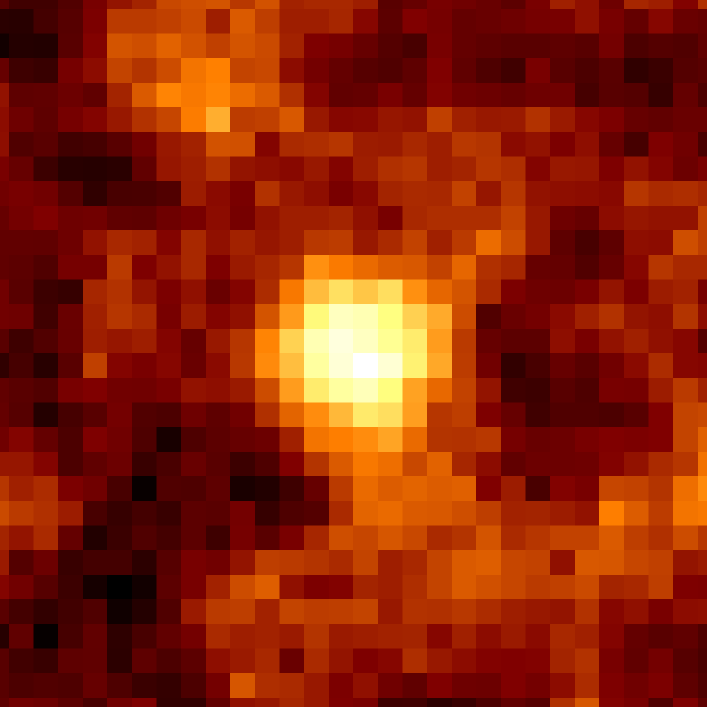}&
\includegraphics[width=.15\linewidth]{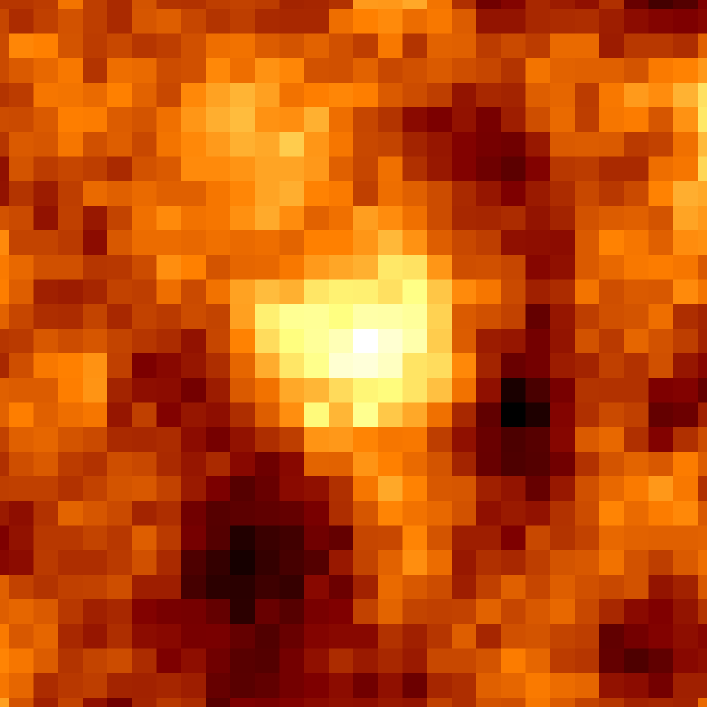}\\

\rotatebox[origin=c]{90}{\makebox[\ImageHt]{\scriptsize $\rm{9.5-10.0}$}}&
\includegraphics[width=.15\linewidth]{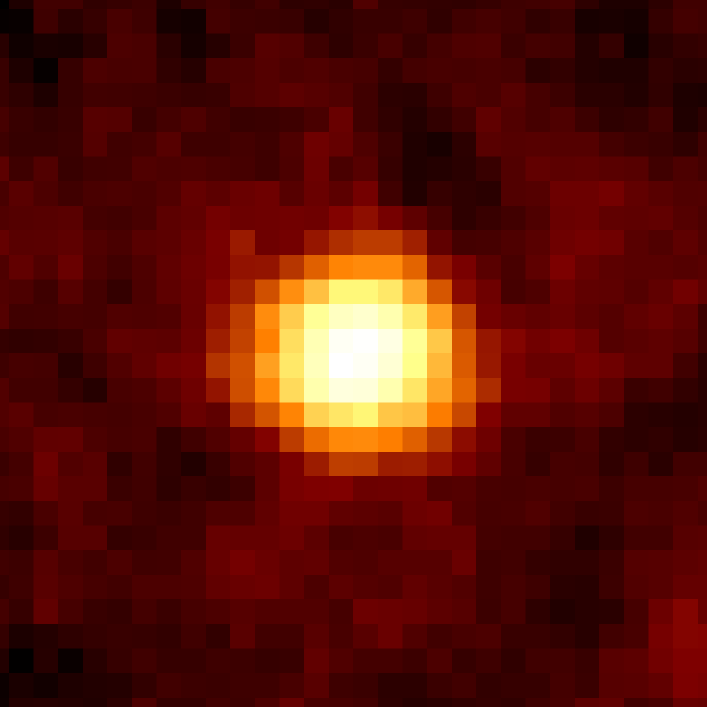}&
\includegraphics[width=.15\linewidth]{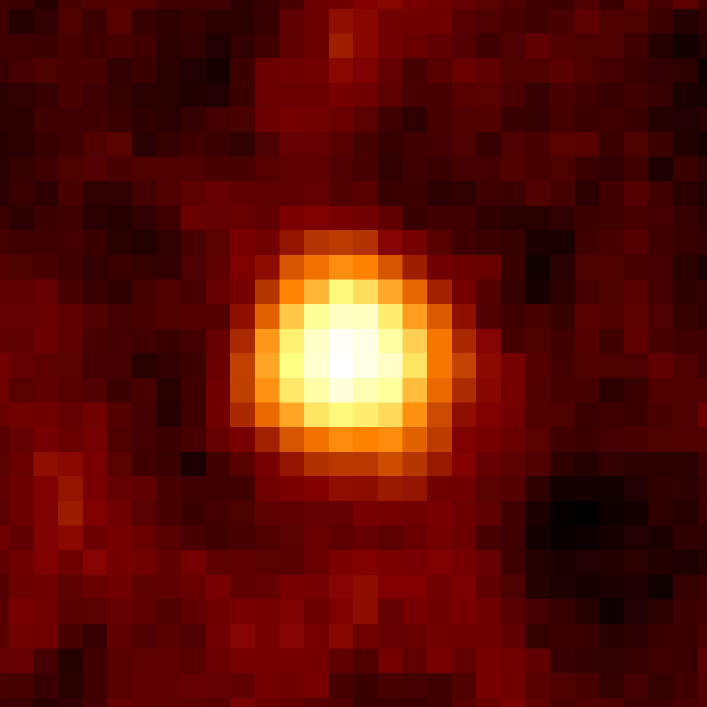}&
\includegraphics[width=.15\linewidth]{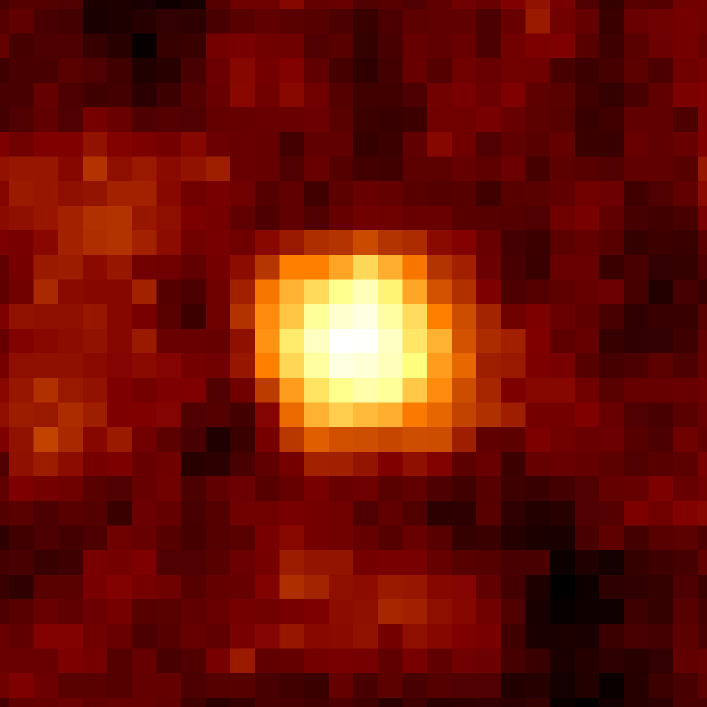}&
\includegraphics[width=.15\linewidth]{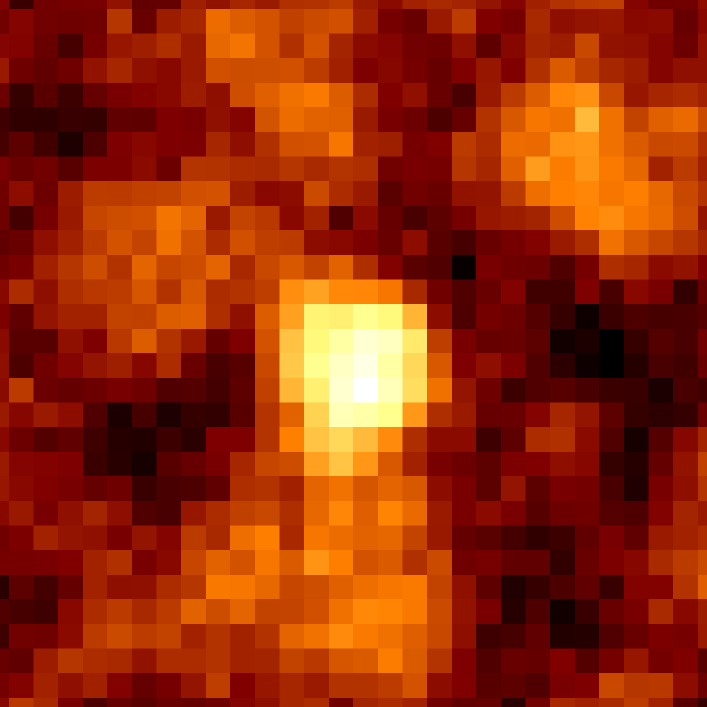}&
\includegraphics[width=.15\linewidth]{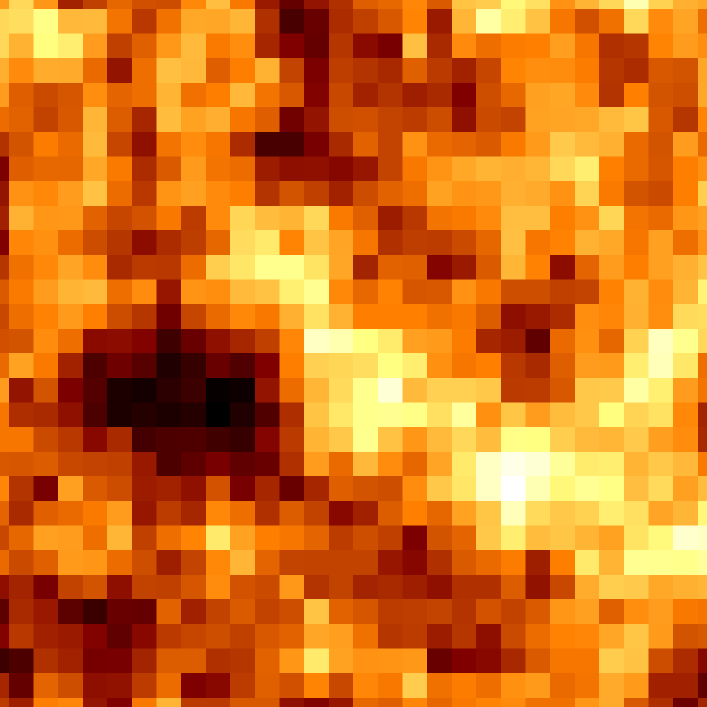}&
\includegraphics[width=.15\linewidth]{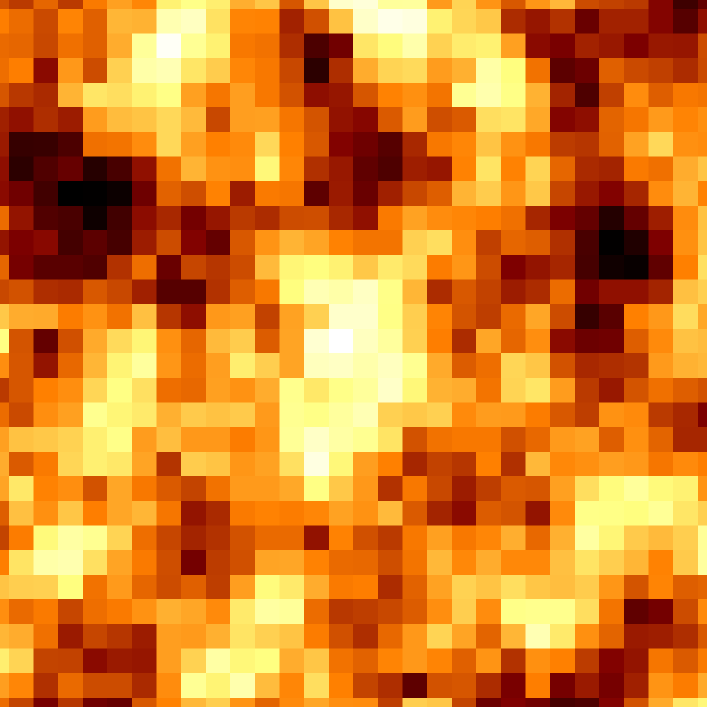}\\

\rotatebox[origin=c]{90}{\makebox[\ImageHt]{\scriptsize $\rm{9.0-9.5}$}}&
{\includegraphics[width=.15\linewidth]{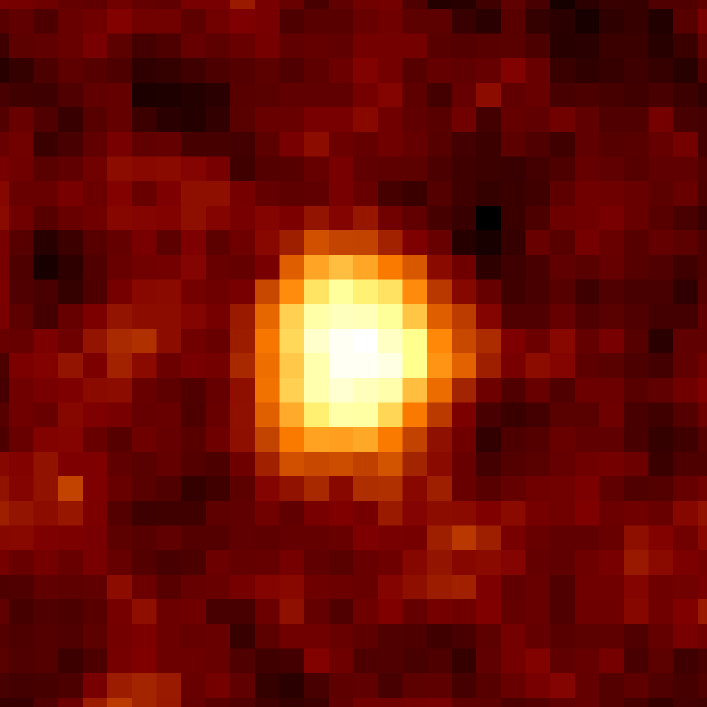}}&
\includegraphics[width=.15\linewidth]{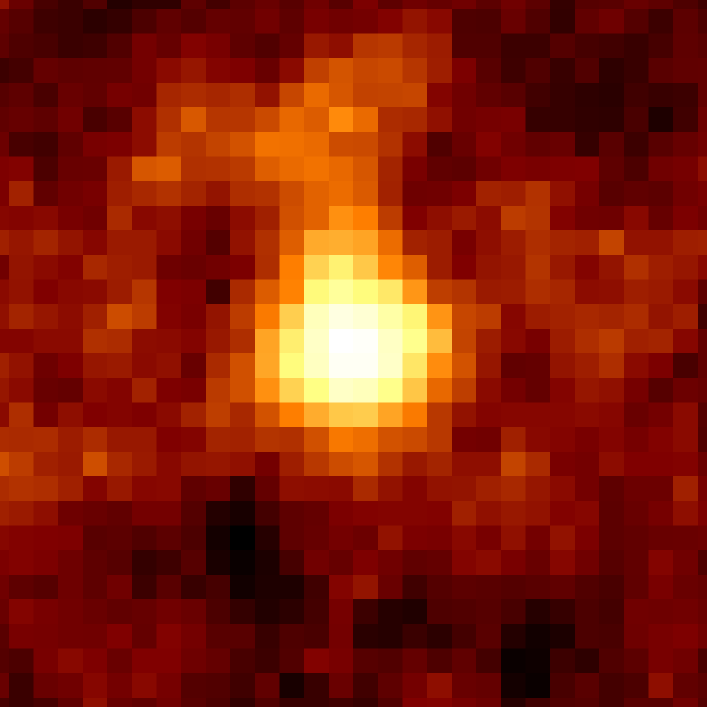}&
\includegraphics[width=.15\linewidth]{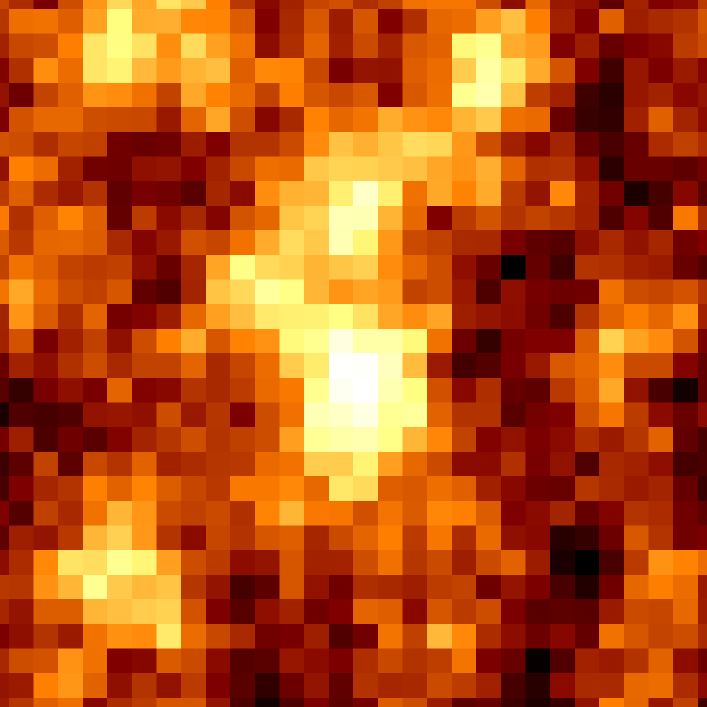}&
\includegraphics[width=.15\linewidth]{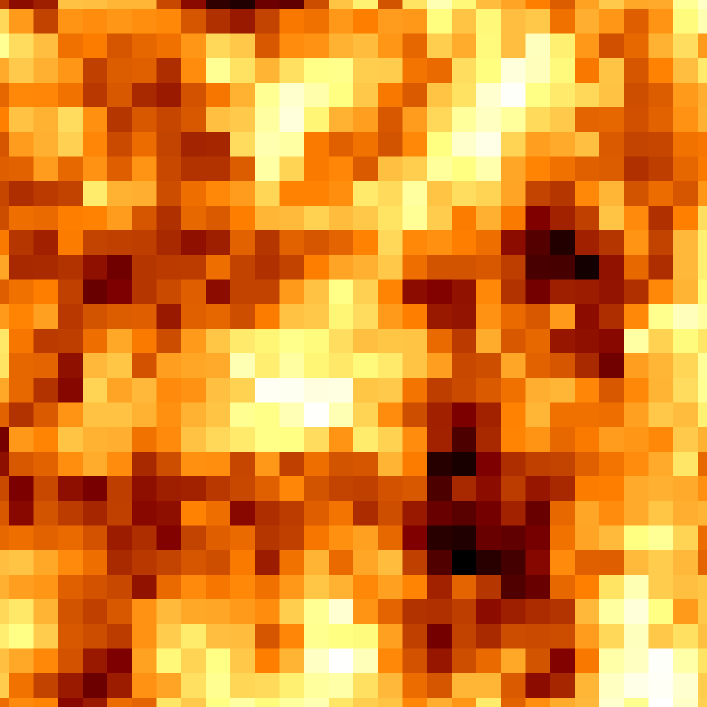}&
\includegraphics[width=.15\linewidth]{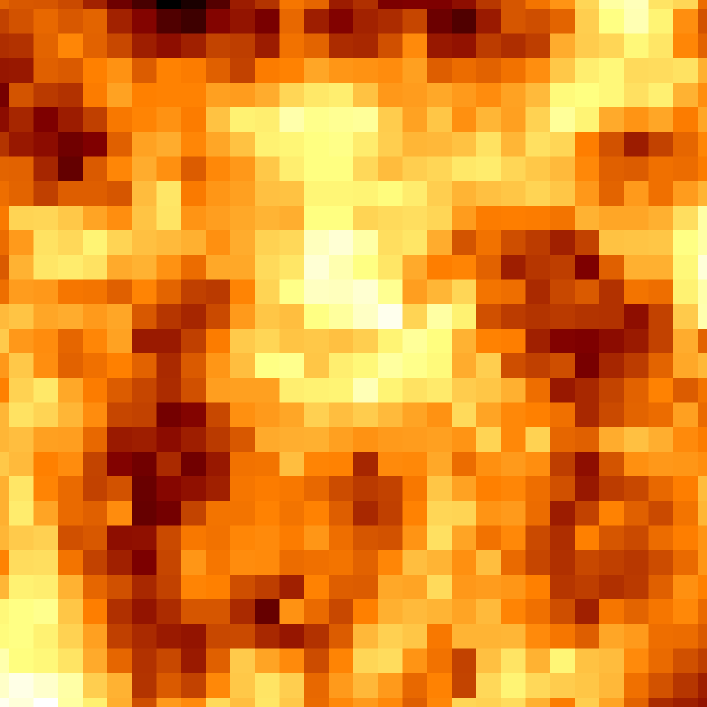}&
\includegraphics[width=.15\linewidth]{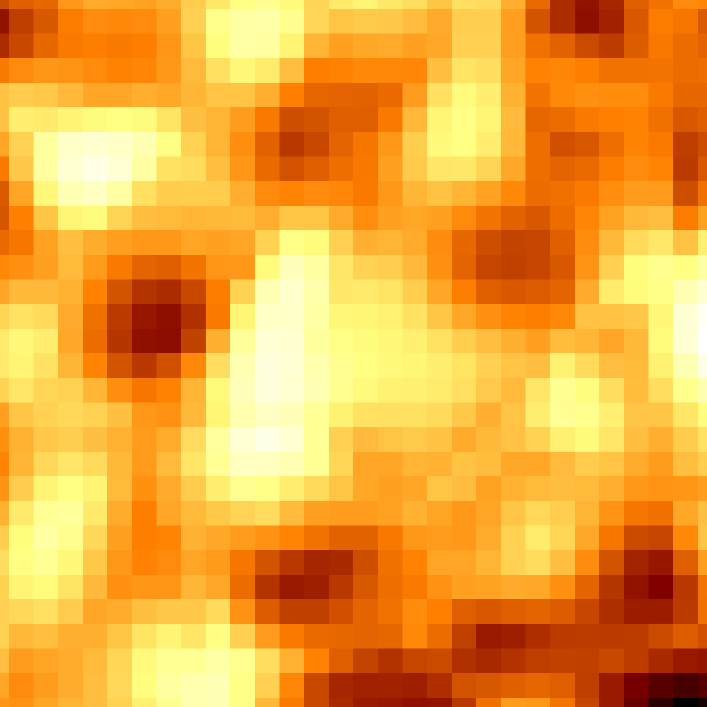}\\

\rotatebox[origin=c]{90}{\makebox[\ImageHt]{\scriptsize $\rm{8.5-9.0}$}}&
\includegraphics[width=.15\linewidth]{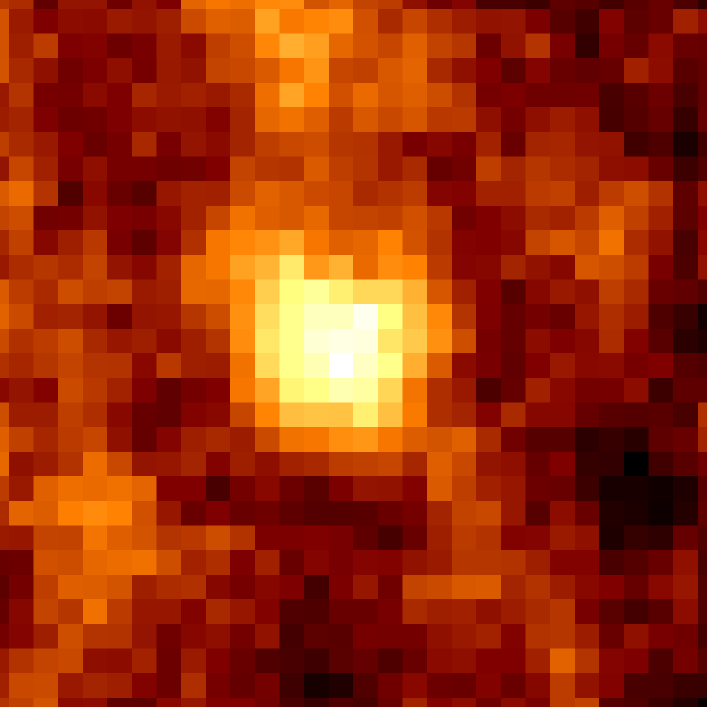}&
\includegraphics[width=.15\linewidth]{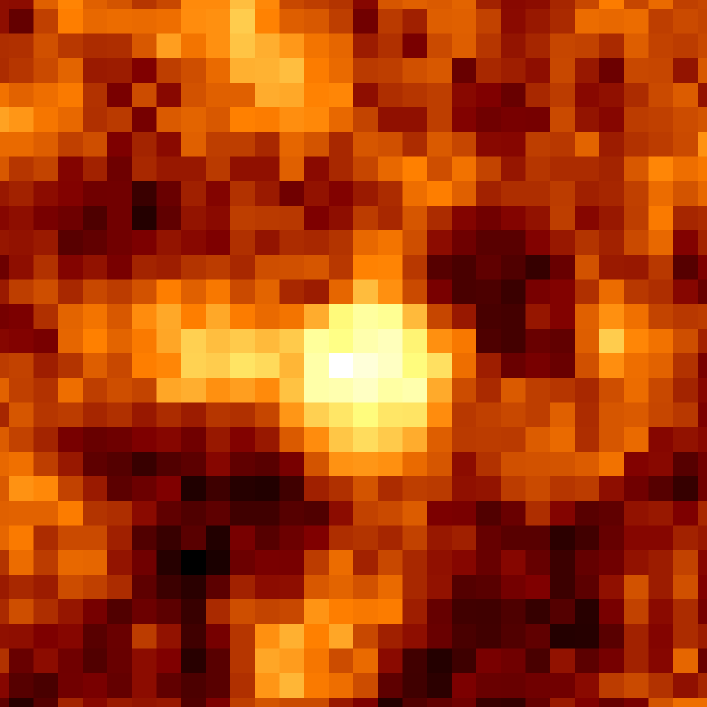}&
\includegraphics[width=.15\linewidth]{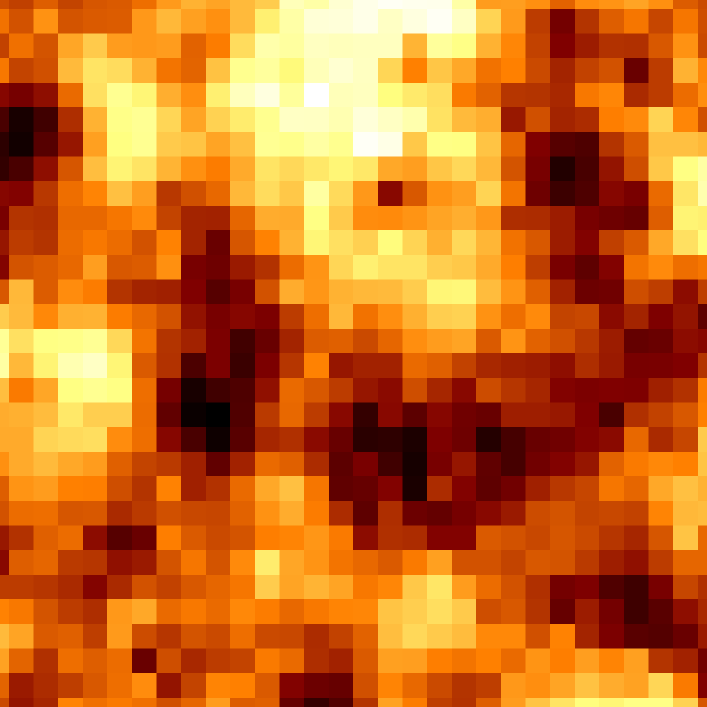}&
\includegraphics[width=.15\linewidth]{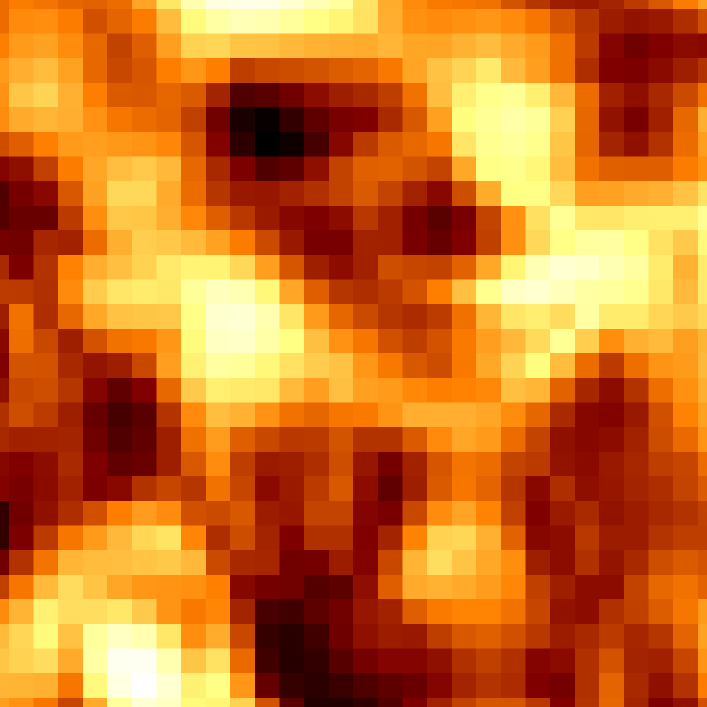}&
\includegraphics[width=.15\linewidth]{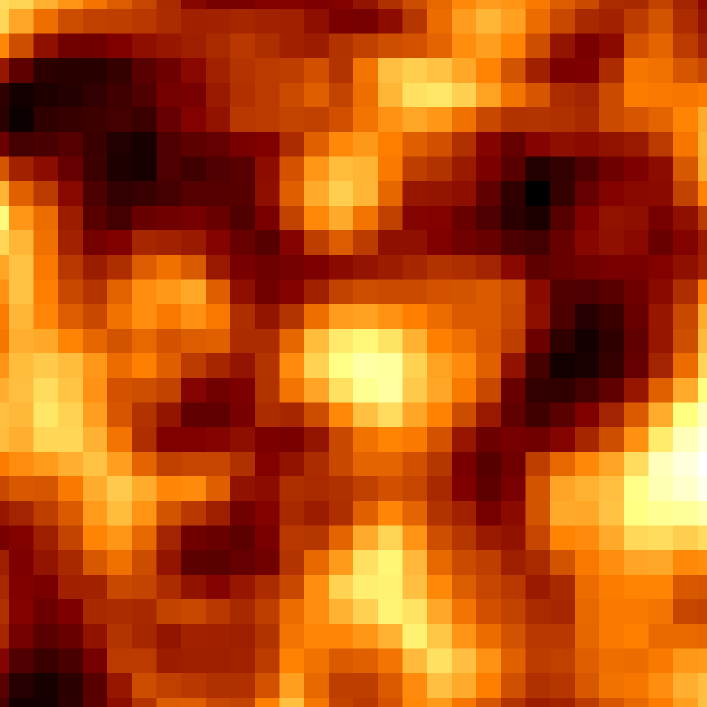}&
\includegraphics[width=.15\linewidth]{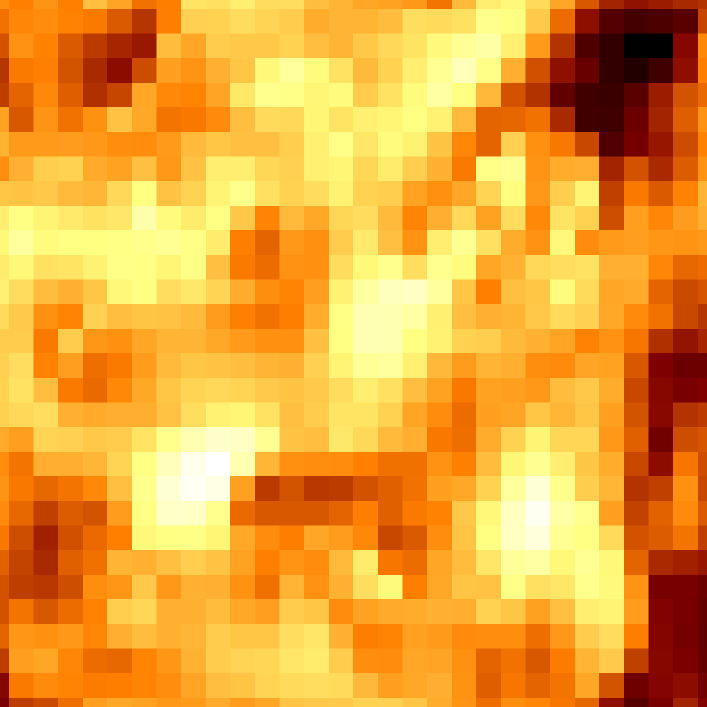}\\

\end{tabular}
\caption{Stacked images (GMRT, 610 MHz) of total intensity for all galaxies. The columns indicate the median stacked 610 MHz total intensity radio images for total galaxies within the range $\rm{\textit{z}\in[0.1-0.3]}$, $\rm{[0.3-0.5]}$, $\rm{[0.5-0.7]}$, $\rm{[0.7-0.9]}$, $\rm{[0.9-1.1]}$, $\rm{[1.1-1.5]}$ for the K-band magnitude mass selected sample. All images have a size of $\rm{\sim 36\times36}$ arcsec$^2$. The rows indicate mass range, $\rm{\textit{M}_{\star}\in[11.0-12.4]}$, $\rm{[10.5-11.0]}$, $\rm{[10.0-10.5]}$, $\rm{[9.5-10.0]}$, $\rm{[9.0-9.5]}$, $\rm{[8.5-9.0]}$ respectively, from top to bottom. All image-scale ranges between 1 and 100 $\rm{\mu}$Jy beam$^{-1}$}. 

\label{all_galaxies_image.fig}
\end{figure}

\begin{figure}
\begin{tabular}{@{}c@{ }c@{ }c@{ }c@{ }c@{ }c@{ }c@{}}
&\textbf{\textit{z}=0.1-0.3} & \textbf{\textit{z}=0.3-0.5} & \textbf{\textit{z}=0.5-0.7} & \textbf{\textit{z}=0.7-0.9} & \textbf{\textit{z}=0.9-1.1} & \textbf{\textit{z}=1.1-1.5}\\
\centering
\rotatebox[origin=b]{90}{\makebox[\ImageHt]{\scriptsize $\rm{11.0-12.4}$}}&
\includegraphics[width=.15\linewidth]{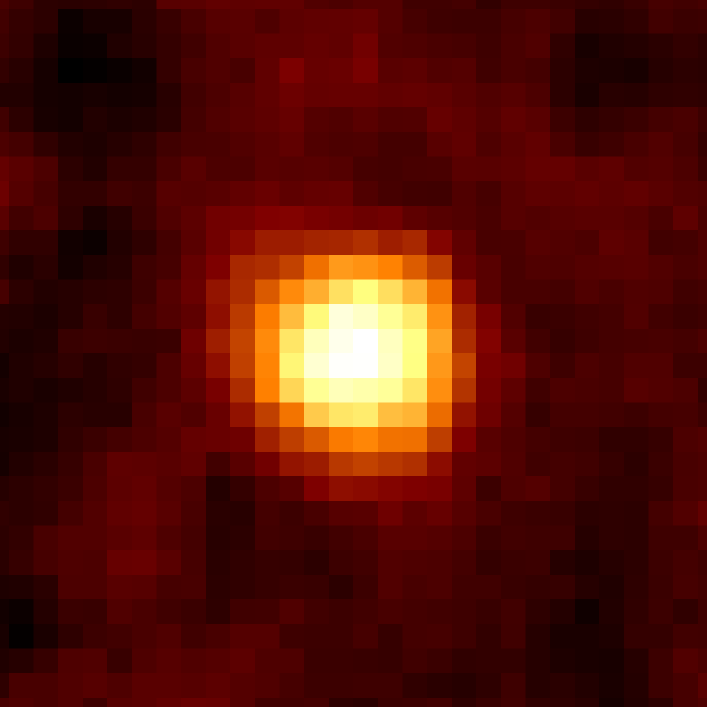}&
\includegraphics[width=.15\linewidth]{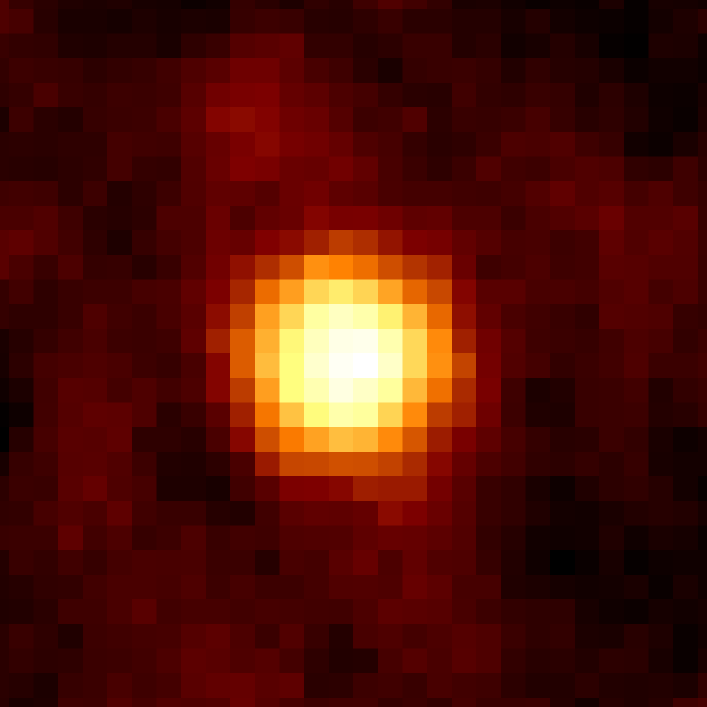}&
\includegraphics[width=.15\linewidth]{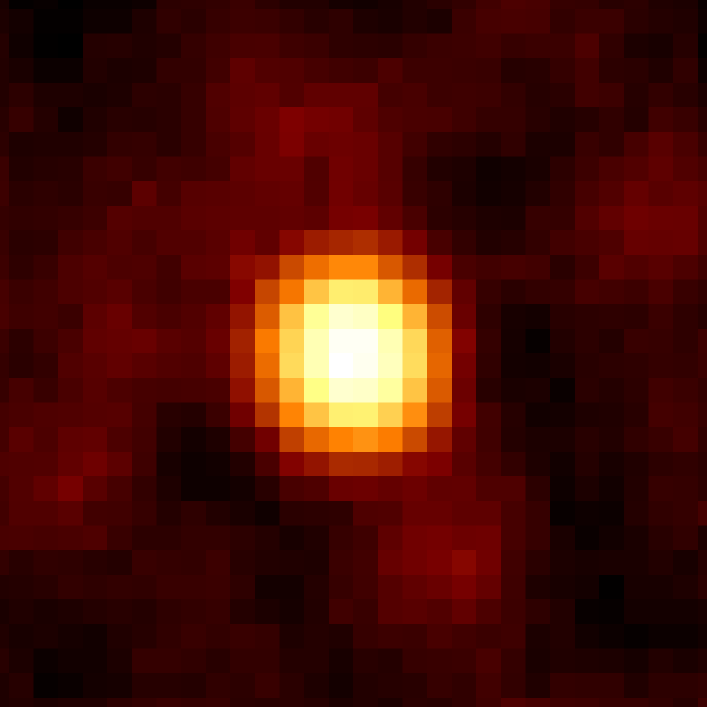}&
\includegraphics[width=.15\linewidth]{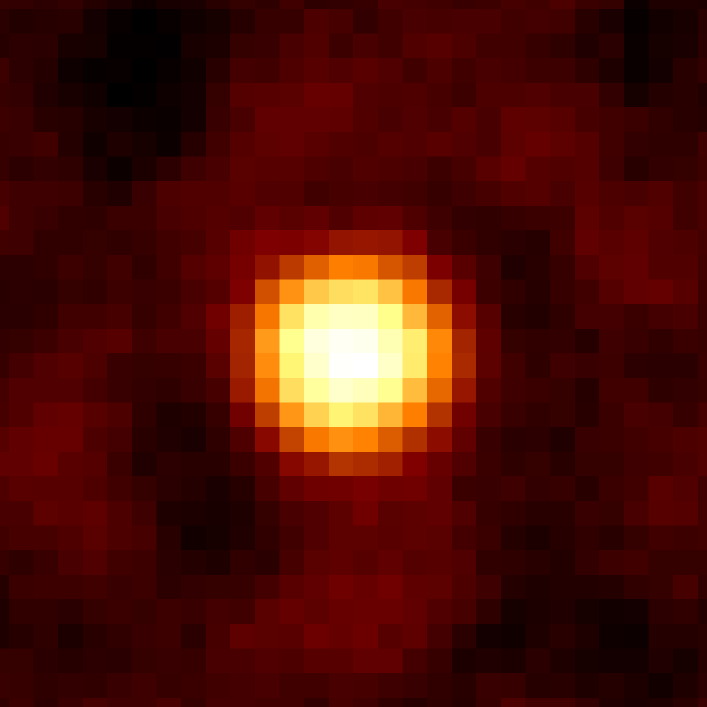}&
\includegraphics[width=.15\linewidth]{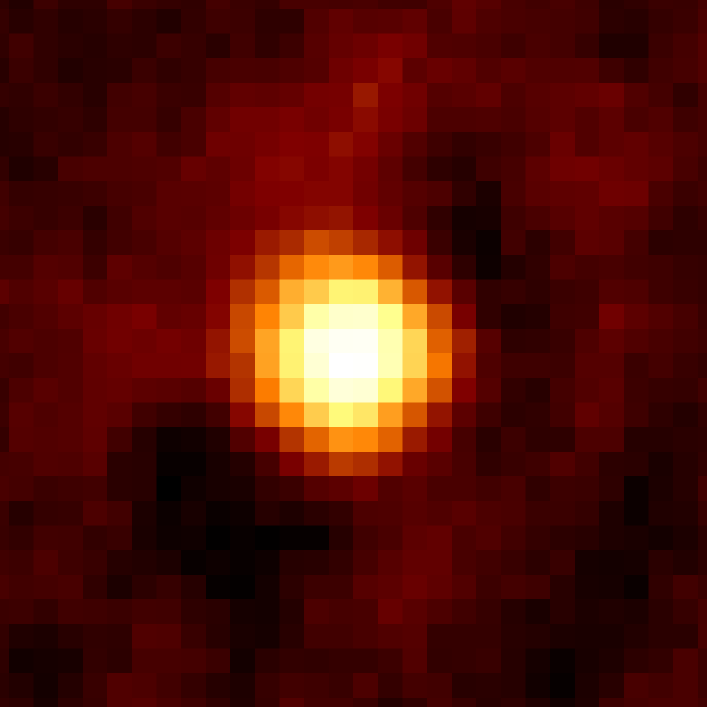}&
\includegraphics[width=.15\linewidth]{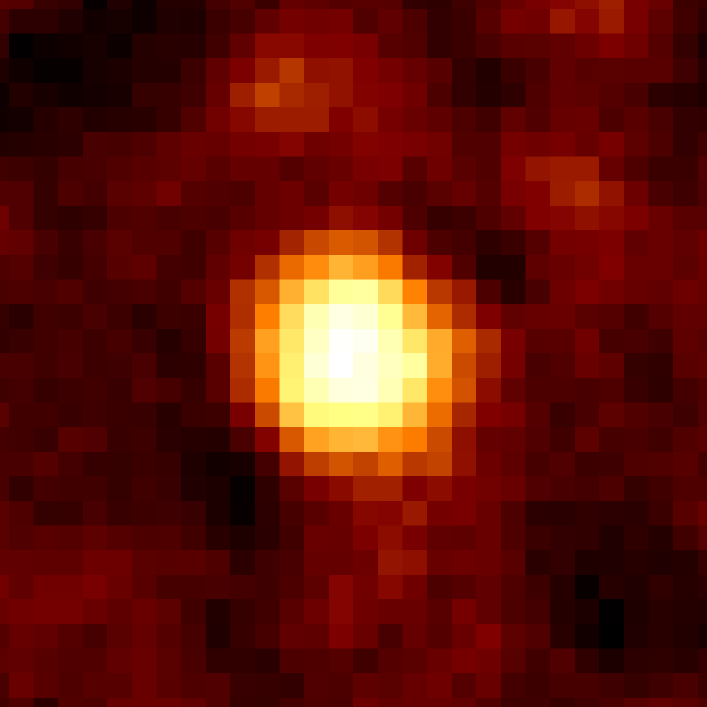}\\

\rotatebox[origin=b]{90}{\makebox[\ImageHt]{\scriptsize $\rm{10.5-11.0}$}}&
\includegraphics[width=.15\linewidth]{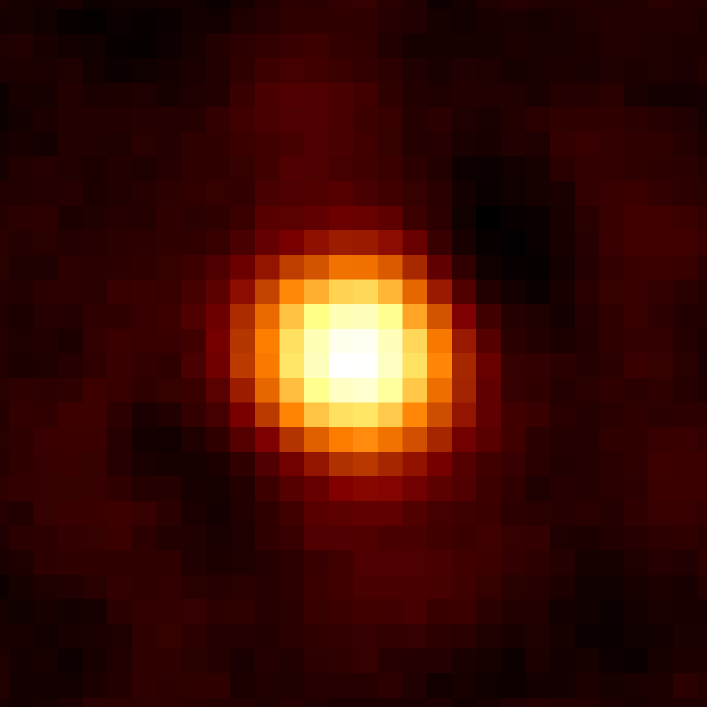}&
\includegraphics[width=.15\linewidth]{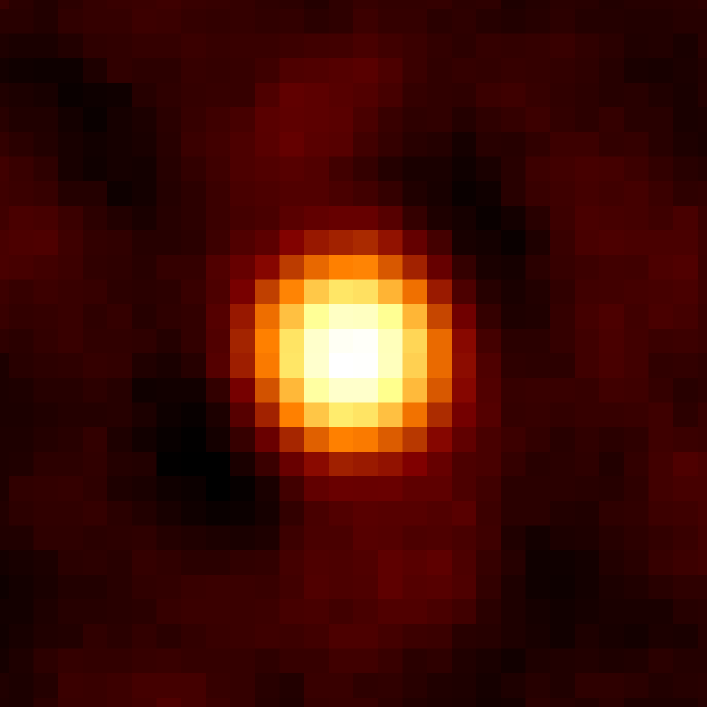}&
\includegraphics[width=.15\linewidth]{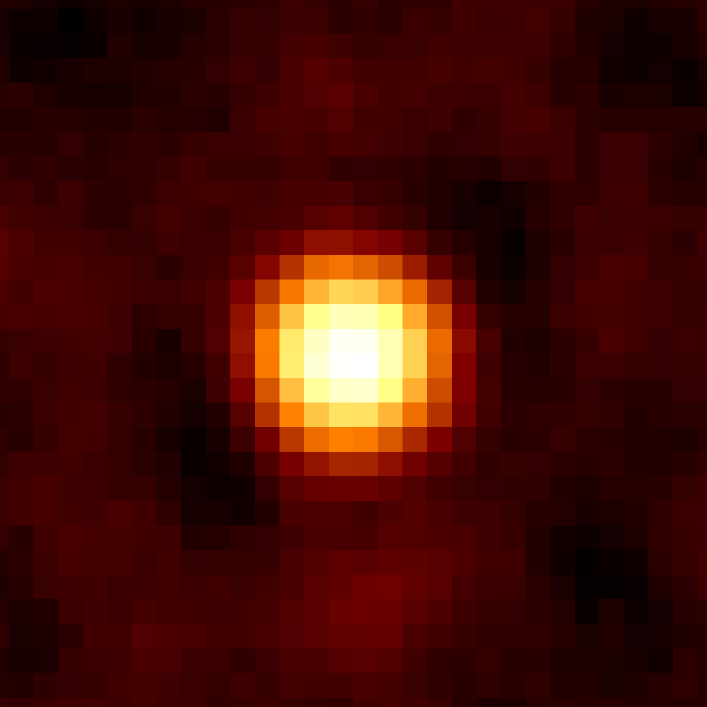}&
\includegraphics[width=.15\linewidth]{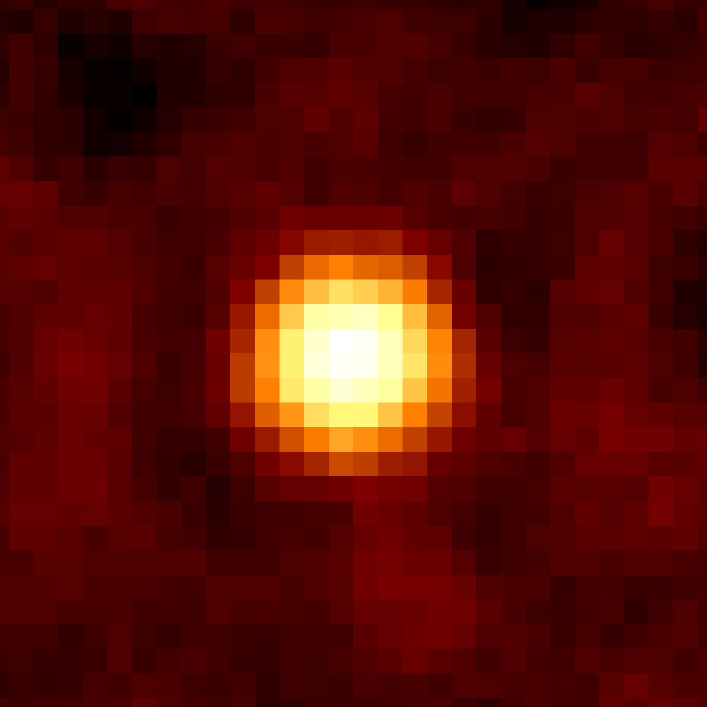}&
\includegraphics[width=.15\linewidth]{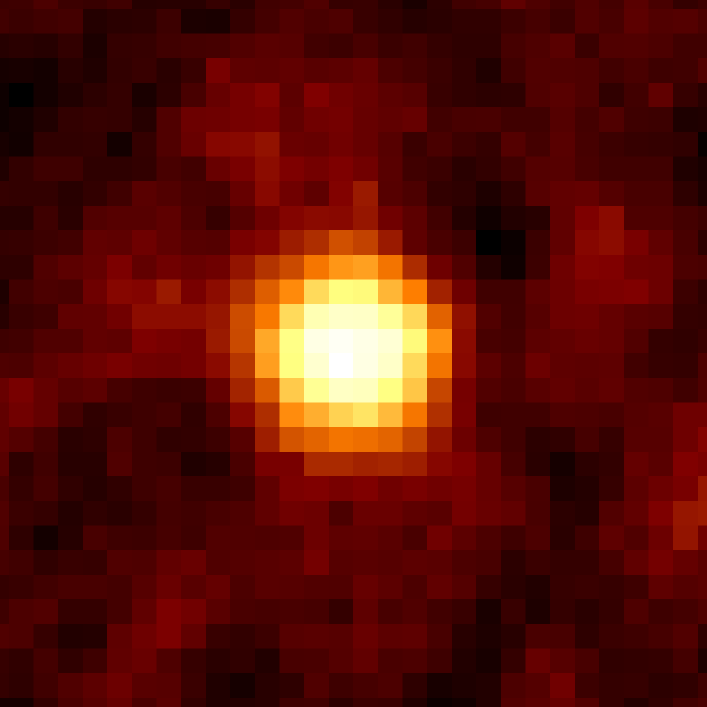}&
\includegraphics[width=.15\linewidth]{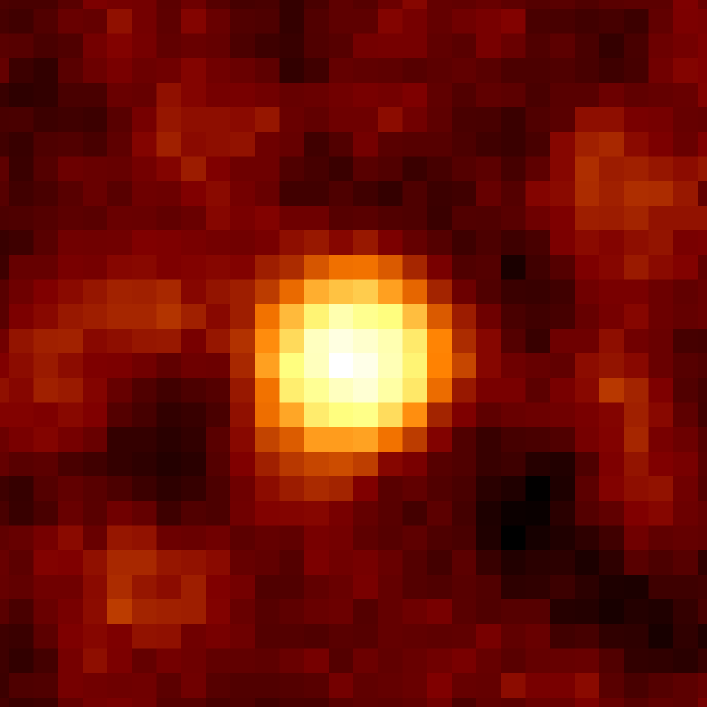}\\

\rotatebox[origin=b]{90}{\makebox[\ImageHt]{\scriptsize $\rm{10.0-10.5}$}}&
\includegraphics[width=.15\linewidth]{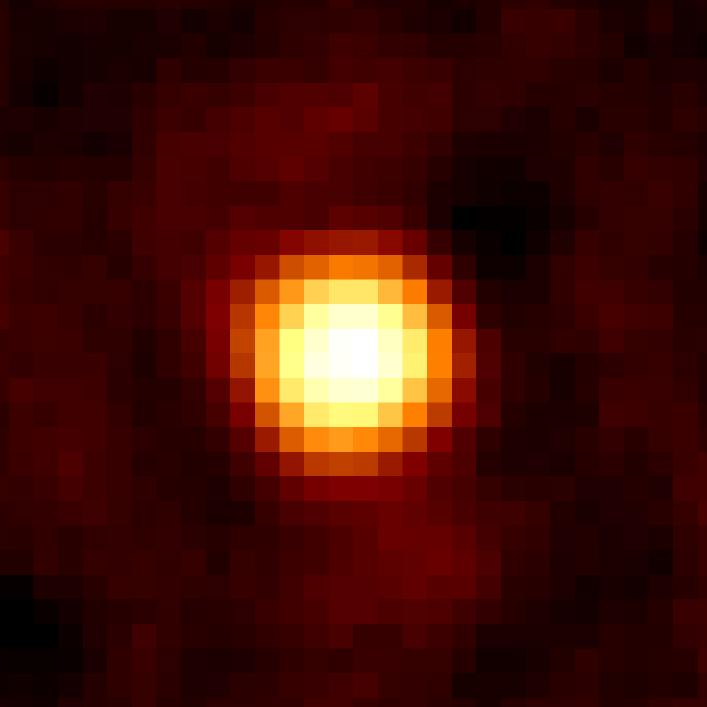}&
\includegraphics[width=.15\linewidth]{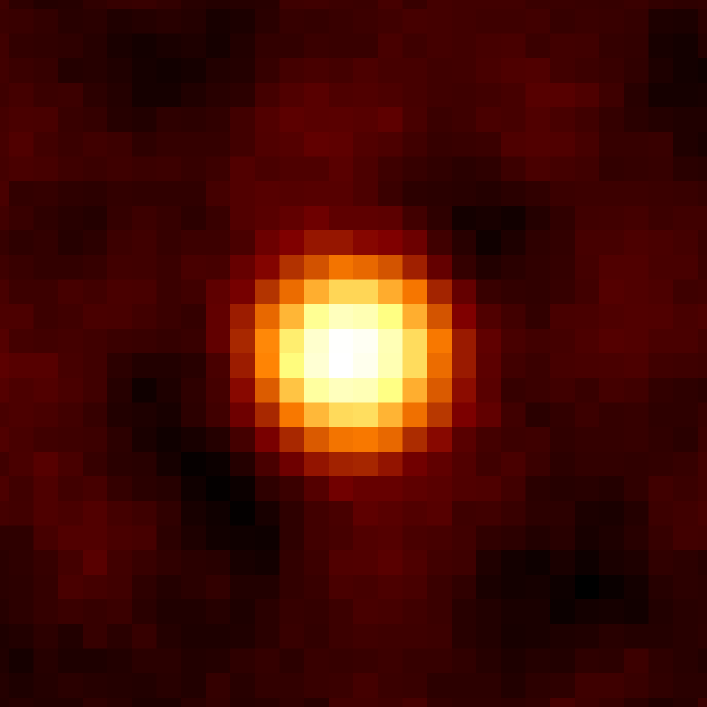}&
\includegraphics[width=.15\linewidth]{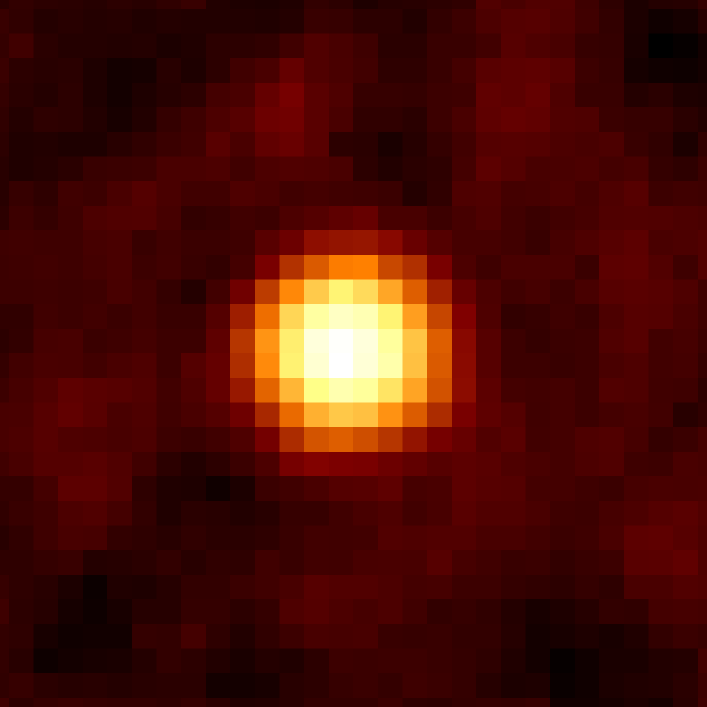}&
\includegraphics[width=.15\linewidth]{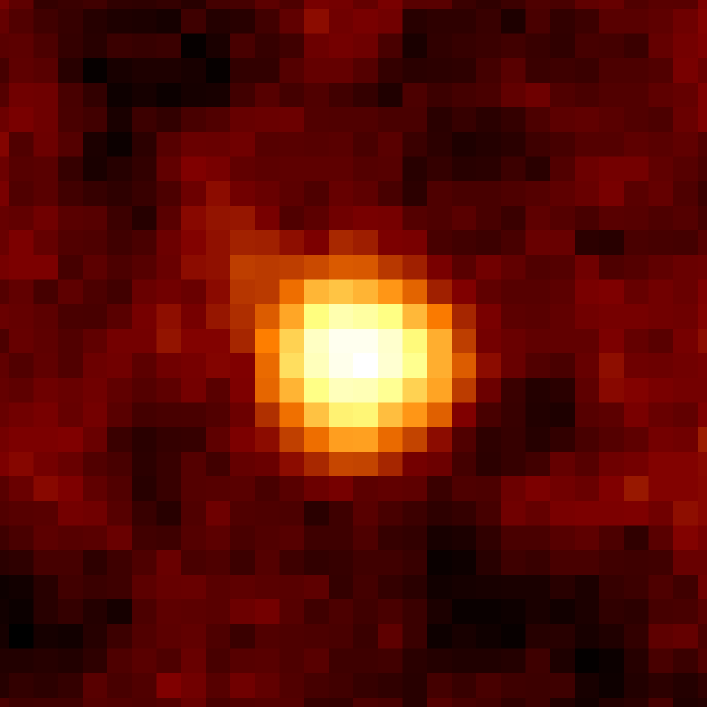}&
\includegraphics[width=.15\linewidth]{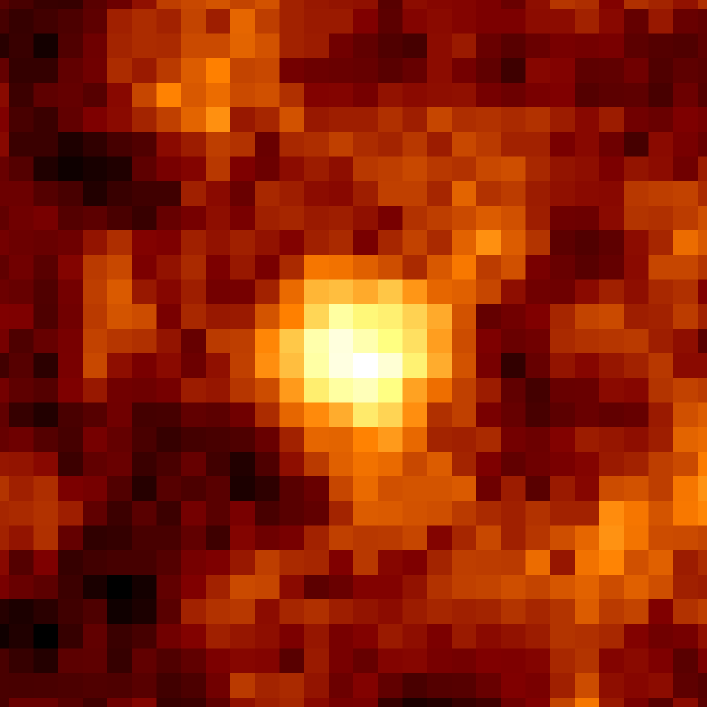}&
\includegraphics[width=.15\linewidth]{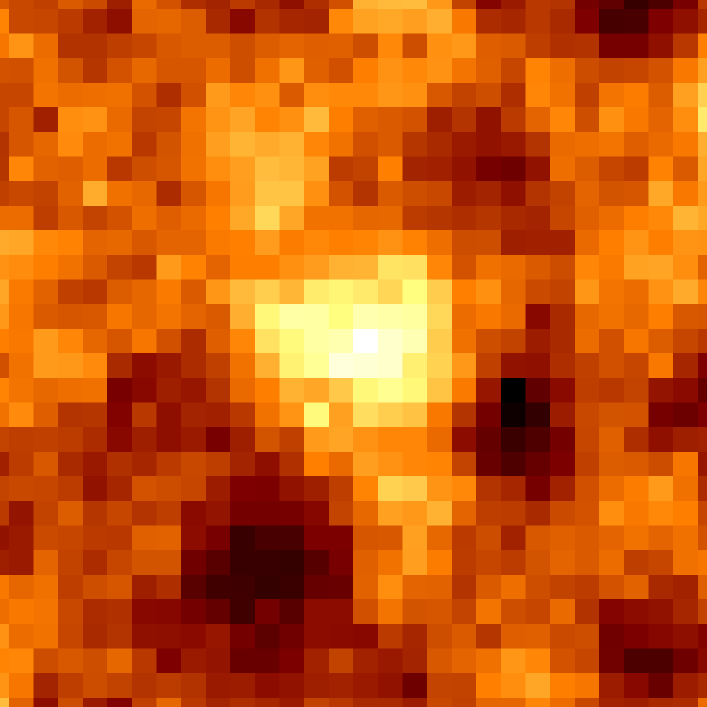}\\

\rotatebox[origin=b]{90}{\makebox[\ImageHt]{\scriptsize $\rm{9.5-10.0}$}}&
\includegraphics[width=.15\linewidth]{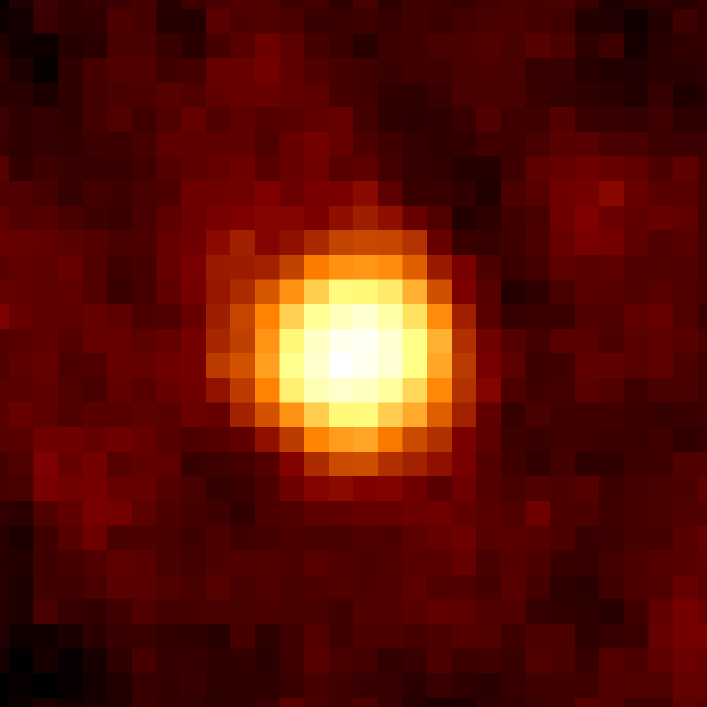}&
\includegraphics[width=.15\linewidth]{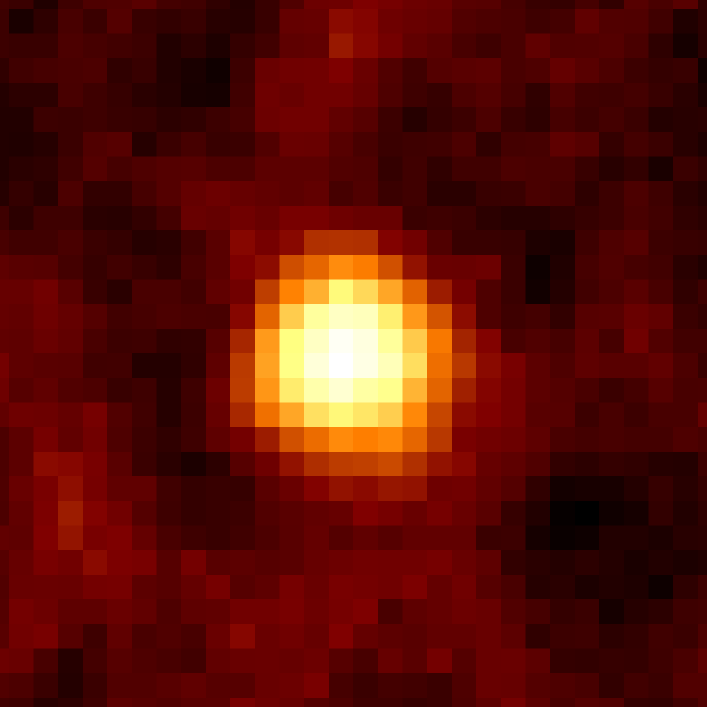}&
\includegraphics[width=.15\linewidth]{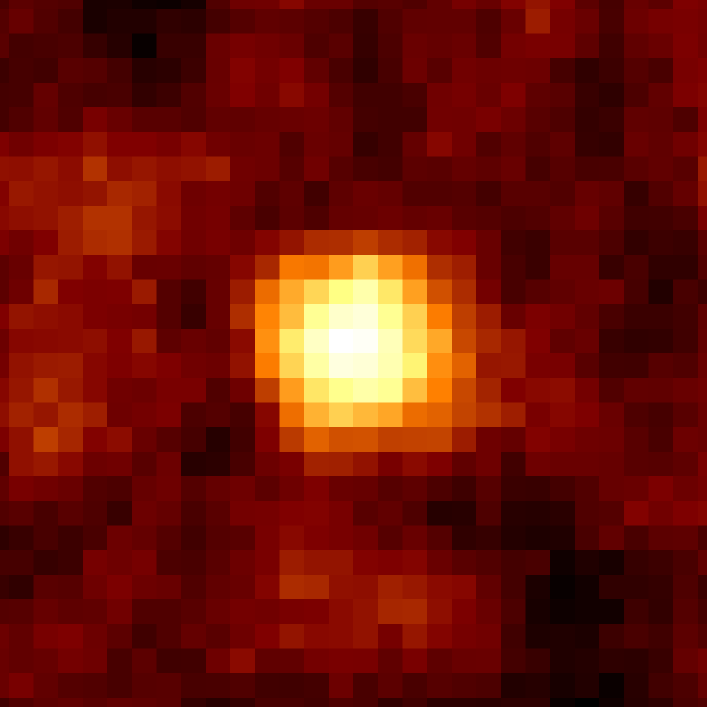}&
\includegraphics[width=.15\linewidth]{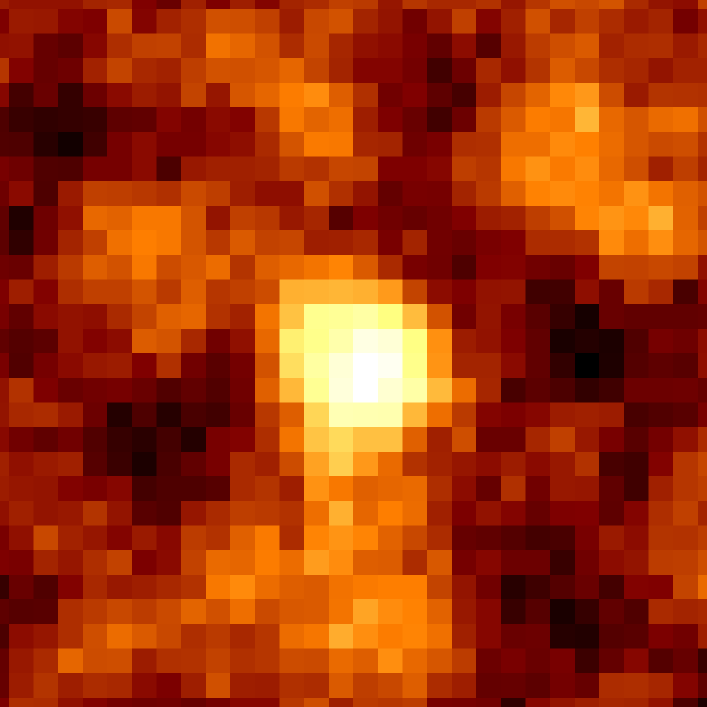}&
\includegraphics[width=.15\linewidth]{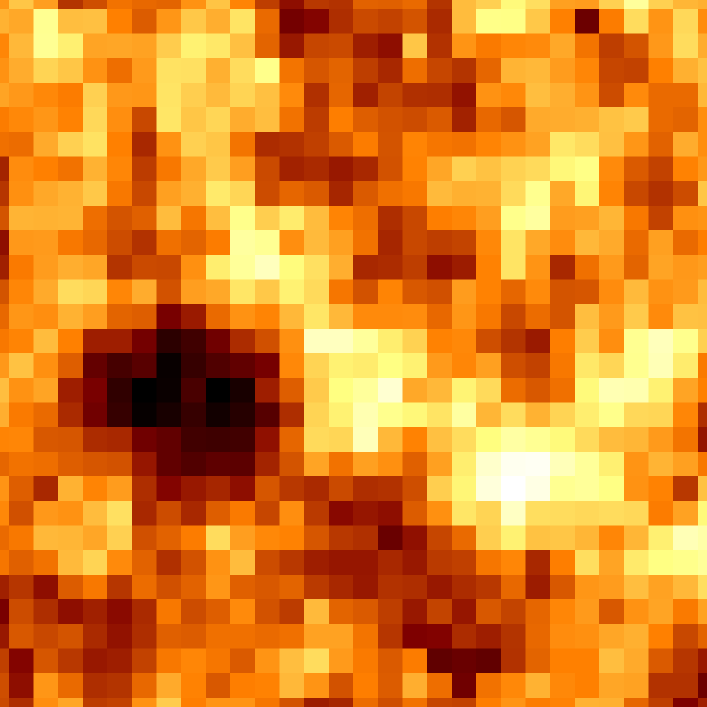}&
\includegraphics[width=.15\linewidth]{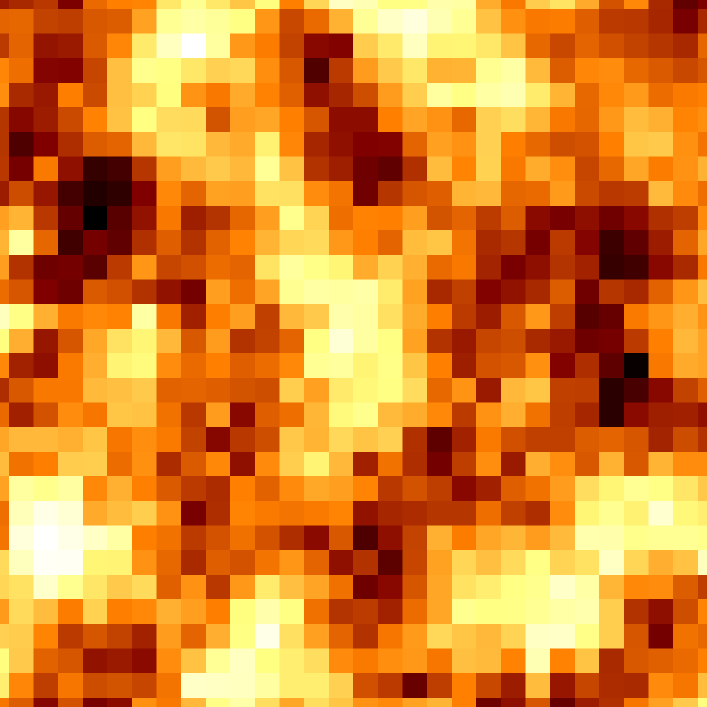}\\

\rotatebox[origin=b]{90}{\makebox[\ImageHt]{\scriptsize $\rm{9.0-9.5}$}}&
\includegraphics[width=.15\linewidth]{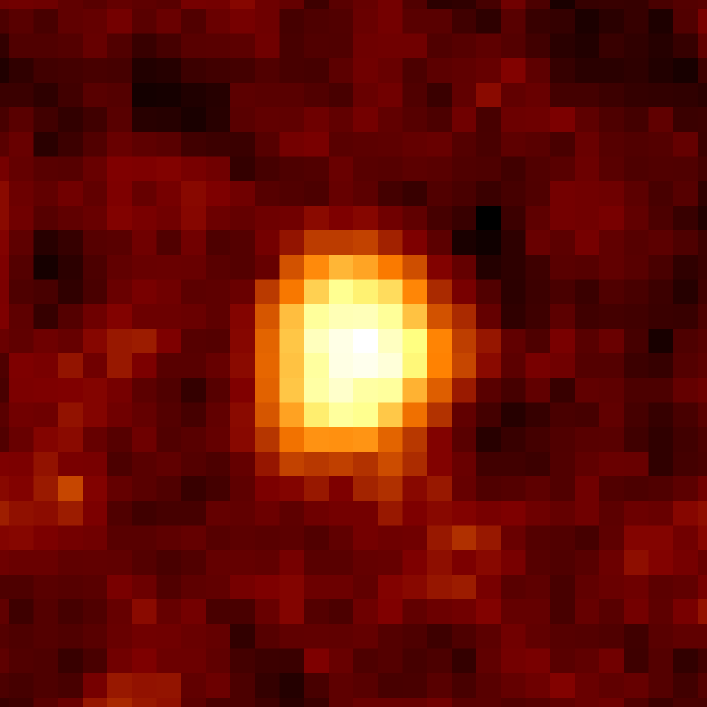}&
\includegraphics[width=.15\linewidth]{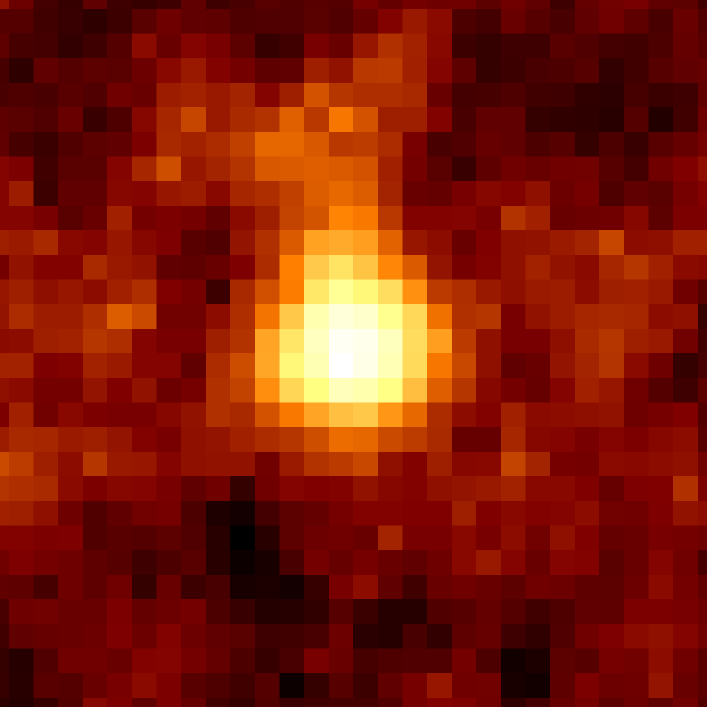}&
\includegraphics[width=.15\linewidth]{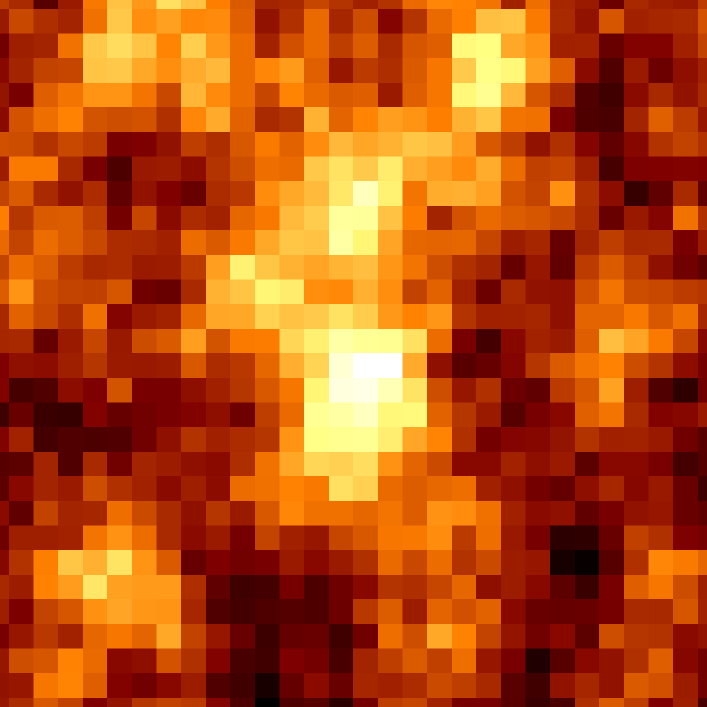}&
\includegraphics[width=.15\linewidth]{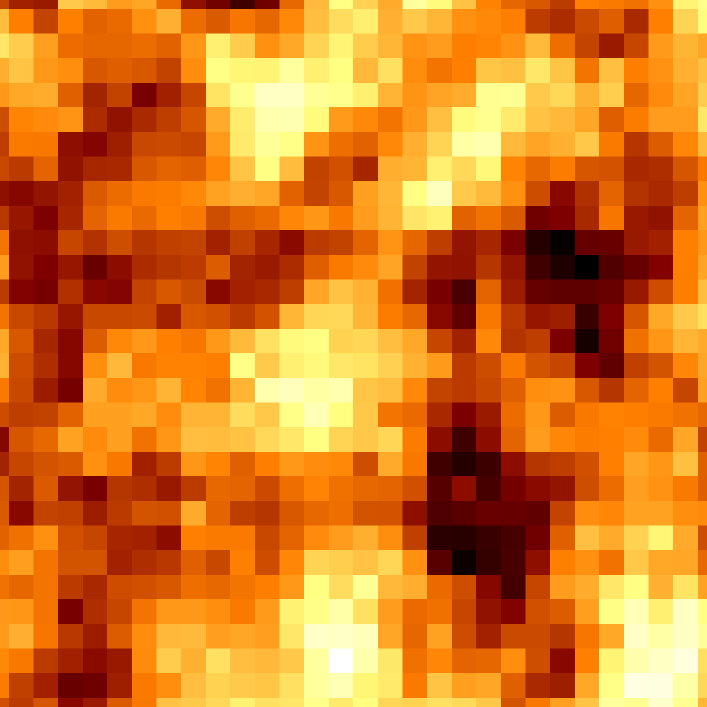}&
\includegraphics[width=.15\linewidth]{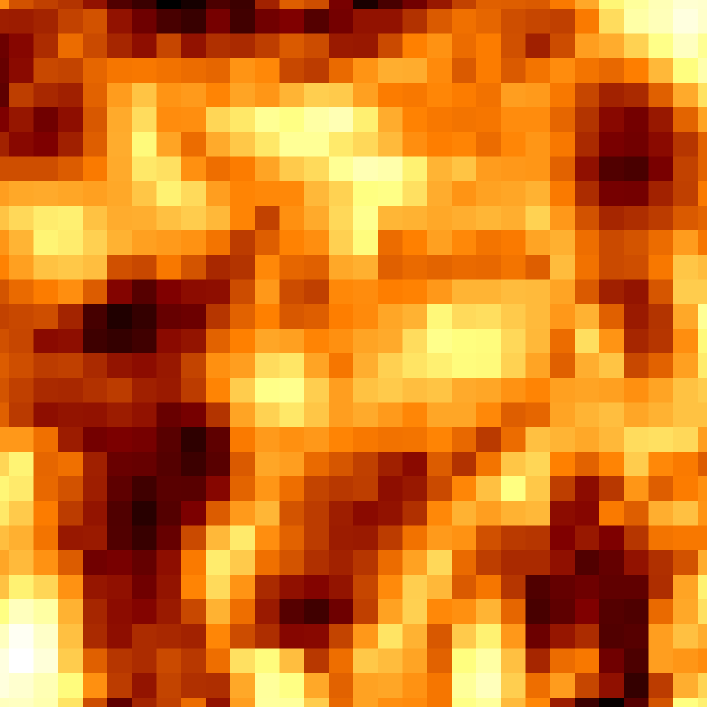}&
\includegraphics[width=.15\linewidth]{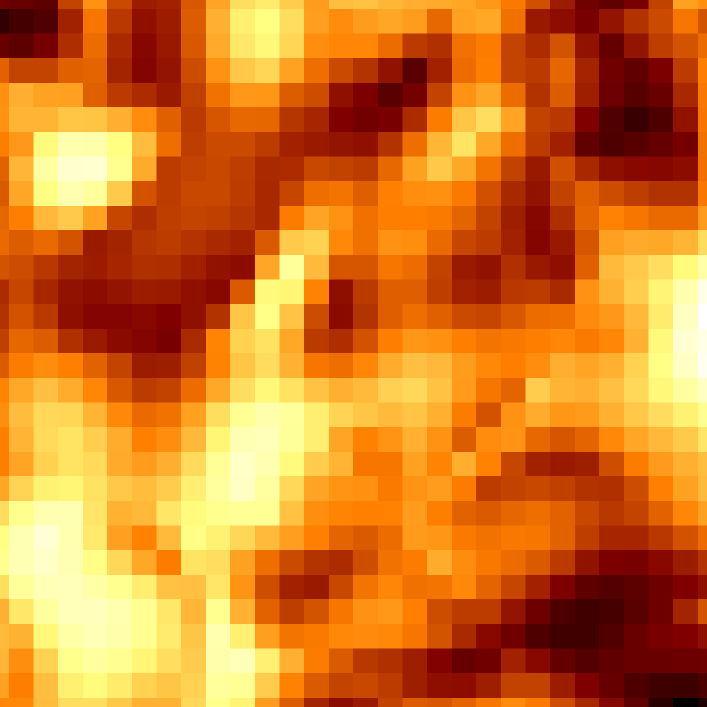}\\

\rotatebox[origin=b]{90}{\makebox[\ImageHt]{\scriptsize $\rm{8.5-9.0}$}}&
\includegraphics[width=.15\linewidth]{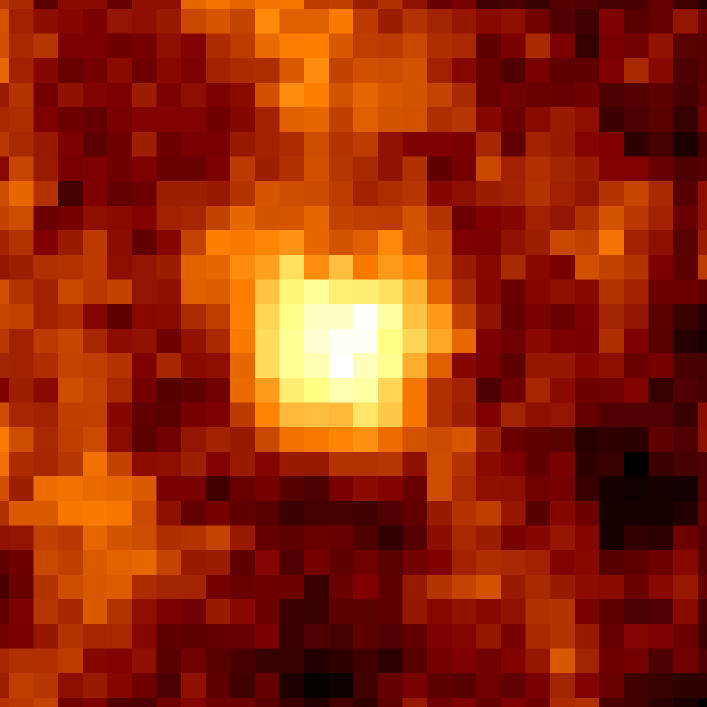}&
\includegraphics[width=.15\linewidth]{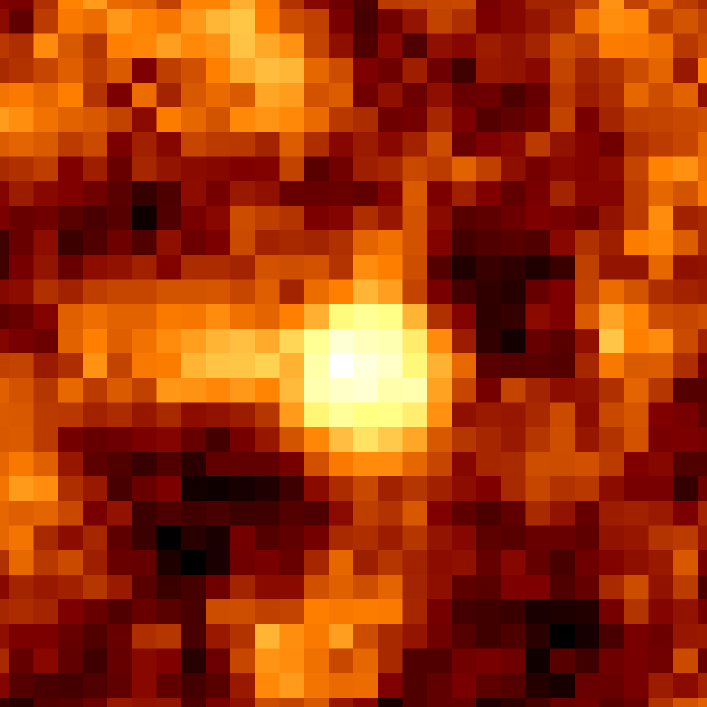}&
\includegraphics[width=.15\linewidth]{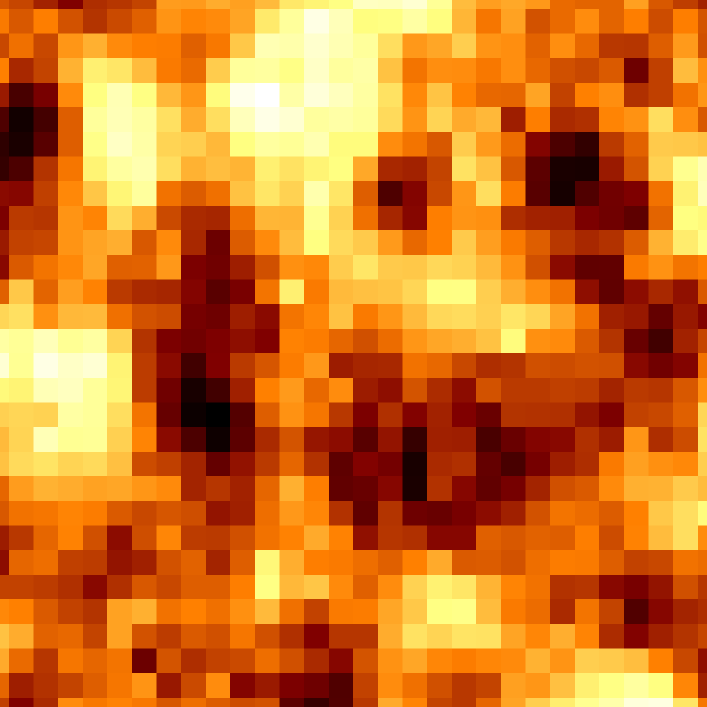}&
\includegraphics[width=.15\linewidth]{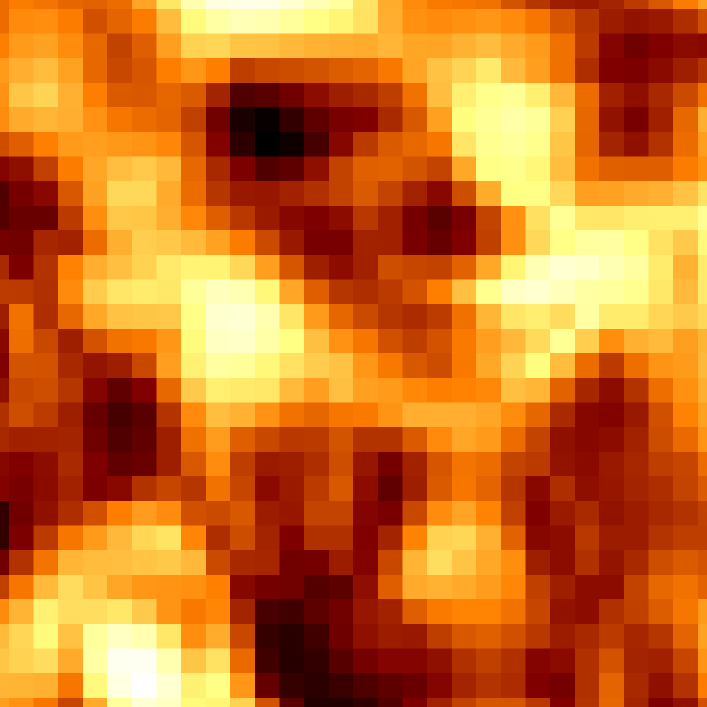}&
\includegraphics[width=.15\linewidth]{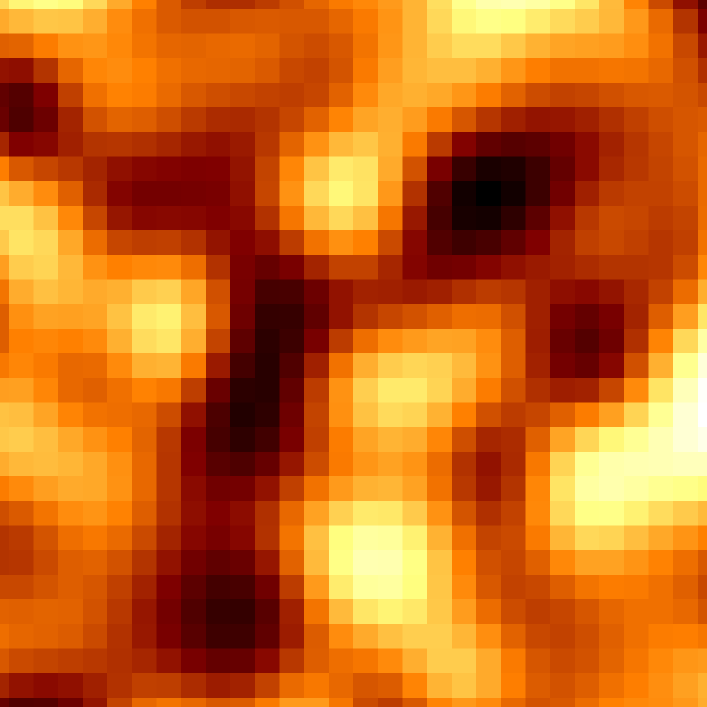}&
\includegraphics[width=.15\linewidth]{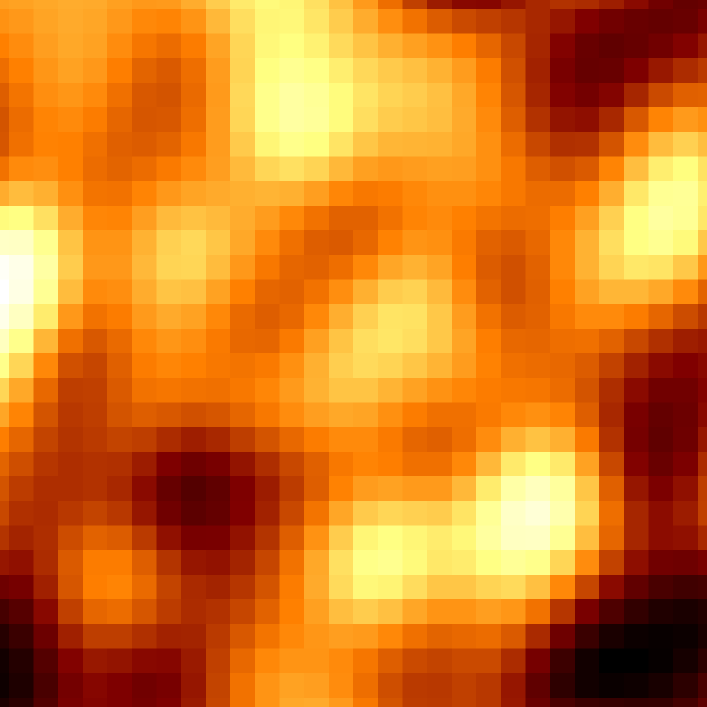}\\

\end{tabular}
\caption{Stacked images (GMRT, 610 MHz) of total intensity for star-forming galaxies. The empty space in the last column (i.e $\rm{\textit{z}\in[1.1-1.5]}$)  represents redshift range where no median stacked 610 MHz image was produced for the selected star-forming galaxy population. See Figure~\ref{all_galaxies_image.fig} for more details.}
\label{sfg_image.fig}
\end{figure}

\begin{figure}
\begin{tabular}{@{}c@{ }c@{ }c@{ }c@{ }c@{ }c@{ }c@{}}
&\textbf{\textit{z}=0.1-0.3} & \textbf{\textit{z}=0.3-0.5} & \textbf{\textit{z}=0.5-0.7} & \textbf{\textit{z}=0.7-0.9} & \textbf{\textit{z}=0.9-1.1} & \textbf{\textit{z}=1.1-1.5}\\
\centering
\rotatebox[origin=b]{90}{\makebox[\ImageHt]{\scriptsize $\rm{11.0-12.4}$}}&
\includegraphics[width=.15\linewidth]{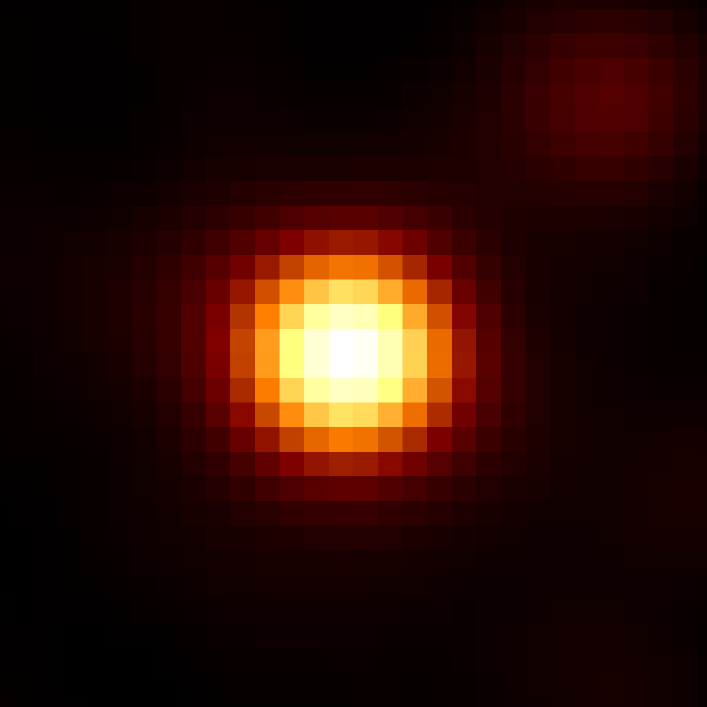}&
\includegraphics[width=.15\linewidth]{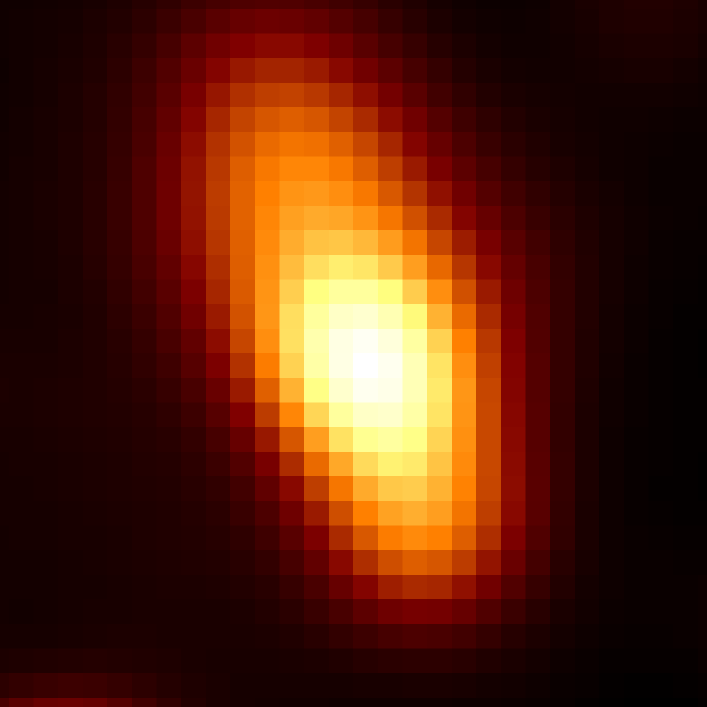}&
\includegraphics[width=.15\linewidth]{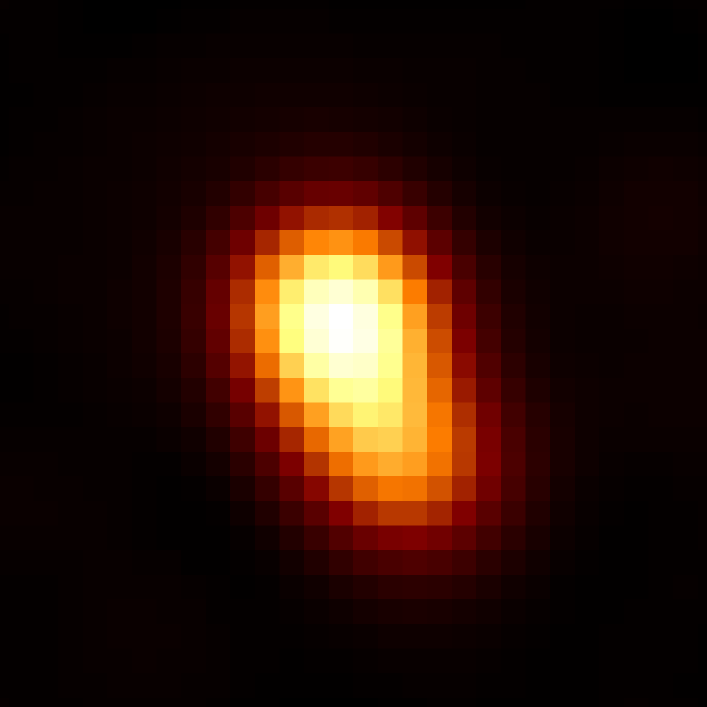}&
\includegraphics[width=.15\linewidth]{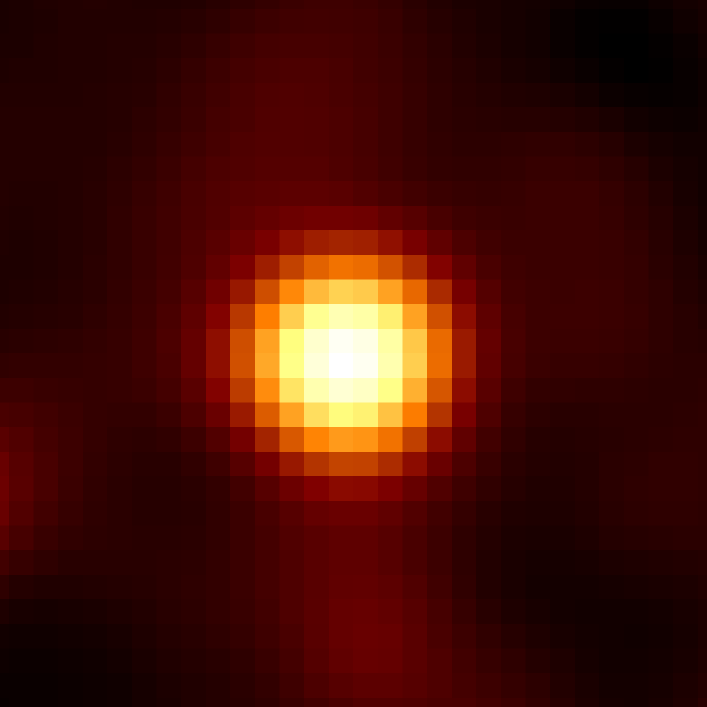}&
\includegraphics[width=.15\linewidth]{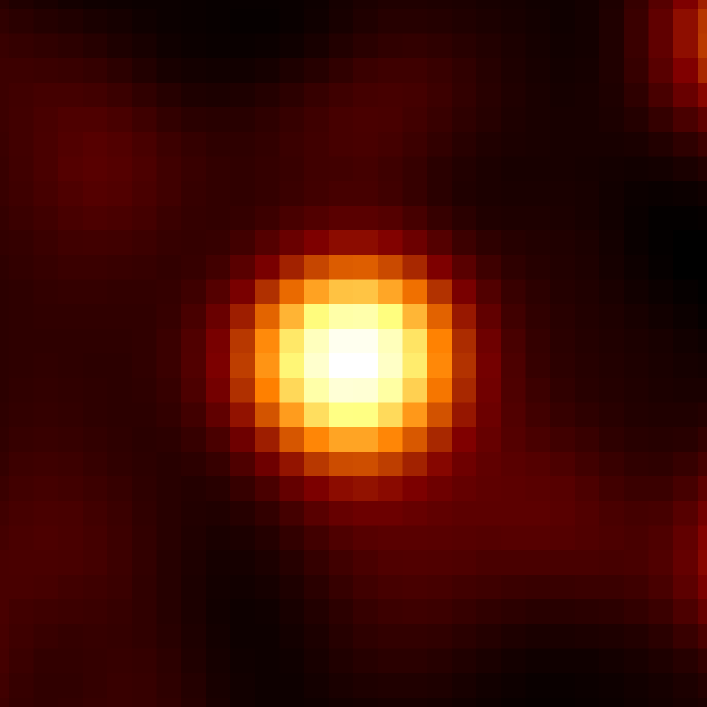}&
\includegraphics[width=.15\linewidth]{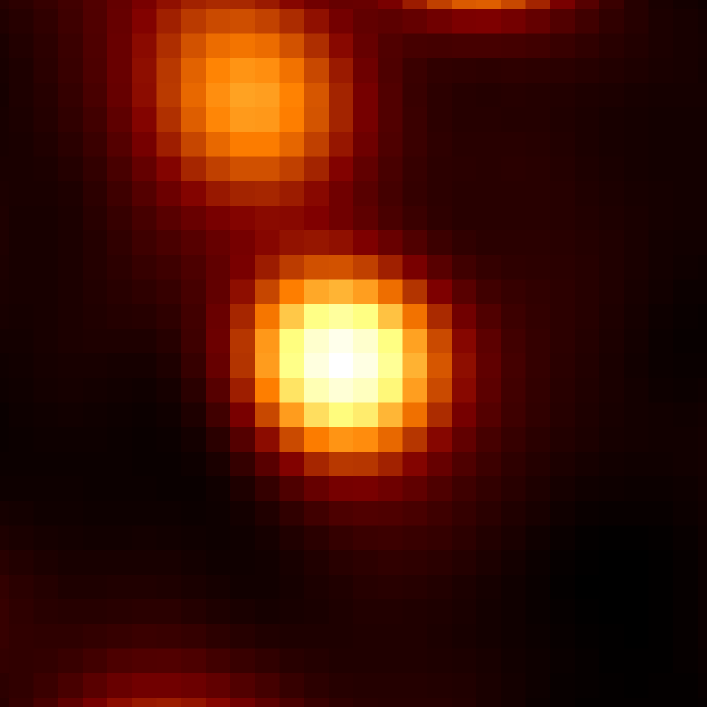}\\

\rotatebox[origin=b]{90}{\makebox[\ImageHt]{\scriptsize $\rm{11.0-12.4}$}}&
\includegraphics[width=.15\linewidth]{figures/stack_mean_z1_s_zbin_sfg05.tx.pdf}&
\includegraphics[width=.15\linewidth]{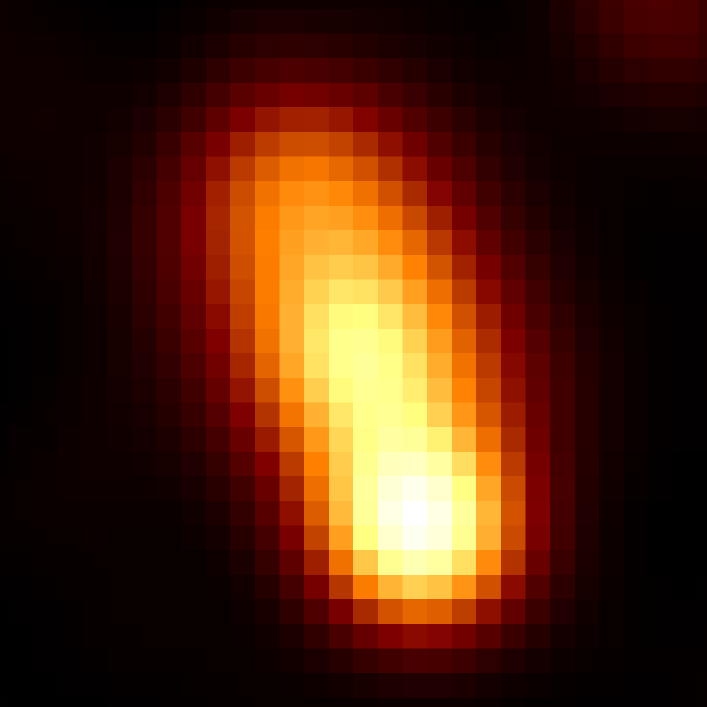}&
\includegraphics[width=.15\linewidth]{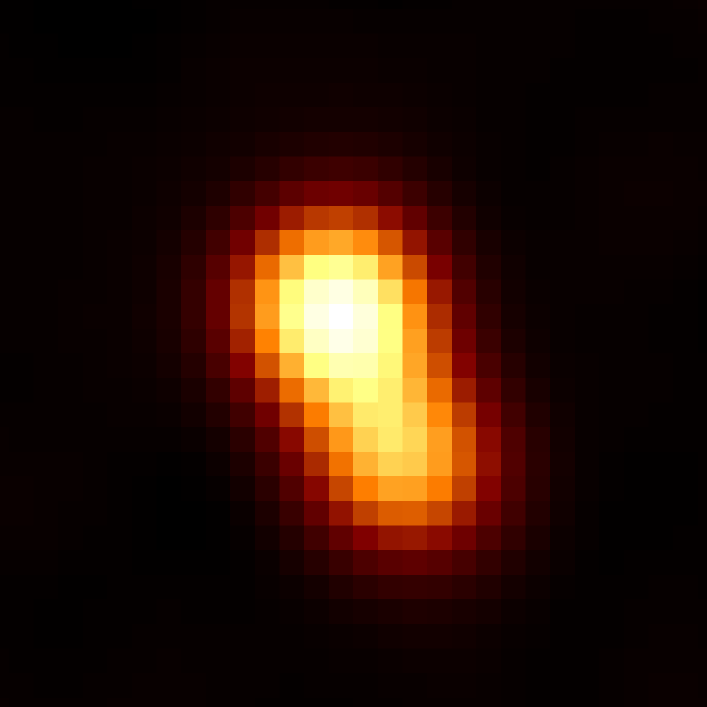}&
\includegraphics[width=.15\linewidth]{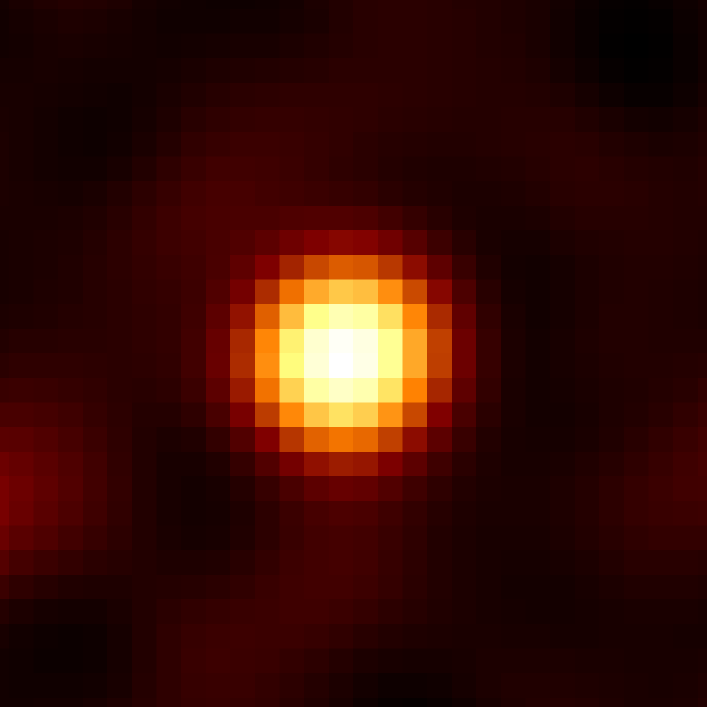}&
\includegraphics[width=.15\linewidth]{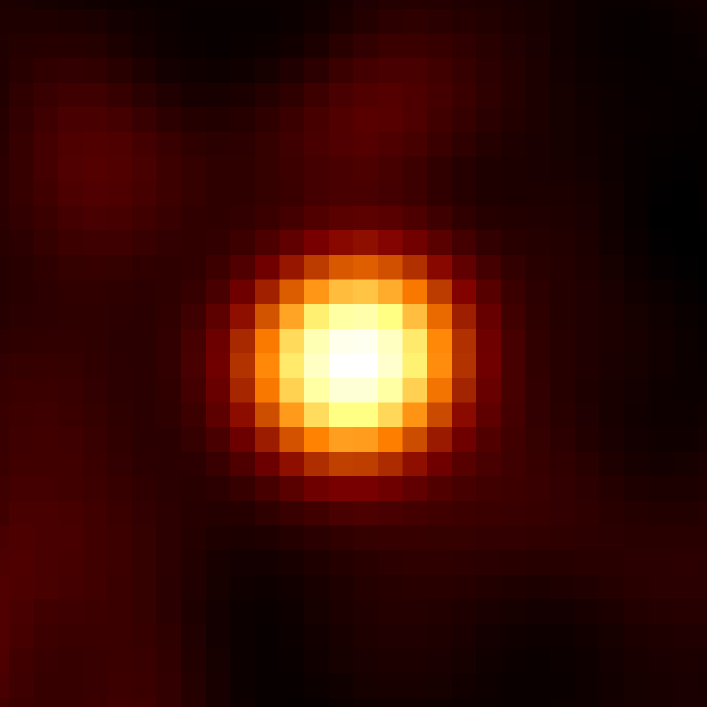}&
\includegraphics[width=.15\linewidth]{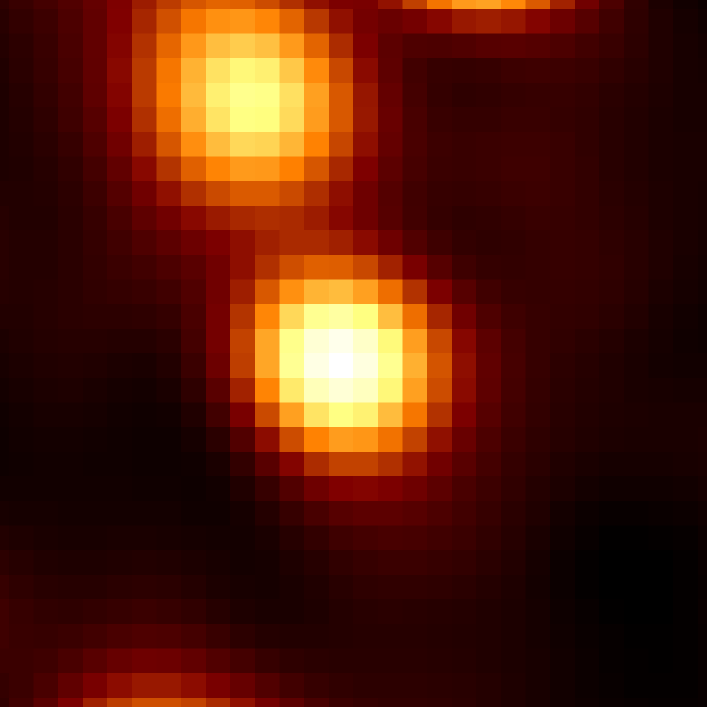}\\

\end{tabular}
\caption{Mean stacked 610 MHz radio images for the same redshift and stellar mass bins for all galaxies (top) and the SFG (bottom) populations, respectively. See Figure~\ref{all_galaxies_image.fig} for more details.}%
\label{mean_stack_fig}
\end{figure}

\begin{figure*}
\includegraphics[width=0.49\textwidth]{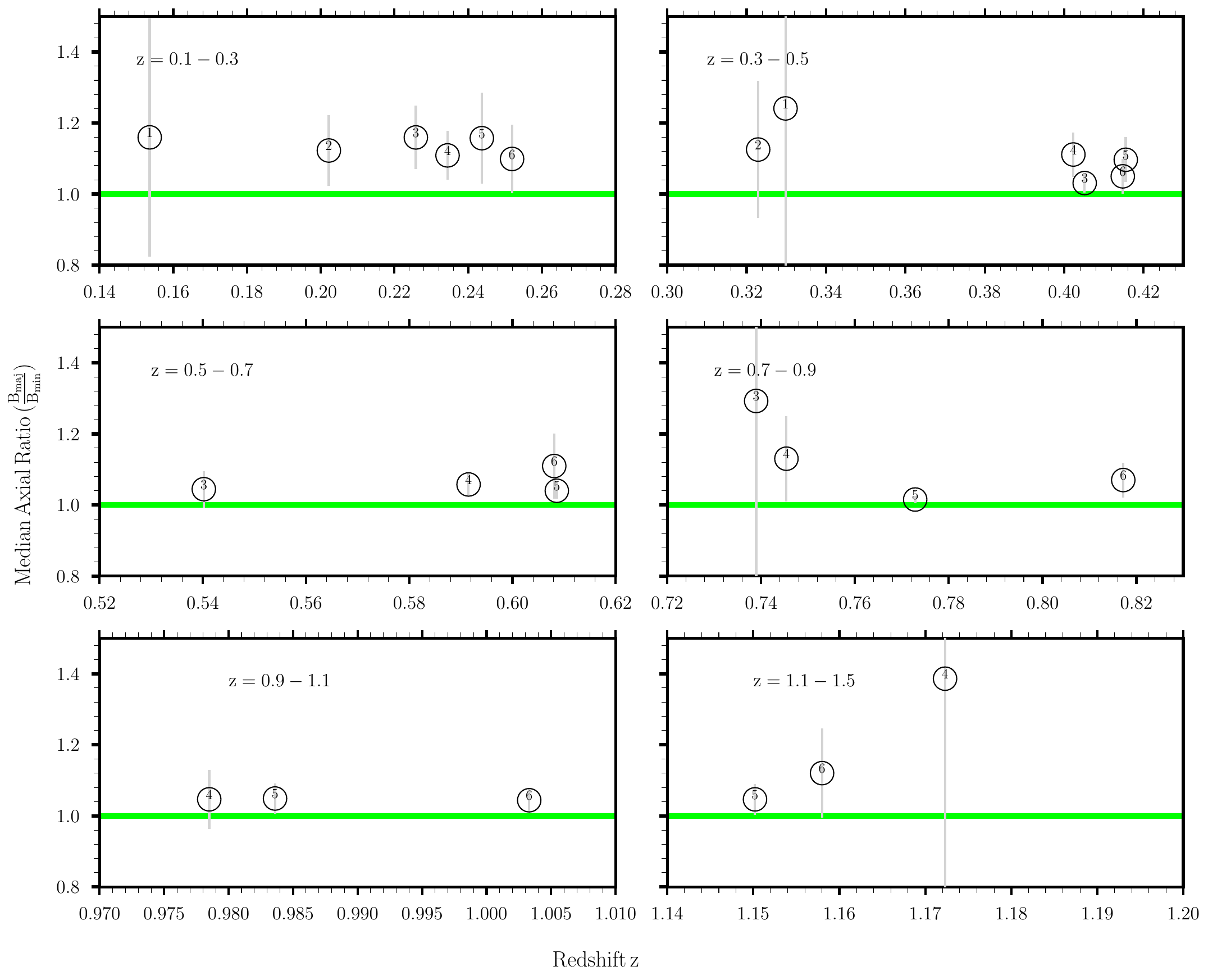}
\includegraphics[width=0.49\textwidth]{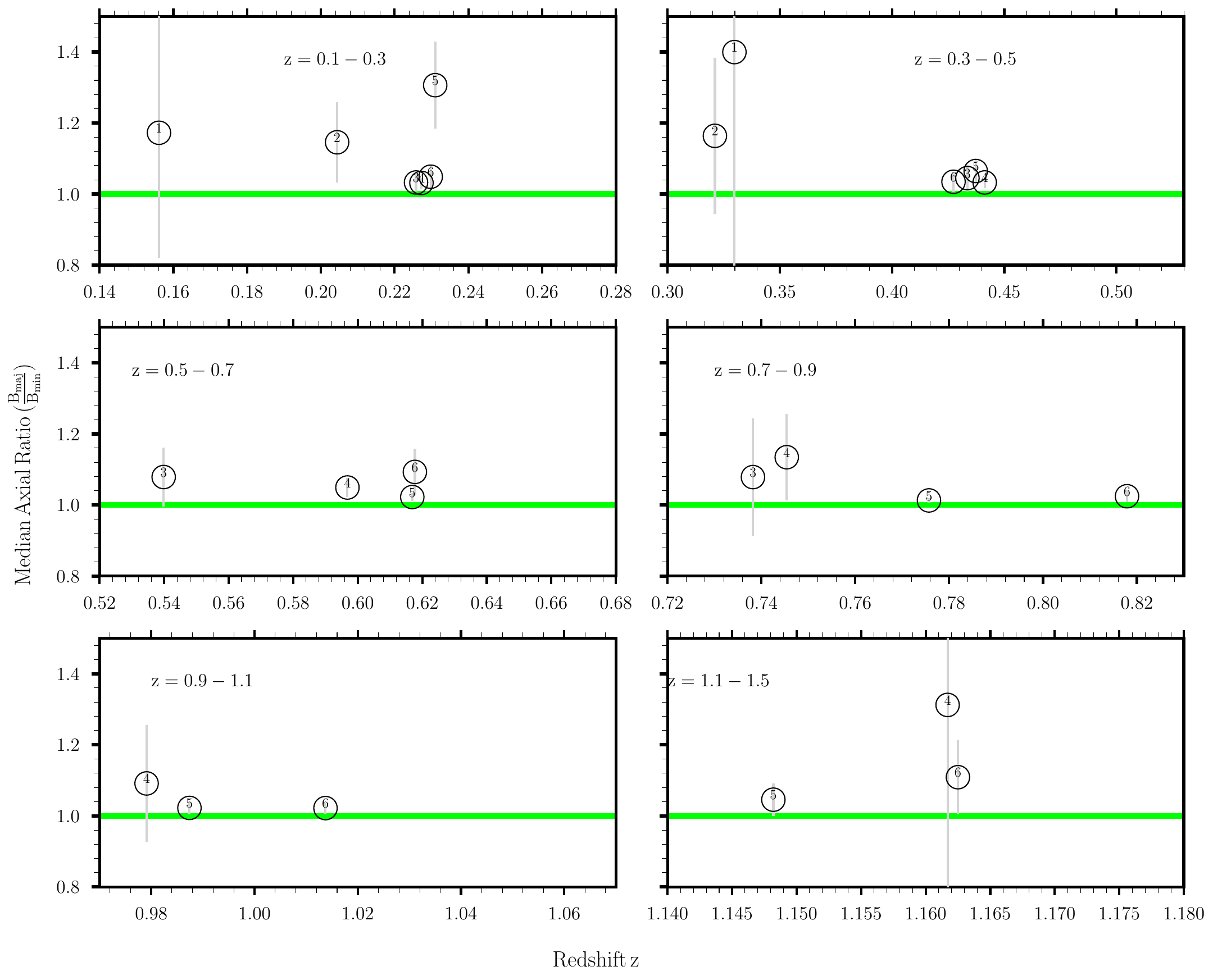}
    \caption{Stacked median axial ratio $\rm{\textit{B}_{maj}/\textit{B}_{min}}$ as a function of median redshift for all galaxies (left) and
star forming galaxies (right) represented as open black circles in each redshift bin. The horizontal solid green  line represent the original axial ratio of the 610 MHz image, where $\rm{\textit{B}_{maj}/\textit{B}_{min}\,=\,1}$. The corresponding stellar mass bins (see Figures~\ref{all_galaxies_image.fig}, ~\ref{sfg_image.fig} and \ref{mass_zbins.fig}) in each redshift range are shown as numbers (i.e. 1 - 6, from low to high stellar mass bins) in the middle of each open black circle. The sources shown here are from the  median stacked images which show a clear detection at their center from which we obtain Gaussian fits using \textsc{PyBDSF} source finder. Notice that these sources coincide with having a median stacked flux density  with SNR $\geq$ 5.0 and in most cases above the mass completeness limit}.

    \label{src_size.fig}
\end{figure*}

\subsection{Estimating the Radio Star Formation Rates (SFR, $\Psi$) }

Here, we calculate the radio luminosity (radio power) of the median stacked images and use it to estimate the radio based SFRs. 

The radio spectrum can be assumed to follow a simple power law $\rm{(S_{\nu}\propto \nu^{\alpha})}$ resulting from the sum of the non-thermal synchrotron and thermal bremsstrahlung components; the power-law index is typically $\rm{\alpha\,\approx\,-0.8}$ for SFGs \citep{1992ARA&A..30..575C,2016MNRAS.461..825G}. 
AGN-dominated sources may have steeper spectral indices \citep{2009MNRAS.397..281I}.

The observed stacked fluxes were converted to rest-frame (emitted) monochromatic luminosities using Equation~\ref{eqn}, which contains a bolometric \textit{K-correction K(z)}:
\begin{equation}
\rm{\textit{L}_{610}\,=\,4\pi d^{2}_\textit{L}S_{610}\textit{K(z)}[1+\textit{z}]^{-1}}
\label{eqn}
\end{equation}
Following the approach by \citet{2011MNRAS.410.1155B}, \textit{K(z)}, which accounts for the shift of the spectrum in relation
to the receiver, assuming  a simple power-law spectrum  to a monochromatic flux is given
$\rm{\textit{K(z)} = [1 + \textit{z}]^{-\alpha}\,where\,\alpha\,=\,-0.8}$.

\begin{figure*}
\includegraphics[width=0.48\textwidth]{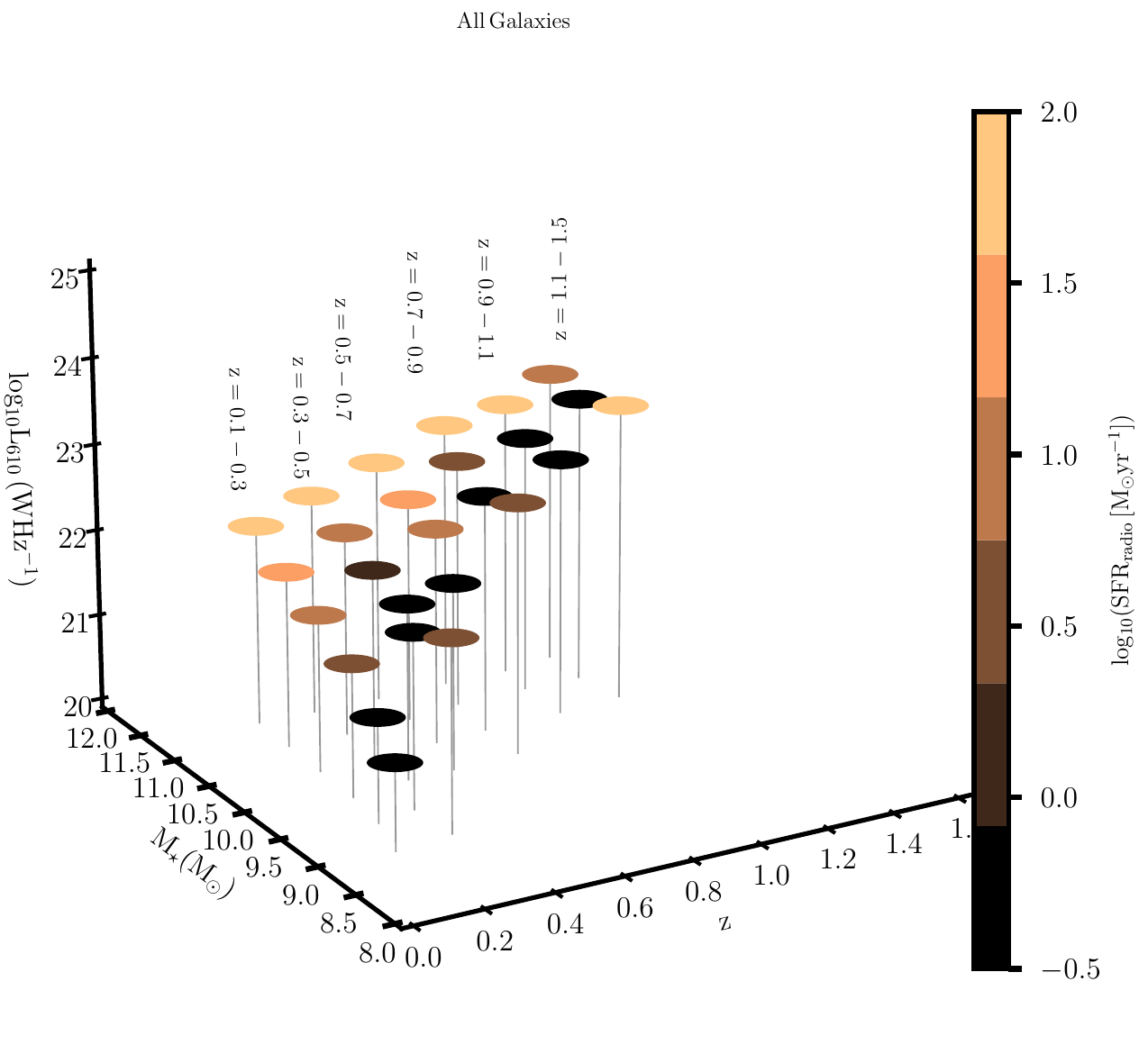}
\includegraphics[width=0.48\textwidth]{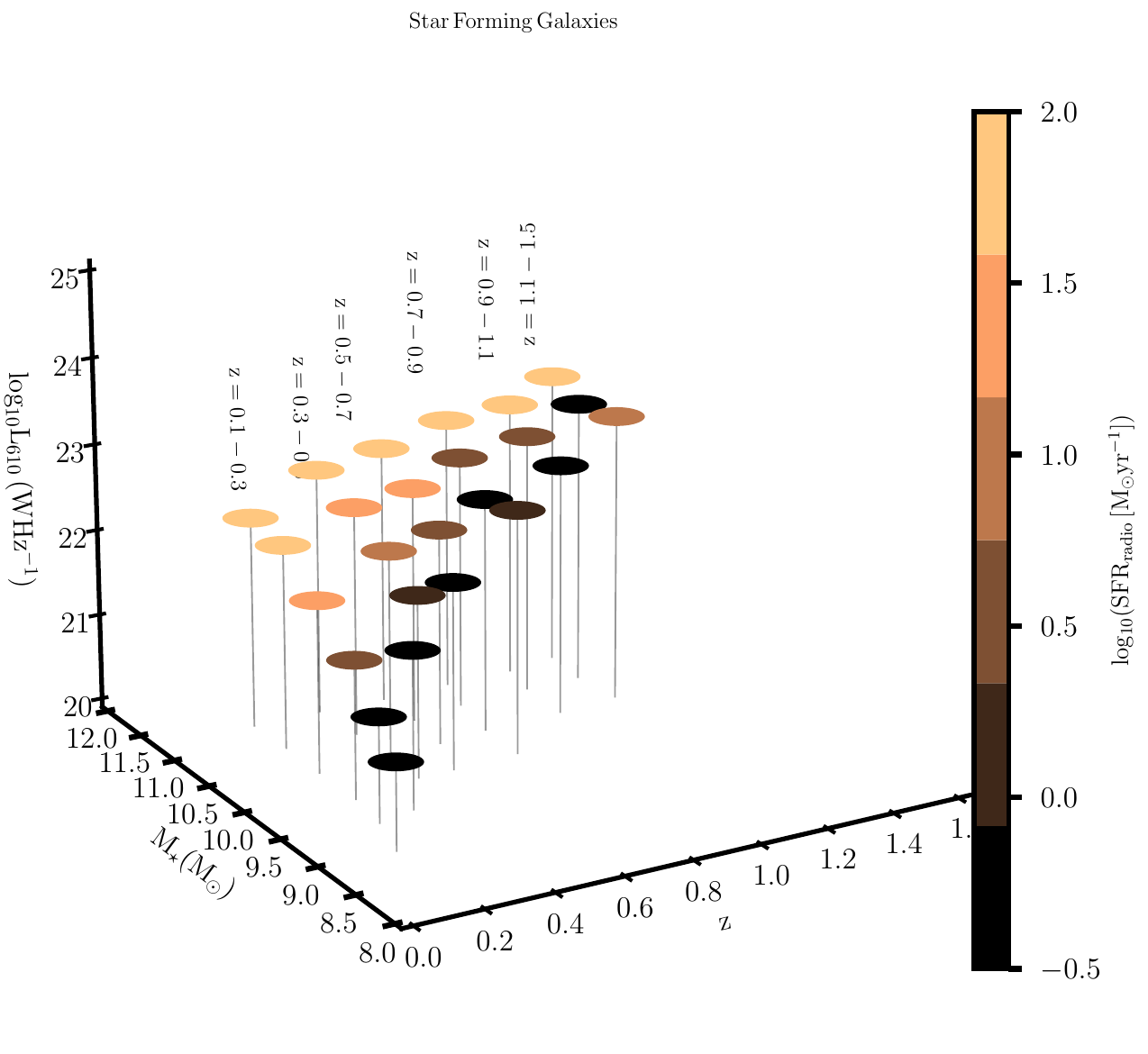}
    \caption{Left: Distribution of the total galaxies as a function of $\rm{\textit{M}_{\star}}$, redshift and stacked radio
power at 610 MHz $\rm{(\textit{L}_{610, MHz)}}$, colour coded by the derived stacked $\rm{SFR_{radio}}$. Right: Distribution of the star-forming driven sources as a function of $\rm{\textit{M}_{\star}}$, redshift and
stacked radio power at 610 MHz $\rm{(\textit{L}_{610, MHz)}}$, colour coded by the derived stacked
$\rm{SFR_{radio}}$.}
    \label{lum_3d.fig}
\end{figure*}
\citet{2003ApJ...586..794B} estimated the SFR, (\textit{$\rm{\Psi}$}), from 1.4 GHz luminosity of galaxies, calibrated from the total infrared SFR for galaxies with $\rm{\textit{L}\leq\,\textit{L}^{\star}}$ (defined as having an infrared luminosity $\rm{\textit{L}_{IR}\sim2\times10^{10}\textit{L}_{\odot}}$). We followed \citealt{2009MNRAS.397.1101G}, and converted this relationship to a 610 MHz equivalent:

\begin{equation}
\rm{\left(\frac{\Psi}{\textit{M}_{\odot}yr^{-1}}\right)\,=\,2.84\times10^{-22}\left(\frac{\textit{L}_{610}}{WHz^{-1}}\right)}
\label{sfr.eqn}
\end{equation}
\
For $\rm{\textit{L}_{610}\,>\textit{L}_{c}}$ (where $\textit{L}_{c}\,=\,3.3\times10^{21}\, WHz^{-1}$ is the luminosity at 610 MHz of a$\sim L_{\star}$ galaxy, with $\Psi\,\simeq1M_{\odot}yr^{-1})$, we can rewrite Equation~\ref{sfr.eqn} as:
\
\begin{equation}
\rm{\left(\frac{\Psi}{\textit{M}_{\odot}yr^{-1}}\right)\,=\,\frac{2.84\times10^{-22}}{0.1+0.9(\textit{L}_{610}/\textit{L}_{c})^{0.3}}\left(\frac{\textit{L}_{610}}{WHz^{-1}}\right)}
\label{sfr_1.eqn}
\end{equation}
The left and right panels of Figure~\ref{lum_3d.fig} shows the distribution of the stacked total and SFGs as a function of redshift, $\rm{\textit{M}_{\star}}$, and stacked radio power ($\rm{\textit{L}_{610}\,MHz}$). The x, y axes represent the $\rm{\textit{z}\,-\,\textit{M}_{\star}}$ plane, while the \textit{z}-axis sets the $\rm{\textit{L}_{610}\,MHz}$ colour-coded by their derived $\rm{SFR_{radio}}$. The relationship given by Equation~\ref{sfr.eqn} and~\ref{sfr_1.eqn} is used to compute the stacked $\rm{SFR_{radio}}$, $\rm{\Psi}$.

Since the SFR is correlated with the stellar mass, a useful quantity to describe the SF regime
of a galaxy is its specific SFR (sSFR), i.e. the SFR divided by the  median stellar mass of the galaxies in the bin. 
\begin{equation}
\rm{sSFR\equiv\,\frac{\Psi}{\textit{M}_{\star}}}
\end{equation}
To explore the specific star-formation we used  the measured radio SFR from the median radio stack divided by the stellar mass. We follow the stellar mass and redshift bins described in subsection~\ref{stackmethod.sec}.

\subsection{Separation of sSFR dependence}

The MS for SFGs correlation, reveals interesting mechanisms of the star formation history (\citealt{2004MNRAS.351.1151B,2007ApJS..173..267S}).
The  MS for SFGs has near-constant slope but shifts towards higher SFRs as the redshift increases (see e.g. \citealt{2011ApJ...739L..40R,2015MNRAS.453.2540J}).
We quantify the relationship between the sSFR and each of $\rm{\textit{M}_{\star}}$ and \textit{z}, following \citet{2011ApJ...730...61K}. 
\begin{equation}
\rm{sSFR(\textit{M}_{\star},\textit{z})\propto\,sSFR(\textit{M}_{\star\,|\textit{z}}) sSFR (\textit{z}|\textit{M}_{\star})\,=\,\textit{M}^{\beta}_{\star}(1 +\textit{z})^\textit{n}}
\end{equation}
We fit the stacked $\rm{sSFR\,-\,\log_{10}\textit{M}_{\star}}$ relation with these two separate functions of $\rm{\textit{M}_{\star}}$ and \textit{z}.
\begin{equation}
\rm{sSFR(\textit{M}_{\star}|\textit{z})=\,c_\textit{M}(\textit{z})M^{\beta}_{\star}} 
\label{powerlaw_mass.eqn}
\end{equation}
 We refer to the index $\rm{\beta}$ also as a slope
since the relation is commonly shown in log space.
\begin{equation}
\rm{sSFR(\textit{z}|\textit{M}_{\star})= c_\textit{z}(\textit{M}_{\star})(1 + \textit{z})^\textit{n}}
\label{powerlaw_redshift.eqn}
\end{equation}

In subsequent sections, we examine the relationship between sSFR, $\rm{M_{\star}}$ and \textit{z}.
We performed bootstrap linear regression fits to each sample \footnote{A resampling method used to estimate the variability of statistical parameters from a dataset which is repeatedly sampled with replacement \citep{JMLR:v20:17-451}}. The dashed lines in Figures~\ref{ssfr_mass.fig} and ~\ref{ssfr_z.fig} depict the best fit to the data in the mass representative $\rm{\beta}$ and redshift \texttt{n} regimes. In our bootstrap linear regression, we do not account for uncertainties associated with the SFR calibration, the photometric redshift, and stellar mass estimates as the large number of objects stacked for each data point ensures that even the joint error budget is statistically reduced to a low level that would not substantially enhance our uncertainty ranges (see \citealt{2011ApJ...730...61K}). We resampled the dataset 1000 times to create new datasets with the same size as the original, and then fitting a linear regression model to each of the resampled datasets.

\subsection{Dependence on Stellar Mass}\label{dep_stellar_mass}

\begin{figure*}
\centerline{\includegraphics[width=0.92\textwidth]{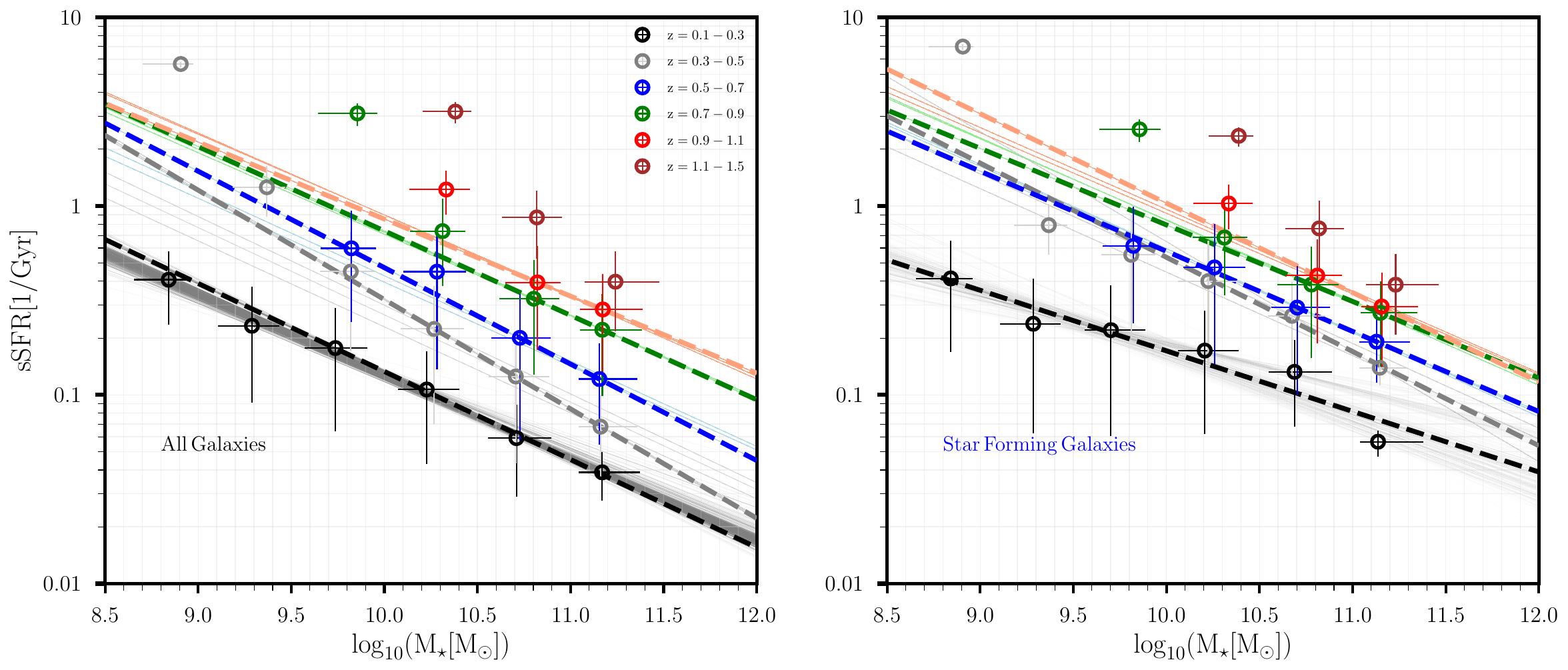}}
    \caption{Radio-stacking based measurement of the sSFR as a function of redshift at a given
stellar mass for all galaxies, (left) and star-forming galaxies (right). Redshift ranges from $\rm{0.1\leq\,\textit{z}\,\leq\,1.5}$. Dashed lines are
two-parameter fits of the form $\rm{c\times(\textit{M}_{\star}/10^{11}\textit{M})^{\beta}}$
to the mass-representative depicted by open circles. Horizontal bars indicate the width of those bins, while the vertical error bars simply reflect the Poisson uncertainties using the prescription of \citet{1986ApJ...303..336G}. The linear regression line that we get from
each bootstrap replicate of the fit to each population as a function of mass are shown as black, grey, blue, red and brown lines following  each redshift bin. }
    \label{ssfr_mass.fig}
\end{figure*}

In Figure~\ref{ssfr_mass.fig}, we show the dependence of sSFR on stellar mass for all galaxies and star-forming galaxy samples. The mass evolution of the sSFRs is well described
by a power law
$\rm{(\textit{M}_{\star}|\textit{z})\propto\,\textit{M}^{\beta}_{\star}}$, as depicted by the solid dashed lines in Figure~\ref{ssfr_mass.fig}. We first consider the whole sample which we refer to as all galaxies and show the redshift-dependent radio-based sSFRs that are distributed in the logarithmic $\rm{sSFR-\textit{M}_{\star}}$ plane. We utilise sources that are above the mass completeness limit in our fitting.  At a given redshift, the sSFR declines with increasing stellar mass for all galaxies. For $\rm{\textit{z}\in[0.1-0.3]}$, the value of the slope \textit{$\rm{\beta_{ALL}=-0.47\,\pm\,0.01}$}. For the redshift bin $\rm{\textit{z}\in[0.3-0.5]}$, we measure the values of the slopes to be \textit{$\rm{\beta_{ALL}=-0.58\,\pm\,0.01}$}. At $\rm{\textit{z}\in[0.5-0.7]}$, we measure the values of the slopes to be \textit{$\rm{\beta_{ALL}=-0.51\,\pm\,0.02}$}. For $\rm{\textit{z}\in[0.7-0.9]}$, the value of the slope is \textit{$\rm{\beta_{ALL}=-0.41\,\pm\,0.02}$}. For $\rm{\textit{z}\in[0.9-1.1]}$, we measure \textit{$\rm{\beta_{ALL}=-0.41\,\pm\,0.02}$}.

For the SFG population, we measure \textit{$\rm{\beta_{SFG}=-0.32\,\pm\,0.05}$} and \textit{$\rm{\beta_{SFG}=-0.50\,\pm\,0.01}$} for the first and second redshift bins, respectively. For reshift bins $\rm{\textit{z}\in[0.5-0.7]}$ and $\rm{\textit{z}\in[0.7 - 0.9]}$ we measure \textit{$\rm{\beta_{SFG}=-0.42\,\pm\,0.02}$} and \textit{$\rm{\beta_{SFG}=-0.41\,\pm\,0.01}$}, respectively. We measure \textit{$\rm{\beta_{SFG}=-0.47\,\pm\,0.03}$} for the fifth redshift bin. We can also infer from the second panel of the plot that the general trend of sSFR decreases with increasing stellar mass for the SFG population.

The ‘mass gradient’ for all galaxies and the SFG sample, i.e. \textit{$\rm{\beta_{ALL}}$} and \textit{$\rm{\beta_{SFG}}$}, is negative in all cases. The steepness of the sSFR with stellar mass is higher at high redshifts for both populations. 
If we ignore the highest redshift bin, for which the slope is poorly constrained, there is a consistent indication that the slope of the relation between sSFR and stellar mass becomes steeper with increasing redshift for both the total and SFG population.
Our results of the individual fits to our data yielding the parameter $\rm{\beta}$ for all and SF galaxies are presented in Table~\ref{tab_med_beta}. Fits have only been applied if more than two data points remained above the mass limit where the individual sample is regarded mass representative.

We observe that the sSFR  is only weakly
dependent on stellar mass, with sSFR decreasing as stellar mass
increases which is consistent with previous work. Our radio-derived SFR  provides a better match to the observed trends in sSFR versus stellar mass in the lowest mass bins, and also in reproducing the low redshift sSFR seen in other wavebands.

\subsection{Dependence on redshift}\label{dep_redshift}
\begin{figure*}
\centerline{\includegraphics[width=0.92\textwidth]{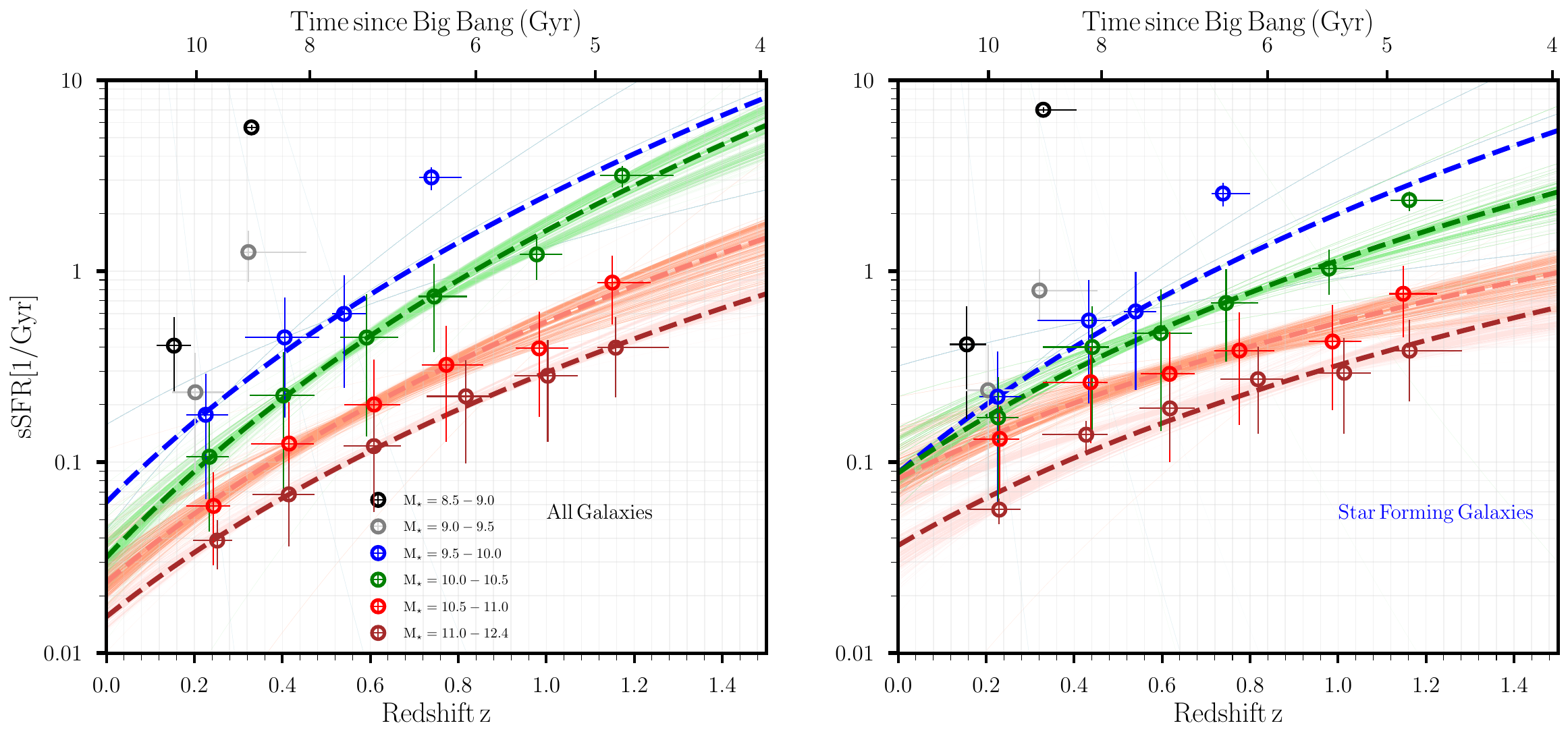}}
    \caption{Radio-stacking based measurement of the sSFR as a function of stellar mass at a
given redshift for all galaxies, (left) and star-forming galaxies (right). Mass ranges from $\rm{10^{8.5}<\textit{M}_{\star}/\textit{M}_{\odot}<10^{12.4}}$. Two-parameter fits
of the form $\rm{c\times(1+\textit{z})^\textit{n}}$ are applied the open circles which are representative samples
for the underlying galaxy population. The error bars follows that of Figure~\ref{ssfr_mass.fig}.}
    \label{ssfr_z.fig}
\end{figure*}

Observational values for  sSFR $\rm{\propto\,(1\,+\,\textit{z})^\textit{n}}$ may vary from $\rm{\textit{n}\,=\,2\,-\,5}$ \citep[see,][]{2007ApJS..173..267S,2011ApJ...730...61K,2019MNRAS.483.3213P}. Results from non-stacked sample by \citet{2014ApJS..214...15S}, indicates that the sSFR evolves with redshift according to a factor
of $\rm{(1\,+\,\textit{z})^{2.5}}$ (we provide a detailed comparison with sSFRs derived from other studies in subsection~\ref{comp.sec}).  
By plotting the same data as a function of redshift rather than in stellar mass classes an evolutionary trend is readily apparent. Figure~\ref{ssfr_z.fig} indicates how sSFRs for our samples evolve with redshift. It is immediately evident that there is a dramatic increase sSFR by a factor of > 100 over the interval $\rm{0.1\leq\,\textit{z}\,\leq\,1.5}$.
This is evident at all masses apart from the lowest mass bin which has poor statistics and 
spans a limited redshift range.
The redshift evolution of the sSFRs is well described
by a power law
$\rm{(\textit{z}|\textit{M}_{\star})\propto(1 + \textit{z})^\textit{n}}$ as depicted by the solid dashed lines in Figure~\ref{ssfr_z.fig} for a given mass
bin. The simple power-law fit provides a good description of the data in most cases. The most massive galaxies have the lowest sSFR, at all epochs for both the total and SFG population. We do not show the fits for mass range $\rm{\textit{M}_{\star}\in[8.5-9.0]\,\&\,[9.0-9.5]}$ because on incompleteness in these bins.
For $\rm{\textit{M}_{\star}\in[9.5-10.0]}$, the value of the slope \textit{n}$\rm{_{ALL}}=5.33\,\pm\,1.51$, whereas \textit{n}$\rm{_{SFG}}=4.50\,\pm\,1.36$. At $\rm{\textit{M}_{\star}\in[10.0-10.5]}$, the slope value increases to \textit{n}$\rm{_{ALL}}=5.69\,\pm\,0.26$ and \textit{n}$\rm{_{SFG}}= 3.70\,\pm\,0.15$ respectively. For $\rm{\textit{M}_{\star}\in[10.5-11.0]}$, we measure \textit{n}$\rm{_{ALL}}=4.51\,\pm\,0.27$ and \textit{n}$\rm{_{SFG}}= 2.71\,\pm\,0.25$ respectively. At the last mass bin, $\rm{\textit{M}_{\star}\in[11.0-12.4]}$, we measure  slopes of \textit{n}$\rm{_{ALL}}=4.25\,\pm\,0.07$ and \textit{n}$\rm{_{SFG}}=3.13\,\pm\,0.34$, respectively.
The redshift-evolution parameter \textit{n} is slightly
higher for all galaxies than for the SFG sample (i.e. \textit{n}$\rm{_{ALL}}$ > \textit{n}$\rm{_{SFG}}$). This implies that at a given stellar mass, redshift evolution is stronger for the full sample than for the SFG sample.
Our results of the individual fits to our data yielding the parameter a $\rm{\textit{n}}$ for all and SF galaxies are presented in Table~\ref{tab_med_n}. Fits have only been applied if more than two data points remained above the mass limit where the individual sample is regarded mass representative.

 
\begin{table*}
 \centering
 
\begin{tabular}{c cc|cc}
\toprule
 &  \multicolumn{2}{c} {All Galaxies}  &  \multicolumn{2}{c} {Star-forming Galaxies}\\ 
{$\bigtriangleup z$}&{$\rm{\log(C_{M,\,All}[1/Gyr])}$}&{$\rm{\beta_{ALL}}$}&{$\rm{\log(C_{M,\,SFG}[1/Gyr])}$}&{$\rm{\beta_{SFG}}$}\\
\midrule
  $\rm{0.1 - 0.3}$&-1.34$\,\pm\,$0.10& -0.47$\,\pm\,$0.01&-1.08$\,\pm\,$0.53 &-0.32$\,\pm\,$0.05 \\
$\rm{0.3 - 0.5}$ &-1.07$\,\pm\,$0.06 & -0.58$\,\pm\,$0.01& -0.77$\,\pm\,$0.09 & -0.50$\,\pm\,$0.01\\
$\rm{0.5 - 0.7}$&-0.83$\,\pm\,$0.11 & -0.51$\,\pm\,$0.02 &-0.66$\,\pm\,$0.11 & -0.42$\,\pm\,$0.02 \\
$\rm{0.7 - 0.9}$&-0.57$\,\pm\,$0.09 & -0.45$\,\pm\,$0.01& -0.50$\,\pm\,$0.05 & -0.41$\,\pm\,$0.01\\
$\rm{0.9 - 1.1}$&-0.47$\,\pm\,$0.12 & -0.41$\,\pm\,$0.02& -0.45$\,\pm\,$0.07 & -0.47$\,\pm\,$0.03\\
$\rm{1.1 - 1.5}$& --& --& --& --\\
 \hline
& $\rm{\langle \beta \rangle}\,=$& $\rm{-0.49\pm0.01}$& $\rm{\langle \beta \rangle}\,=$& $\rm{-0.42\pm0.02}$\\
\hline
 \end{tabular}
 \caption{Table summarizing the two parameter fits to the mass dependence of the sSFR. We applied a power-law fit of the form $\rm{c\times(\textit{M}_{\star}/10^{11}\textit{M})^{\beta}}$ (Equation~\ref{powerlaw_mass.eqn}, see also \citet{2011ApJ...730...61K})  to the radio-stacking-based sSFRs as a function of mass within any redshift slice. All the slopes have been computed in a mass complete range.}
 \label{tab_med_beta} 
 \end{table*}

\begin{table*}
 \centering
 \begin{tabular}{c cc |cc}
\toprule
 &  \multicolumn{2}{c} {All Galaxies}  &  \multicolumn{2}{c} {Star-forming Galaxies}\\ 
{$\bigtriangleup\log(M_{\star}[M_{\odot}])$}&{$\rm{\log(C_{z,\,All}[1/Gyr])}$}&{$\rm{\textit{n}_{ALL}}$}&{$\rm{\log(C_{z,\,SFG}[1/Gyr])}$}&{$\rm{\textit{n}_{SFG}}$}\\
\midrule
  $\rm{8.5 - 9.0}$& --& --& --& --\\
$\rm{9.0 - 9.5}$& --& --& --& --\\
$\rm{9.5 - 10.0}$&-1.21$\,\pm\,$0.15&5.33$\,\pm\,$1.51&-1.06$\,\pm\,$0.12&4.50$\,\pm\,$1.36\\
$\rm{10.0 - 10.5}$&-1.50$\,\pm\,$0.03&5.69$\,\pm\,$0.26&-1.06$\,\pm\,$0.04&3.70$\,\pm\,$0.15 \\
$\rm{10.5 - 11.0}$&-1.63$\,\pm\,$0.04&4.51$\,\pm\,$0.27&-1.09$\,\pm\,$0.05&2.71$\,\pm\,$0.25\\
$\rm{11.0 - 12.4}$&-1.81$\,\pm\,$0.01&4.25$\,\pm\,$0.07&-1.44$\,\pm\,$0.09&3.13$\,\pm\,$0.34\\
 \hline
& $\rm{\langle \textit{n} \rangle}\,=$& $\rm{4.94\,\pm\,0.53}$& $\rm{\langle \textit{n} \rangle}\,=$& $\rm{3.51\,\pm\,0.52}$\\
\hline
 \end{tabular}
 \caption{Table summarizing the two parameter tits to the redshift evolution of the sSFR. We applied a power-law fit of the form $\rm{c\times(1+\textit{z})^\textit{n}}$ (Equation~\ref{powerlaw_redshift.eqn}) to the radio-stacking-based sSFRs as a function of redshift within any mass bin. }
 \label{tab_med_n} 
 \end{table*}

\section{Discussion}\label{disc.sec}

Our results on the  sSFR-mass relation steepening with redshift are in broad agreement with those based on
 far-infrared stacking experiments that found almost flat relations up to $\rm{\textit{z}\sim1.5}$. 

 On the whole, there is a good agreement between our fits and  recent work, mostly probing high-\textit{z} observations. This is justification that our fits as functions of redshift with a power-law do provide nearly as good fits compared to the literature.
We are not able to definitively rule
out a possible “plateauing” of the sSFR in the redshift range
explored here. In inference, our results seem to favor a scenario where the sSFRs  will continue to increase until at least $\rm{\textit{z}\sim 3}$, as found by studies from the literature, if we were to probe a broader redshift range. 
Redshift dependence of the $\rm{\textit{sSFR}\,-\,\textit{M}_{\star}}$ relation is more uncertain \citep{2013A&A...556A..55I,2015A&A...575A..74S,2018A&A...615A.146P} and studying the redshift evolution of these different populations through stacking  thus provide complementary insights into the host properties of these sources. In this section, we discuss our results and compare with sSFRs derived from other studies including those from radio stacking experiments.

\subsection{Comparison with $\mathit{\beta}$ and $\mathit{n}$ of {\small\textsc{s}}SFRs derived from previous studies} \label{comp.sec} 
We compare our results to
those  previous studies conducted at 1.4 GHz and the authors have considered more than one SFR indicator.  
In subsections~\ref{dep_stellar_mass} and ~\ref{dep_redshift} we found that sSFR decreases with stellar mass (downsizing; see
Figure~\ref{ssfr_mass.fig}) and increases with redshift (see Figure~\ref{ssfr_z.fig}) for all galaxies and SFG populations. Additionally, $\rm{\beta_{ALL}}$ and $\rm{\beta_{SFG}}$ for the dependence on stellar mass are all negative whereas their values become steeper with increasing redshift.
Radio-based measurements of the $\rm{\textit{sSFR}-\textit{M}_{\star}}$ relation have been studied by previously in the literature.

\begin{figure}
\centering
\includegraphics[width=0.5\textwidth]{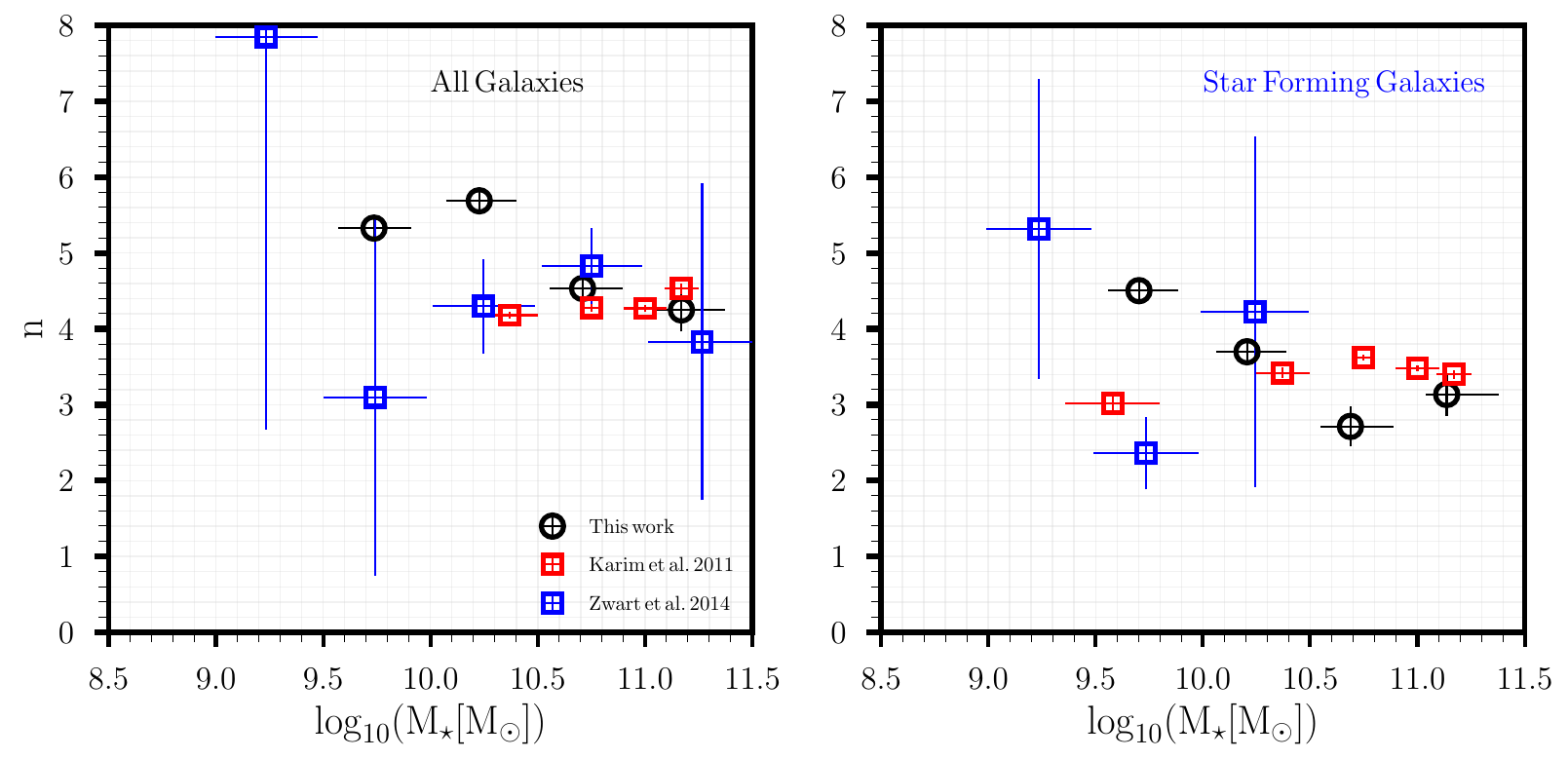}
\\
\includegraphics[width=0.5\textwidth]{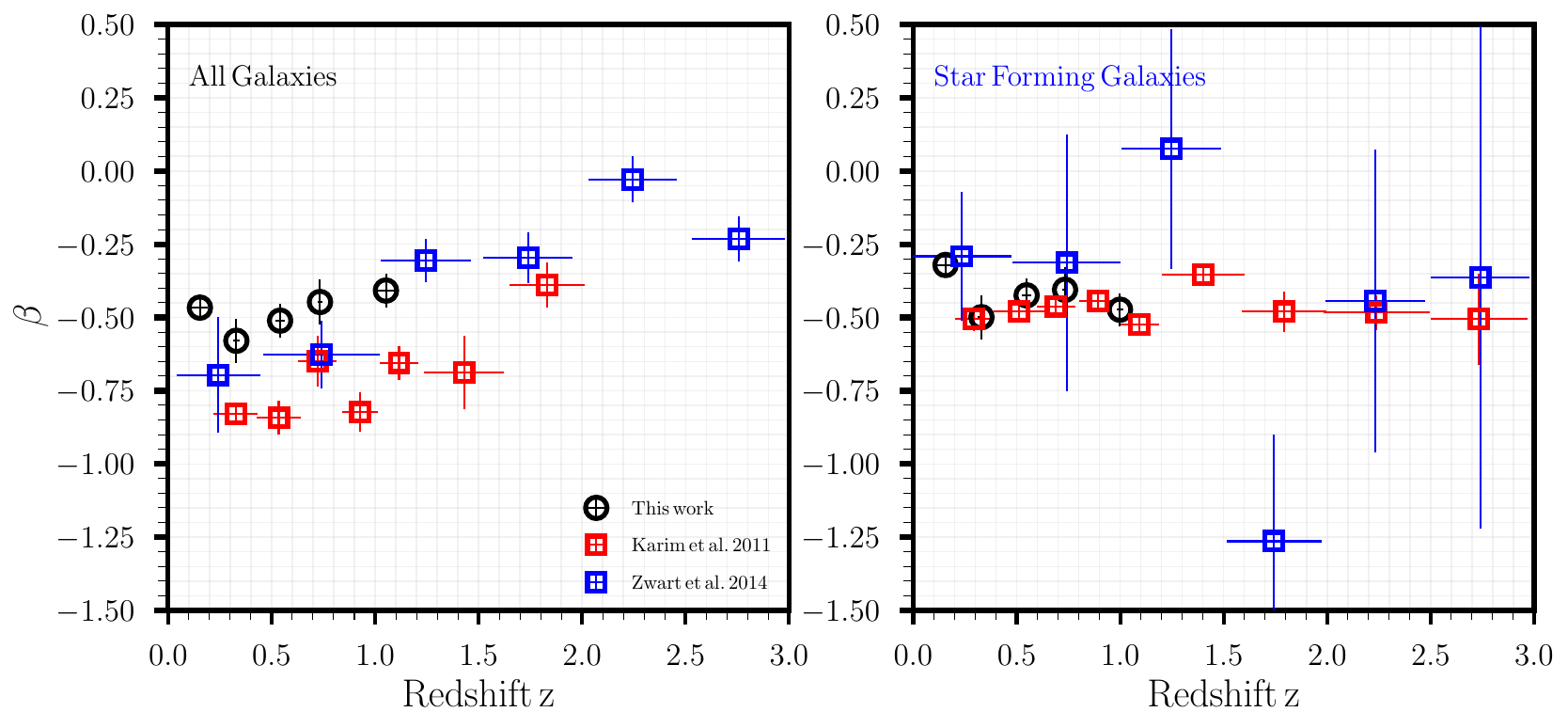}
    \caption{Top: Comparison of gradient $\rm{\beta}$ against stellar mass as a function of redshift for the total (left) and the star-forming galaxy population (right). The open red squares represents measurements from \citealt{2011ApJ...730...61K}, whereas the open blue squares represent measurements from \citet{2014MNRAS.439.1459Z}.
    Bottom: Comparison of redshift evolution parameter of \textit{n} sSFR as a function of stellar mass for all galaxies (left panel) and the SFG population (right panel).}
    \label{mass_grad.fig}
\end{figure}

We compare with our measurements with the mass dependent slope estimates from  \citet{2011ApJ...730...61K} and \citet{2014MNRAS.439.1459Z}. The top panel of Figure~\ref{mass_grad.fig} presents  the comparison of gradient $\rm{\beta}$ against stellar mass as a function of redshift for the total (left) and the star-forming galaxy population (right). These studies form the literature and
our work share some methodological similarities (e.g., the use of a mass and \textit{K}-selected samples and a radio-stacking approach) and should therefore be directly comparable. However, there are some technical differences in the exact implementation of the
image stacking as already discussed (see subsection~\ref{stackmethod.sec}). 
It is important to also point out that the calibration
of the individual radio star formation rates and binning of this work are different from these previous studies. 
However, in terms of the evolution of the sSFR sequence both studies show a reasonable agreement with our work. 
The differences that arise may be attributed to our study
tracing 610 MHz rather than 1.4GHz, which these past studies were conducted.

\citet{2011ApJ...730...61K} concluded that the sSFR sequence itself tends to flatten toward lower masses and \textit{z} > 1.5.
They inferred this  might be explained by an upper
limiting threshold where average SF systems already reach
levels of star formation that qualify them to double their
mass within a dynamical time. Their data show that there is a tight correlation with power-law dependence, $\rm{sSFR \propto \textit{M}_{\star}^{\beta}}$, between sSFR and stellar mass at all epochs. Excluding quiescent galaxies  from their analysis, a shallow index $\rm{\beta_{SFG}\approx -0.4}$ fits the correlation for star-forming sources. For their total population the sSFR-mass gradient $\rm{\beta}$ becomes steeper with $\rm{\beta_{ALL}\approx -0.67}$.
The sSFR–mass gradients $\rm{\beta}$ found by \citet{2014MNRAS.439.1459Z} become less steep with redshift (from $\rm{\beta \approx -0.75}$ to $\rm{\beta \approx -0.25}$ out to $\rm{\textit{z}\simeq2}$) for the full and
elliptical samples, but show no dependence with redshift ($\rm{\beta \approx-0.5}$) for the starburst and irregular galaxies for their stacked deep (17.5 $\mu$Jy) VLA radio observations.

Studies based on IR SFRs such as \citet{2010A&A...518L..25R} found that the sSFR-mass relation steepens with redshift for all galaxies, becoming almost flat at $\rm{\textit{z}<1.0}$ and reaching a slope of $\rm{\textit{n}\,=\, -0.50^{+0.13}_{-0.16}}$ at $\rm{\textit{z}\sim 2}$.  Moreover, they also show that the most massive galaxies have the lowest sSFRs at any redshift.
Further implying that they have formed their stars earlier and more rapidly than their low mass counterparts which corresponds with our findings. \citet{2010MNRAS.405.2279O} in their analysis of sSFR  activity of galaxies and their evolution near the peak of the cosmic far-infrared background at 70 and 160$\rm{\mu}$m found a trend $\rm{sSFR\propto \textit{M}_{star}^{\beta}}$ with $\rm{\beta \sim-0.38}$. They found a stronger trend for early type galaxies
($\rm{\beta \sim -0.46}$) than late type galaxies ($\rm{\beta \sim -0.15}$).

The bottom panel of Figure~\ref{mass_grad.fig} shows a comparison of the redshift evolution parameter recorded of \textit{n} sSFR as a function of stellar mass for all galaxies (left) and the SFG (right) population derived from Figure~\ref{ssfr_z.fig}. Although we observe that all measured sSFRs (i.e. total galaxies and SFG) increase with redshift, massive galaxies have the lowest sSFRs. The sSFRs span a smaller
range at high redshift, with massive galaxies evolving
faster compared to low mass galaxies, decreasing their sSFR at earlier epochs. Our results are in broad agreement with those based on
radio-stacking which find almost flat relations up to $\rm{\textit{z}\sim2}$ \citep[see,][]{2009MNRAS.394....3D,2009ApJ...698L.116P}. 
\citet{2011ApJ...730...61K} noted that at redshift 0.2 < \textit{z} < 3 both populations show a strong and mass-independent decrease in their sSFR toward the present epoch 
where $\rm{\textit{n}\sim4.3}$ for all galaxies and $\rm{\textit{n}\sim3.5}$ for star-forming sources. \citet{2014MNRAS.439.1459Z} reported that the redshift evolution of sSFR  is much faster for their full sample than their starburst sample. \citet{2010MNRAS.405.2279O} found that the
sSFR evolves as $\rm{(1 + \textit{z})^\textit{n}}$ with $\rm{\textit{n}\,=\,4.4\pm0.3}$ for galaxies with $\rm{10.5<\log_{10}\textit{M}_{\star}/\textit{M}_{\odot}<12.0}$. For early type galaxies, they found that the average evolution in this mass range is stronger ($\rm{\textit{n}\sim 5.7}$) but decreases to higher mass.

Our SFG sample comprises sources that do not satisfy quiescent galaxy criterion, spectroscopically identified as AGN, satisfies the \cite{2012ApJ...748..142D} IR AGN criterion, and have an X-ray counterpart.  This does not explicitly make our sample immune from AGN contamination. It is expected that AGN  reside in massive star-forming
galaxies \citep[see,][]{2003MNRAS.346.1055K,2012ApJ...753L..30M,2013ApJ...764..176J,2013A&A...560A..72R}. As such, AGN contamination could be a major reason to doubt that 610 MHz radio emission might be a reliable star formation tracer. \citet{2017FrASS...4...35P} emphasize that for flux density $\lesssim$ 1 mJy, the faint radio sky is populated by both non-jetted AGN and a quickly decreasing
fraction of jetted AGN (see also \citealt{2016A&ARv..24...13P}).
Studies by \citet{2007ApJ...670..156D} conducted  at mid- and far-IR to submillimeter, radio, and rest-frame UV wavelengths, measured
contamination to SFRs from X-ray-emitting AGN. 
They used radio stacking to investigate trends for radio undetected sources and found that the $\rm{\textit{L}_{IR}}$ estimated from $24\,\mu m$ exceeds on average by an order of magnitude the same quantity derived from radio. This was attributed to the additional radio emission from an AGN, as suggested also by \citet{2007ApJ...660..167D}, mostly in low redshift galaxies.
\citet{2022ApJ...925...74J} highlighted the importance AGN selection effects on the distributions of host galaxy properties. They combined a study of X-ray and IR AGN at $\rm{\textit{z}\approx2}$ and compared the star formation and morphological properties of AGN and non-AGN host galaxies.
Their studies revealed that non-AGN SFGs on the main sequence and X-ray AGN have similar median star formation properties.
Classification of sources as either SFG or AGN, appears to be a more complex problem and in reality, not all sources will give unambiguous results over all criteria. Hence, we do not reject  AGN contamination contribution argument to the radio emission in the selected SFGs.
Studies by \citet{2022ApJ...929...53I} reveal that the frequency of AGN hosted by transitional, from SFGs to quiescent galaxies, depends significantly on how the AGN are selected. 

In summary, our measurements for both the sSFR-mass and 
sSFR-redshift evolution are largely consistent between the VIDEO (i.e. \citealt{2014MNRAS.439.1459Z}), COSMOS (i.e. \citealt{2011ApJ...730...61K}) and our 610 MHz GMRT data set of the ELAIS-N1. Although for the sSFR-redshift evolution parameter, n, we measure is slightly higher in our data. We also measure slightly steeper mass gradient, $\rm{\beta}$.



\subsection{Comparison of our MS for SFGs to previous studies}
\begin{figure}
\centerline{\includegraphics[width=0.45\textwidth]{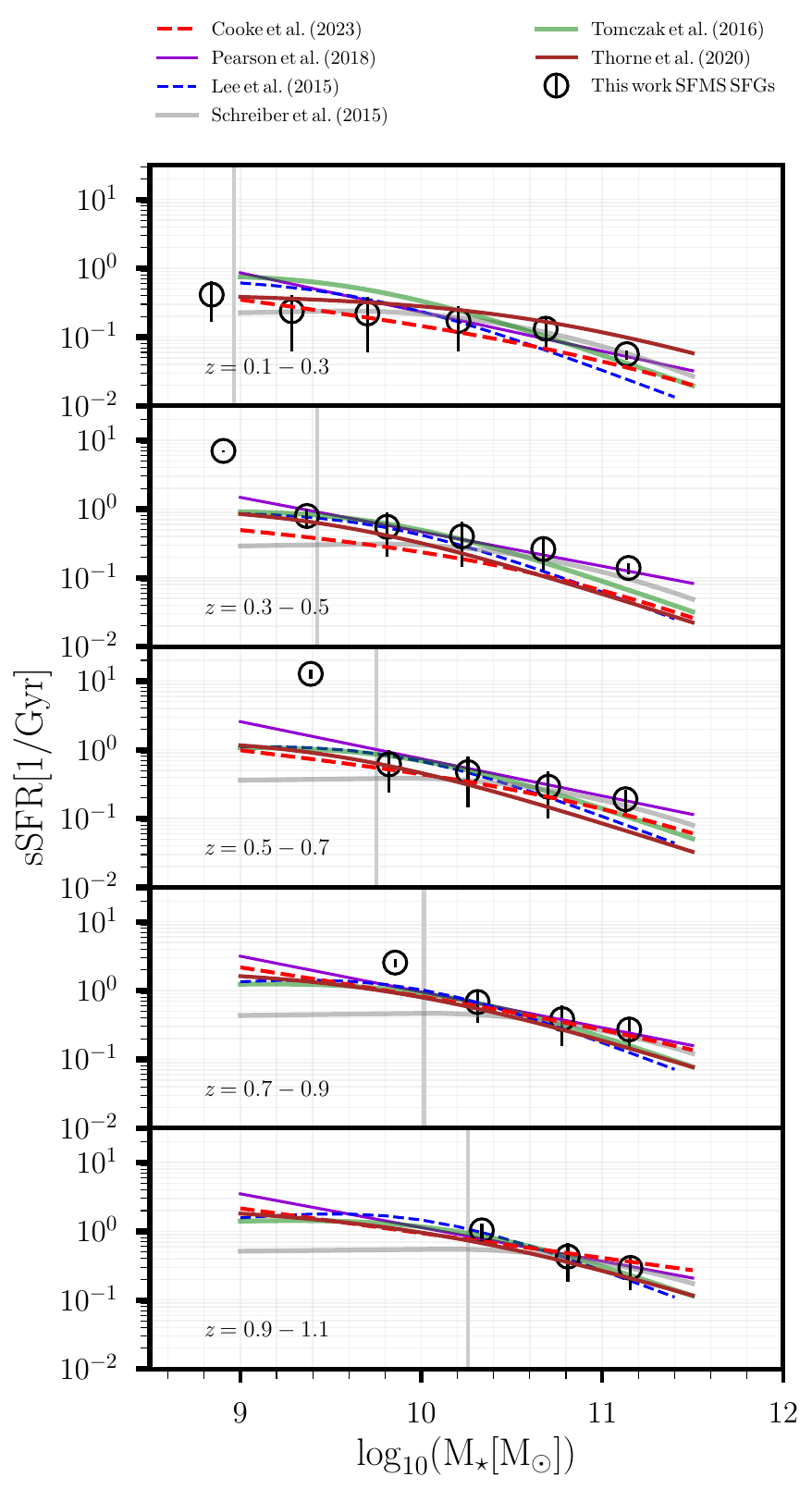}}
    \caption{Comparison of the radio-stacked based measurement of sSFR for SFGs to the MS trends observed by \citet{2015ApJ...801...80L}, \citet{2015A&A...575A..74S}, \citet{2016ApJ...817..118T}, \citet{2018A&A...615A.146P}, \citet{2021MNRAS.505..540T}, \citet{2023ApJ...942...49C}, shown in each panel as dash blue curves, solid grey curves, solid violet lines, solid green curves, solid brown curves, and dash red curves respectively. The solid grey vertical lines in each panel represents the mass completeness limit, $\rm{\textit{M}_{lim}}$. }
    \label{ssfr_mass_lee.fig}
\end{figure}

The UV, IR and radio wavelengths have been used to characterize  the star formation properties for different classes of sources in literature, by investigating their star formation rates. 
We provide in Figure~\ref{ssfr_mass_lee.fig} a comparison of the our radio-stacked based measurement of sSFR evolution with other works in literature. We compare the evolution of the sSFR derived for five different redshift  bins to the MS trends observed by other studies in the literature. These measurements from previous studies were conducted in the \textit{SFR-M$_{\star}$} plane. 
To illustrate the scientific value of our stacked data, we convert to the \textit{sSFR-M$_{\star}$} plane in order to easy comparison.  The solid grey vertical lines in each panel represents the mass completeness limit, $\rm{\textit{M}_{lim}}$, above which we perform the fitting in subsections~\ref{dep_stellar_mass} and ~\ref{dep_redshift}.
\citet{2015ApJ...801...80L} used rest-frame color-color diagram $\mathit{(NUV\,-\,r)}$ versus $\mathit{(r\,-\,K)}$, to study the MS in the COSMOS field at $\rm{0.3<\textit{z}<1.3}$. 
\citet{2015A&A...575A..74S} conducted stacking analysis of \textit{UVJ} selected galaxies in the deep Herschel PACS maps of the CANDELS fields. They demonstrated that galaxies at $\rm{\textit{z} = 4}$ to 0 of all stellar masses follow the MS for SFGs.
\citet{2016ApJ...817..118T} performed stacking analysis of \textit{UVJ} selected galaxies combining mean IR luminosity with mean \textit{NUV} luminosity to derive SFR. We compare their fits, solid green curves, in similar redshift range to our stacked points. 
\citet{2018A&A...615A.146P} selected SFGs following the \textit{UVJ} selection as in \citet{2014ApJ...795..104W} and traced the MS over $\rm{0.2\,\leq\,\textit{z}\,<\,6.0}$ and $\rm{10^{9.0}<\,{\textit{M}_{\star}}/{\textit{M}_{\odot}}<10^{11.0}}$. Their simple two-parameter power law fits are shown as solid violet lines in each panel in Figure~\ref{ssfr_mass_lee.fig}. \citet{2021MNRAS.505..540T} used multiwavelength photometry from the \textit{Deep Extragalactic VIsible Legacy Survey} \citep[DEVILS,][]{2018MNRAS.480..768D} and measured stellar masses and SFR for galaxies in the COSMOS field mapping the evolution of the \textit{SFR-M$_{\star}$} relation for $\rm{0\,<\,\textit{z}\,<\,4.25}$ redshift range. Their fits, which is obtained by adapting the parameterisation from \citet{2015ApJ...801...80L} and  adding an slope to freely model SFR at high stellar masses are shown as solid brown curves.
\citet{2023ApJ...942...49C} investigated the relationship between \textit{SFR-M$_{\star}$} of SFGs in the COSMOS field from $\rm{0\,<\,\textit{z}\,<\,3.5}$. The fitted MS curves measured from their  construction of \textit{FUV–FIR} SEDs of stellar mass-selected sample are shown in the individual panels of Figure~\ref{ssfr_mass_lee.fig} as dash red curves. We adopt the \citet{2003PASP..115..763C} initial mass function (IMF) since all but the \citet{2015A&A...575A..74S} data points used this IMF. In the case of \citet{2015A&A...575A..74S} who applied a \citet{1955ApJ...121..161S} IMF, we multiplied by constant factors of 0.62$\rm{\textit{M}_{\star,S}}$  to convert to Chabrier.
Deep and wide-area radio surveys, such as our 610 MHz data, are powerful tools to study a range of source populations.
These comparisons demonstrate that the measurements from the literature are consistent  with our derived radio-stacked measurements for SFGs. Similar to the IR, the radio emission at 610 MHz is equivalently a good tracer of the SFR in SFGs.

\section{Conclusions and Future Work}\label{conc.sec}
We combined deep multi-wavelength optical and infrared observations from the LoTSS survey with deep 610 MHz GMRT observations to conduct a stacking analysis of star-forming galaxies between $\rm{0.1\leq\,\textit{z}\,\leq\,1.5}$.
The depth of
our 610 MHz data represents a potentially very useful tool to
address the role of SFGs in galaxy evolution. We have stacked deep, below the $\sim$40 $\mu$Jy sensitivity of the 610 MHz GMRT radio observations
at the positions of \textit{K}-selected sources in the ELAIS-N1 field (for \textit{K}-band < 22.7, sensitive
to $\rm{0.1\leq\,\textit{z}\,\leq\,1.5}$).
We remove sources quiescent galaxies, and suspected of hosting active AGN from all samples based on optical, X-ray and IR indicators. 
Using median image stacking technique that is best applied in the radio regime where
the angular resolution is high, we have measured stellar mass-dependent average star
formation rates in the redshift range $\rm{0.1\leq\,\textit{z}\,\leq\,1.5}$.
Our principal findings can be summarized as follows:
\begin{itemize}
%
\item We use a combination of rest-frame \textit{u - r} color, optical spectroscopy, X-ray information, and IR colours to  separate quiescent galaxies and AGN-driven sources from  SFGs of redshift and mass-selected galaxies. 
\item We used
median single-pixel stacking, converting the stacked radio fluxes to SFRs. We apply the \citet{2003ApJ...586..794B}
 relationship between radio luminosity and SFR, calibrated
from local galaxies, and successfully apply it to high-redshift, high-SFR galaxies, and for the first time study the relationship between radio stacked sSFR, stellar mass and redshift using deep 610 MHz data.
\item We subdivided our sample into stellar-mass and redshift bins and fit the sSFRs as a separable function of stellar mass and redshift in each bin. We found that sSFR falls with stellar mass for both our full and SFG samples. Hence the ‘downsizing’ scenario is supported by our 610 MHz data because we measure $\rm{\beta\,<\,0}$, implying that galaxies tend to form their stars more actively at higher redshifts.
\item 
We report an average of mass slope $\rm{\langle \beta_{All} \rangle\,=\, -0.49\pm0.01}$ for all galaxies and  $\rm{\langle \beta_{SFG} \rangle\,=\, -0.42\pm0.02}$ for the SFG population.
\item We report a strong increase of the
sSFR with redshift, for a given stellar mass, that is best parameterized by a power
law $\rm{\propto(1\,+\,\textit{z})^{4.94}}$ for all galaxies. The SFG population is is best parameterized by a power
law $\rm{\propto(1\,+\,\textit{z})^{3.51}}$ .
\item The sSFR appears to flatten at $\textit{z} > 1.0$ for $\rm{\textit{M}_{\star}\,>\, 10^{10.5}\textit{M}_{\odot}}$. The $\rm{sSFR - \textit{M}_{\star}}$ relation that is steeper at low masses than at high-masses (i.e. a flattening is present). Furthermore, that most massive galaxies in both the full sample and the SFG sample consistently exhibit the lowest sSFRs at all redshifts. 
\item We compare our stacked sSFR estimates to previous measurements in the \textit{sSFR-M$_{\star}$} plane, and the  evolution of the MS. We find good agreement with these previous measurements. This result opens the possibility of using the radio bands at low frequency to estimate the SFR even in the hosts of quiescent galaxies and  bright AGN. 
\end{itemize}

In view of the wealth of multi-wavelength information provided by the LoTSS catalogue, there are still exist significant opportunities to expand this work. 
A more comprehensive science analysis through stacking will include:
\begin{itemize}
\item Surveys at low frequencies, where extensive surveys exist, present different and complementary views on radio sources to that of high
frequency surveys.  We aim to further compare  our findings at 610 MHz, with results from LOFAR 150 MHz.


\item
 To undertake the exploitation of the radio luminosity functions (RLFs) of these distinct galaxy populations measured above and below the detection threshold of these surveys, using a Bayesian model-fitting technique. Extending this technique to study the cosmic star formation rate density (SFRD) at high redshifts. 
\end{itemize}
Future radio surveys will be dominated by galaxies substantially fainter than those in this current sample.
The prospects for studying the faint radio sky are very bright, as we are being rapidly flooded with survey data by  SKA pathfinders.  In conjunction with other multi-wavelength facilities, such as Euclid \citep{2018LRR....21....2A}  and the Vera C. Rubin Legacy Survey of Space and Time (LSST; \citealt{2019ApJ...873..111I}),
these projects that will survey the sky vastly
 faster than it is possible with existing radio telescopes. 

\section*{Acknowledgements}
We would like to thank the anonymous referee for their careful comments which led to a highly improved paper.
This research was supported by the Korea Astronomy and Space Science Institute under the R\&D program (Project No. 2022186804) supervised by the Ministry of Science and ICT.
JMS acknowledges the support of the Natural Sciences and Engineering Research Council of Canada (NSERC), 2019-04848. CHIC acknowledges the support of the Department of Atomic Energy, Government of India, under the project  12-R\&D-TFR-5.02-0700.
EFO would like to acknowledge the hospitality of the Inter-University Institute for Data Intensive Astronomy (IDIA) which is a partnership of the University of Cape Town, the University of Pretoria, the University of the Western Cape and the South African Radio Astronomy Observatory.
MV acknowledges financial support from the Inter-University Institute for Data Intensive Astronomy (IDIA),
a partnership of the University of Cape Town, the University of Pretoria, the University of the Western Cape
and the South African Radio Astronomy Observatory, and from the South African Department of Science and Innovation's
National Research Foundation under the ISARP RADIOSKY2020 Joint Research Scheme (DSI-NRF Grant Number 113121)
and the CSUR HIPPO Project (DSI-NRF Grant Number 121291).
We acknowledge the use of the ilifu cloud computing facility - \url{https://www.ilifu.ac.za}, a partnership between the University of Cape Town, the University of the Western Cape, the University of Stellenbosch, Sol Plaatje University, the Cape Peninsula University of Technology and the South African Radio Astronomy Observatory. The ilifu facility is supported by contributions from the Inter-University Institute for Data Intensive Astronomy (IDIA - a partnership between the University of Cape Town, the University of Pretoria, the University of the Western Cape and the South African Radio Astronomy Observatory), the Computational Biology division at UCT and the Data Intensive Research Initiative of South Africa (DIRISA). We thank Ben Keller for sharing his PASTA stacking code with us before its public release and for his helpful advice on the installation of the code on the ilifu cloud computing facility at the Inter-University Institute for Data Intensive Astronomy (IDIA).
\section*{Data Availability}
The derived data generated in this research will be shared upon reasonable request to the corresponding author. The LOFAR science-ready multi-wavelength data is available from \url{https://lofar-surveys.org/deepfields_public_en1.html}. 
\section*{Software}
This work relies on \textsc{Python} programming language (\url{https://www.python.org/})
 The Python Astronomical Stacking Tool Array (\texttt{PASTA})  program Developed at the University of Calgary by Ben Keller and Jeroen Stil\, is available at \url{https://github.com/bwkeller/PASTA}. We used astropy ( \url{https://www.astropy.org/}; \citet{2013A&A...558A..33A, 2018AJ....156..123A}), numpy (\url{https://numpy.org/}), matplotlib (\url{https://matplotlib.org/}).



\bibliographystyle{mnras}
\bibliography{example} 

 \appendix
\section{The colour-mass diagram}\label{append.colourmass}
\begin{figure}
\centering
\centerline{\includegraphics[width=0.5\textwidth]{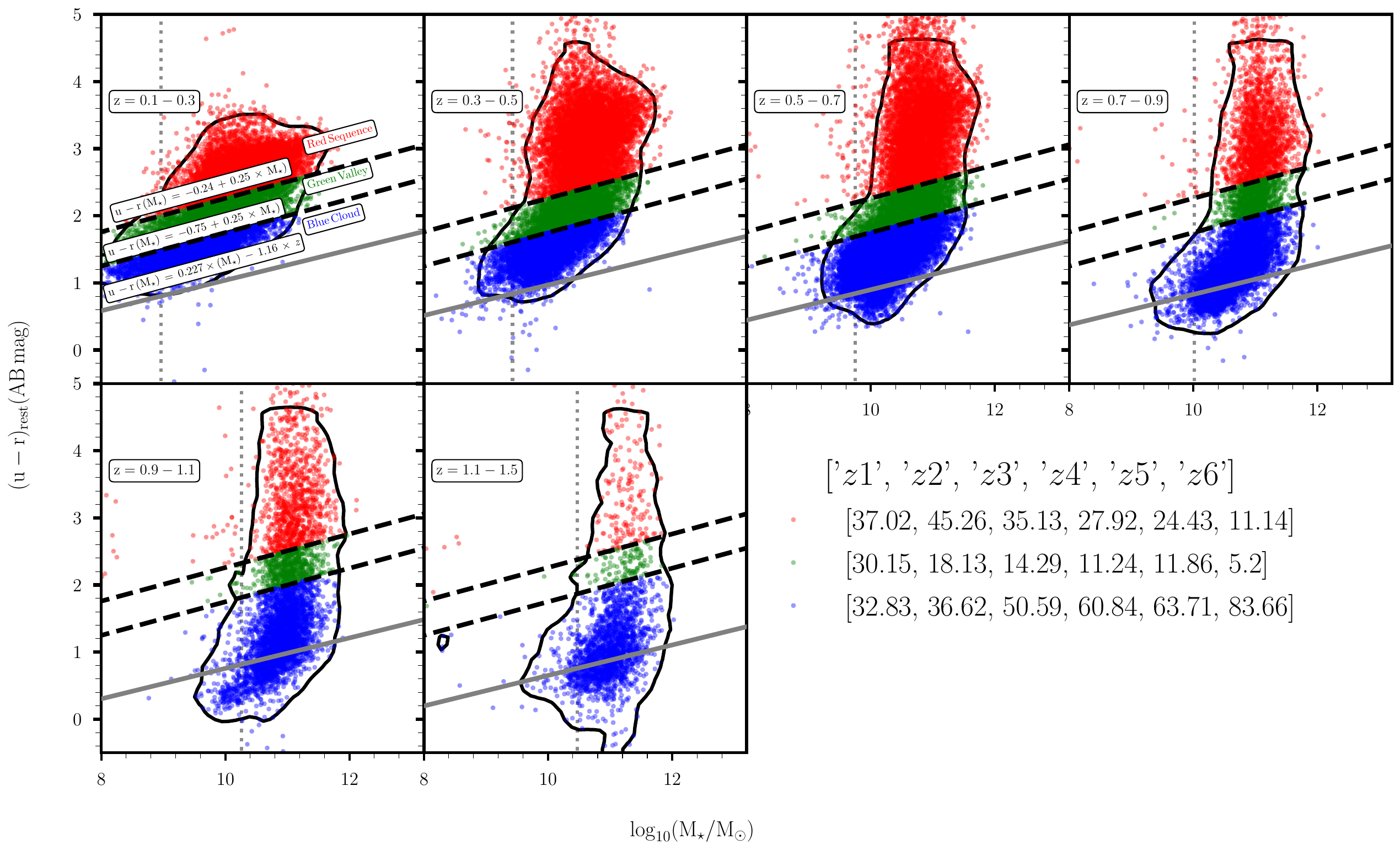}}
    \caption{Same as Figure~\ref{ur_mass.fig}.  The colours indicate the red, quiescent/passively evolving galaxies are at the top, in the red sequence with the green dots indicating the "green valley" is the transition zone in between. We show the red and blue sequence splitting line analogous to \citet{2006A&A...453..869B} redshift evolution up to $\rm{\textit{z}\,=\,1.5}$ as grey solid lines in each panel. The contours enclose those galaxies located in the $90\%$ confidence interval of the data points. The dotted grey vertical lines in each panel represents the mass completeness limit, $\rm{\textit{M}_{lim}}$.
    }
\label{ur_mass_.fig}
\end{figure}

In Figure~\ref{ur_mass_.fig}, we show  the colour-mass distribution diagram same as Figure~\ref{ur_mass.fig}.  We show the red and blue sequence splitting lines analogous to \citet{2006A&A...453..869B} redshift evolution up to $\rm{\textit{z}\,=\,1.5}$ as grey solid lines in each panel. We replaced the \textit{U} with \textit{u} and \textit{V} with \textit{r} colors. We note that these colors are not equivalent and therefore the redshift evolution of the colour-mass diagram will be different from the literature.  Using the \textit{u - r} colors, the $\rm{\textit{M}_{\star}\,-\,(\textit{u}\,-\,\textit{r})_{rest}}$ plane is given by:  $\rm{(\textit{u\,-\,r})_{rest}\,>\,0.227\log_{10}M_{\star}\,-\,1.16\,-\,0.352z}$, analogous to \textit{U - V} colors. 
The colours indicate the red, quiescent/passively evolving galaxies are at the top, in the red sequence region. The green dots indicating the "green valley" is the transition zone in between. The blue dots indicate the galaxies that reside in the blue star-forming cloud region.
Consequently, we can infer from Figure~\ref{ur_mass_.fig} that the red and blue sequence splitting line analogous to \citet{2006A&A...453..869B} (see grey solid lines) reside in the blue cloud region. Hence our inability to simply adopt this line to separate blue SFGs from red passively evolving galaxies.
For this reason, we use only the \citet{2014MNRAS.440..889S}
criteria (see main text) to separate our sources. We indicate the corresponding percentage of red, green and blue galaxies for each redshift bin in Figure~\ref{ur_mass_.fig}. The number of red and green galaxies decreases with redshift from \textit{z1} to \textit{z2}. Conversely, the number of blue galaxies increases with redshift from \textit{z1} to \textit{z2}. Our final selection of SFGs combines galaxies residing in the green valley and the blue cloud region in Figure~\ref{ur_mass_.fig}.
 
\section{STACKING ANALYSIS: DETAILS AND TESTS}

\begin{figure}
\begin{tabular}{@{}c@{ }c@{ }c@{ }c@{ }c@{ }c@{ }c@{}}
&\textbf{\textit{z}=0.1-0.3} & \textbf{\textit{z}=0.3-0.5} & \textbf{\textit{z}=0.5-0.7} & \textbf{\textit{z}=0.7-0.9} & \textbf{\textit{z}=0.9-1.1} & \textbf{\textit{z}=1.1-1.5}\\
\centering
\rotatebox[origin=b]{90}{\makebox[\ImageHt]{\scriptsize $\rm{8.5-9.0}$}}&
\includegraphics[width=.15\linewidth]{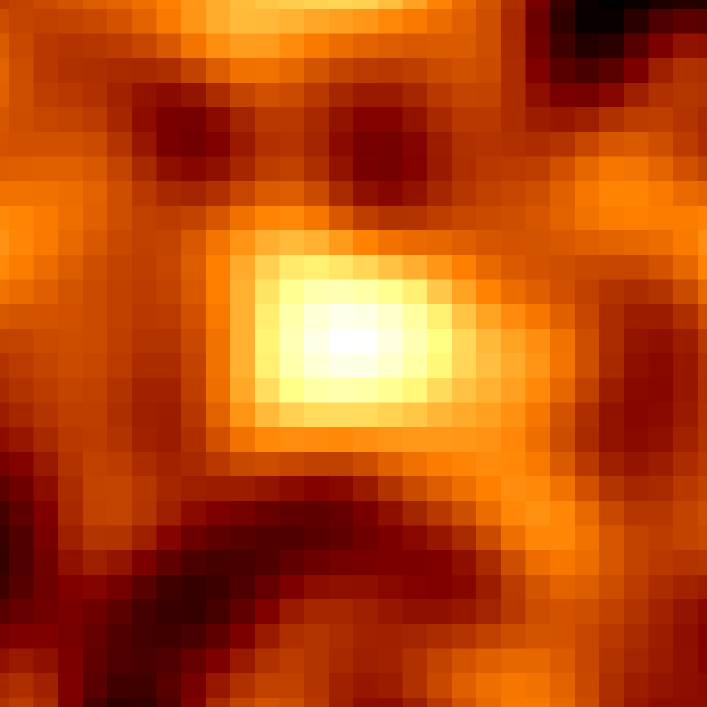}&
\includegraphics[width=.15\linewidth]{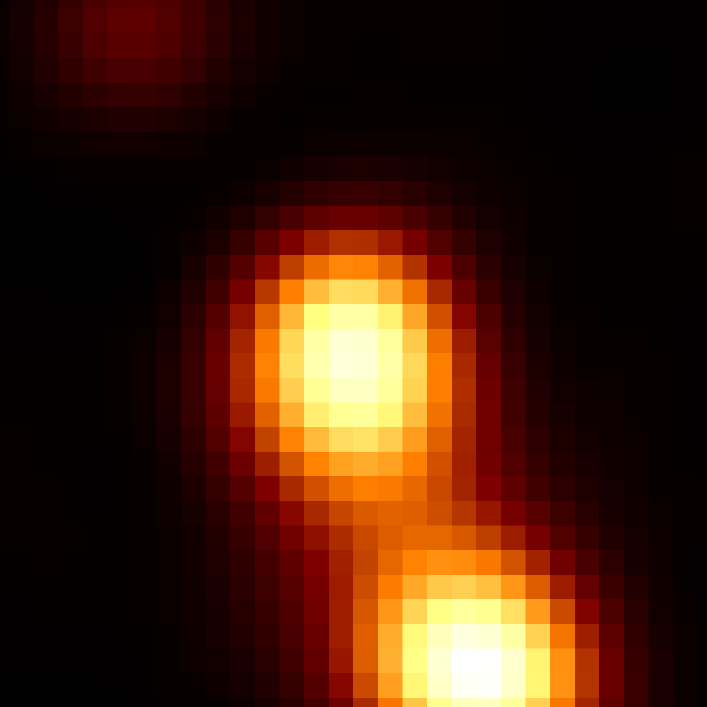}&
\includegraphics[width=.15\linewidth]{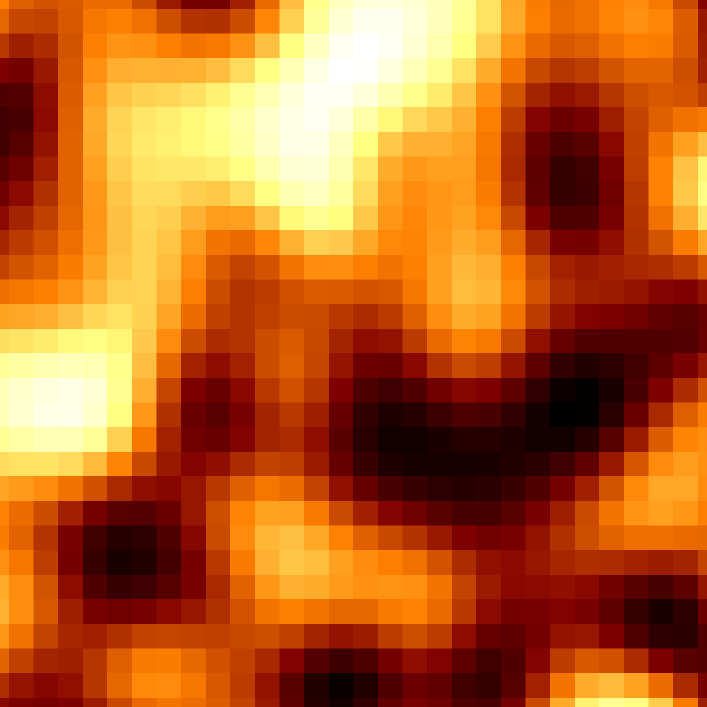}&
\includegraphics[width=.15\linewidth]{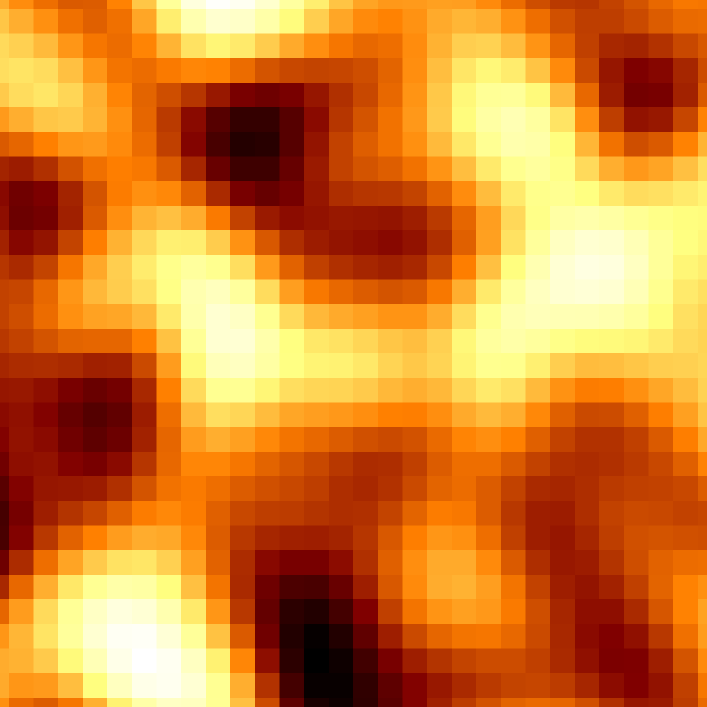}&
\includegraphics[width=.15\linewidth]{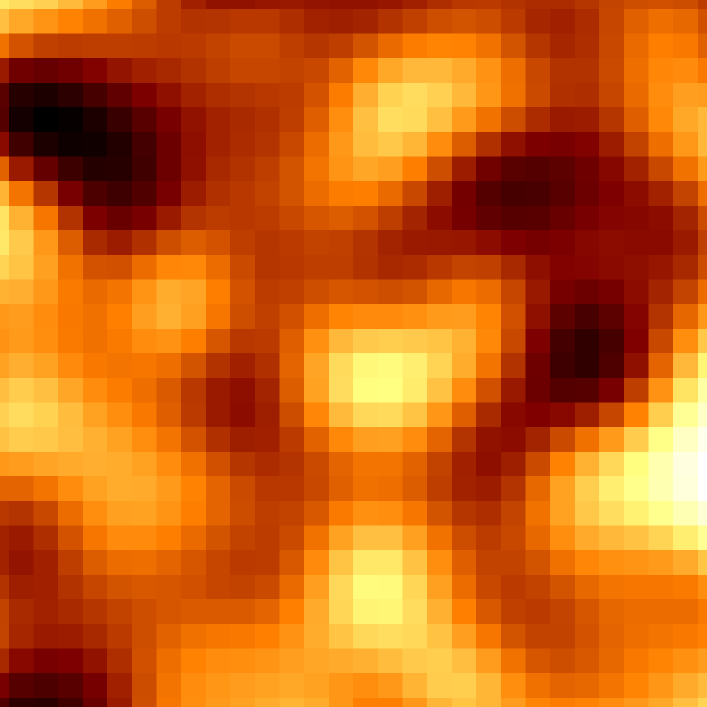}&
\includegraphics[width=.15\linewidth]{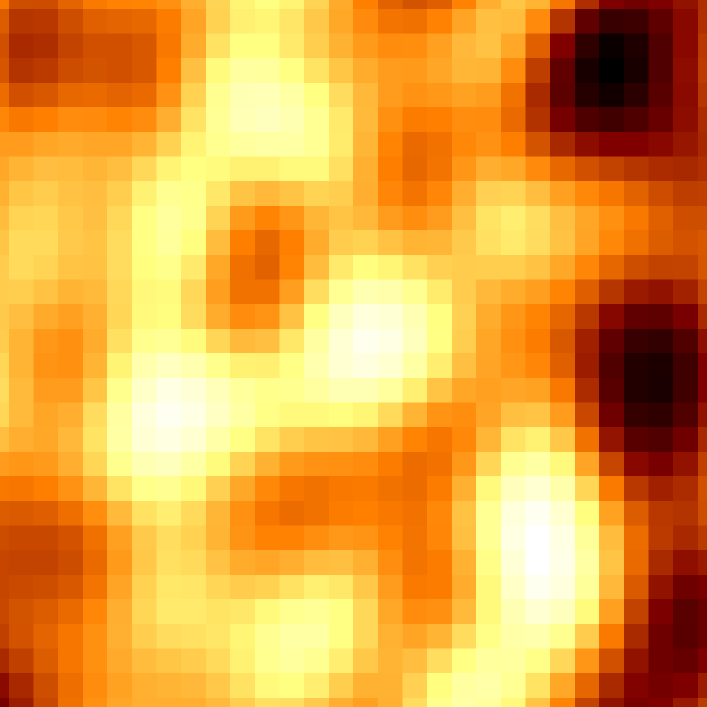}\\

\rotatebox[origin=b]{90}{\makebox[\ImageHt]{\scriptsize $\rm{8.5-9.0}$}}&
\includegraphics[width=.15\linewidth]{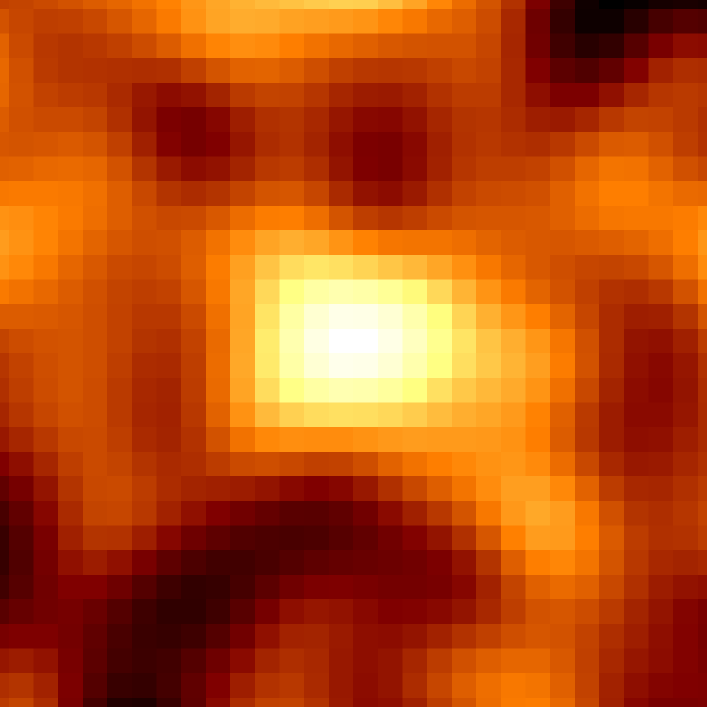}&
\includegraphics[width=.15\linewidth]{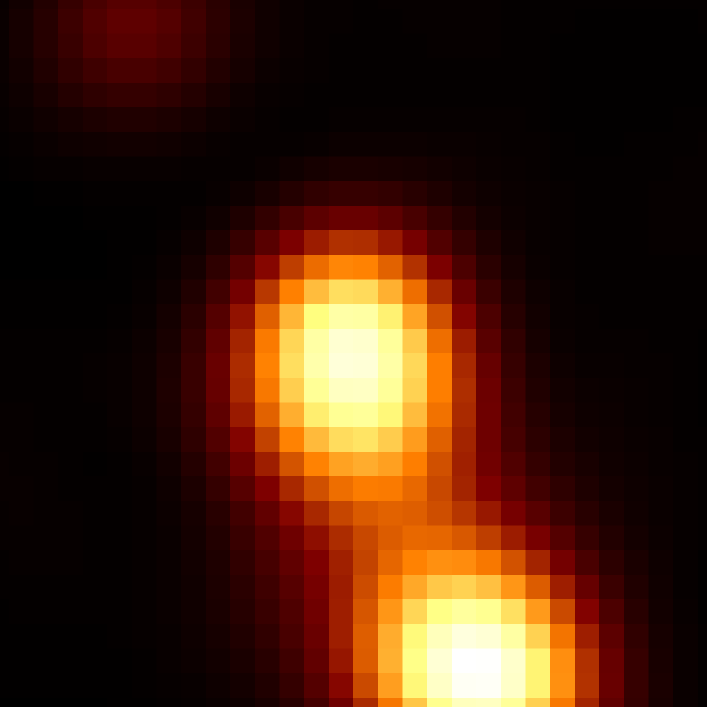}&
\includegraphics[width=.15\linewidth]{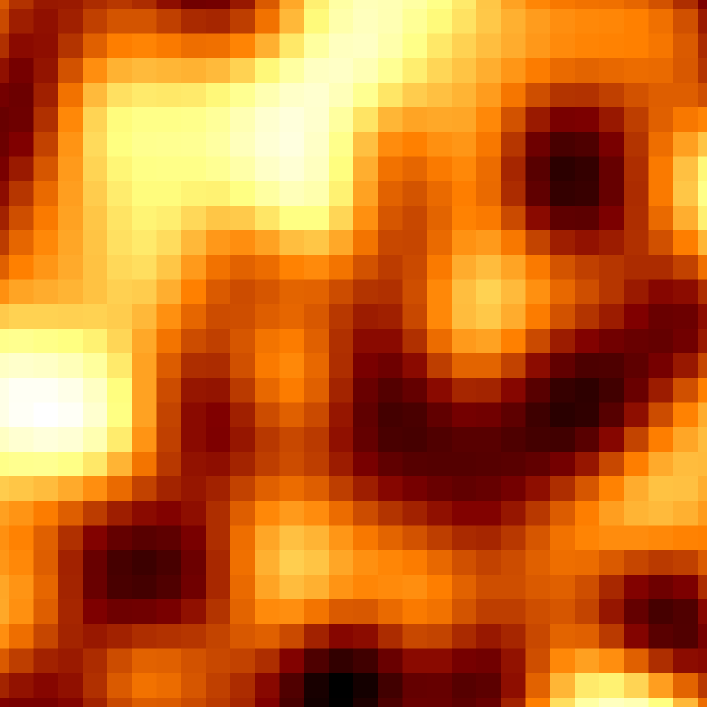}&
\includegraphics[width=.15\linewidth]{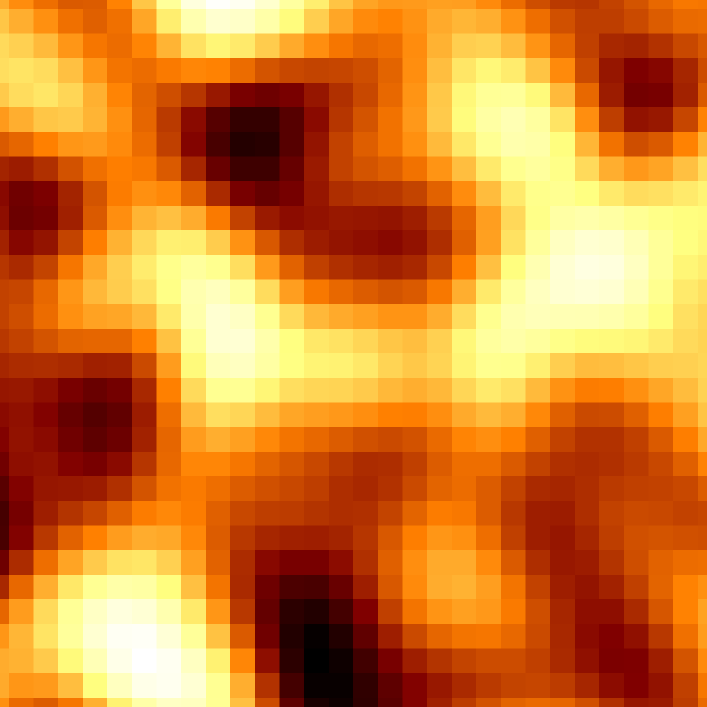}&
\includegraphics[width=.15\linewidth]{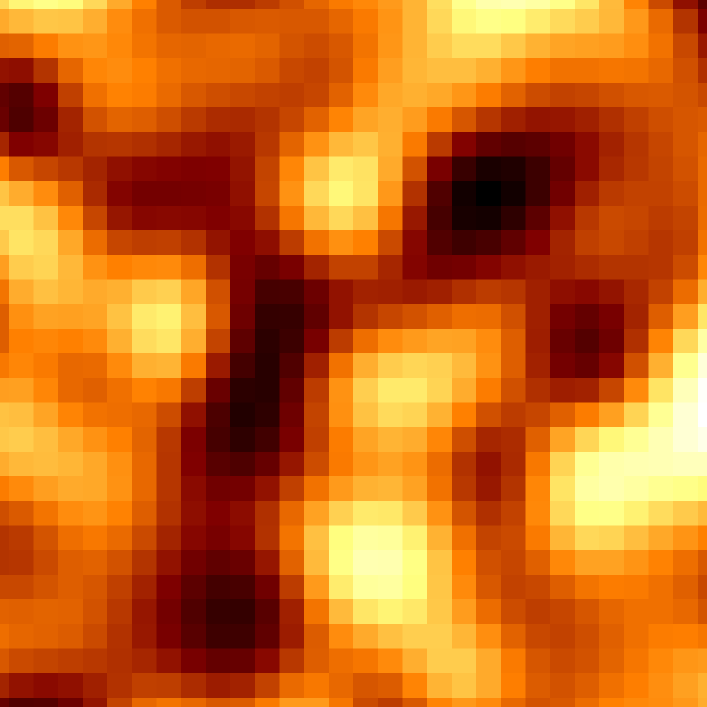}&
\includegraphics[width=.15\linewidth]{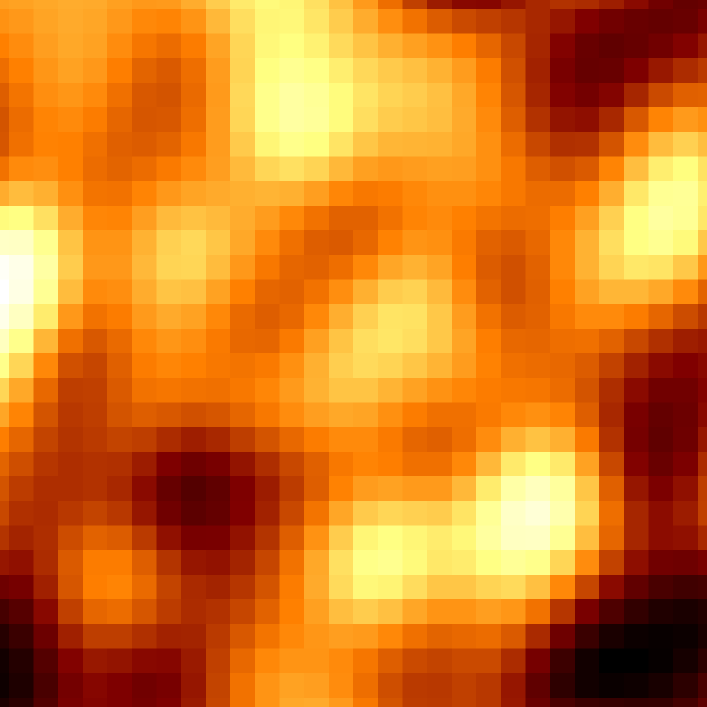}\\

\end{tabular}
\caption{Mean stacked 610 MHz radio images for the same redshift and stellar mass bins for all galaxies (top) and the SFG (bottom) populations, respectively. See Figure~\ref{all_galaxies_image.fig} for more details.}%
\label{mean_stack_fig_append}
\end{figure}

We discuss some of the issues concerning stacking radio images, including the choice of whether to represent the average as the mean or the median, and in particular, the method to measure total flux. 
A median stacking analysis is mostly preferable to mean
stacking because the median analysis is more stable and robust to small numbers of bright sources. The main problem with a mean stacking analysis is that it is very sensitive to bright outliers, which contaminate on-source flux measurements and introduce considerable noise from nearby, bright neighbors \citep[][]{2010ApJ...717..175L,2012MNRAS.421.3027B}. Many studies that use mean stacking avoid this problem by removing bright sources from all images before stacking. Figure~\ref{mean_stack_fig_append} shows mean stacked 610 MHz radio images for the same redshift and stellar mass bins for the total (top) and the SFG (bottom) populations, respectively. We verify that there is no potential bias in the median stacked
fluxes by performing a null test. We draw 
null stacks (numbers of random sky positions ) with equal to the number of sources in each stellar mass and redshift bin. We use the same stacking procedure as when the real sources are used. The results of the null tests fluctuate around zero in for both the total and SFG sources. These random stacks centered around zero indicates that the contribution to our stacked fluxes
of source confusion due to the blending of faint sources is negligible \citep[see,][]{2017MNRAS.472.2221S,2022A&A...663A.153R}.

The assumption that the pixel value at the position of each catalogue object gives the correct radio flux density for that source is generally good, though it requires a small correction to give the total integrated flux of the source.  In general, radio images have pixel units of Janskys (Jy) per beam and do adhere to the convention whereby each pixel value is equal to the flux density of a point source located at that position \citep[see][]{2009MNRAS.394..105G,2011MNRAS.410.1155B}. The integrated-flux
correction is calculated from the stacked \say{postage-stamp} image of
the total and SFG sample. This image is created by cutting out a 30 pixel
square$^{3}$ centred on each source and stacking the images by taking
the median value of each pixel. The integrated flux is calculated
using the  \textsc{PyBDSF} source finder \citep{2015ascl.soft02007M}, and the correction is simply the ratio of this to the value of the central pixel in the image following \cite{2011MNRAS.410.1155B}.

\begin{figure}
\centering
\includegraphics[width =
0.45\textwidth]{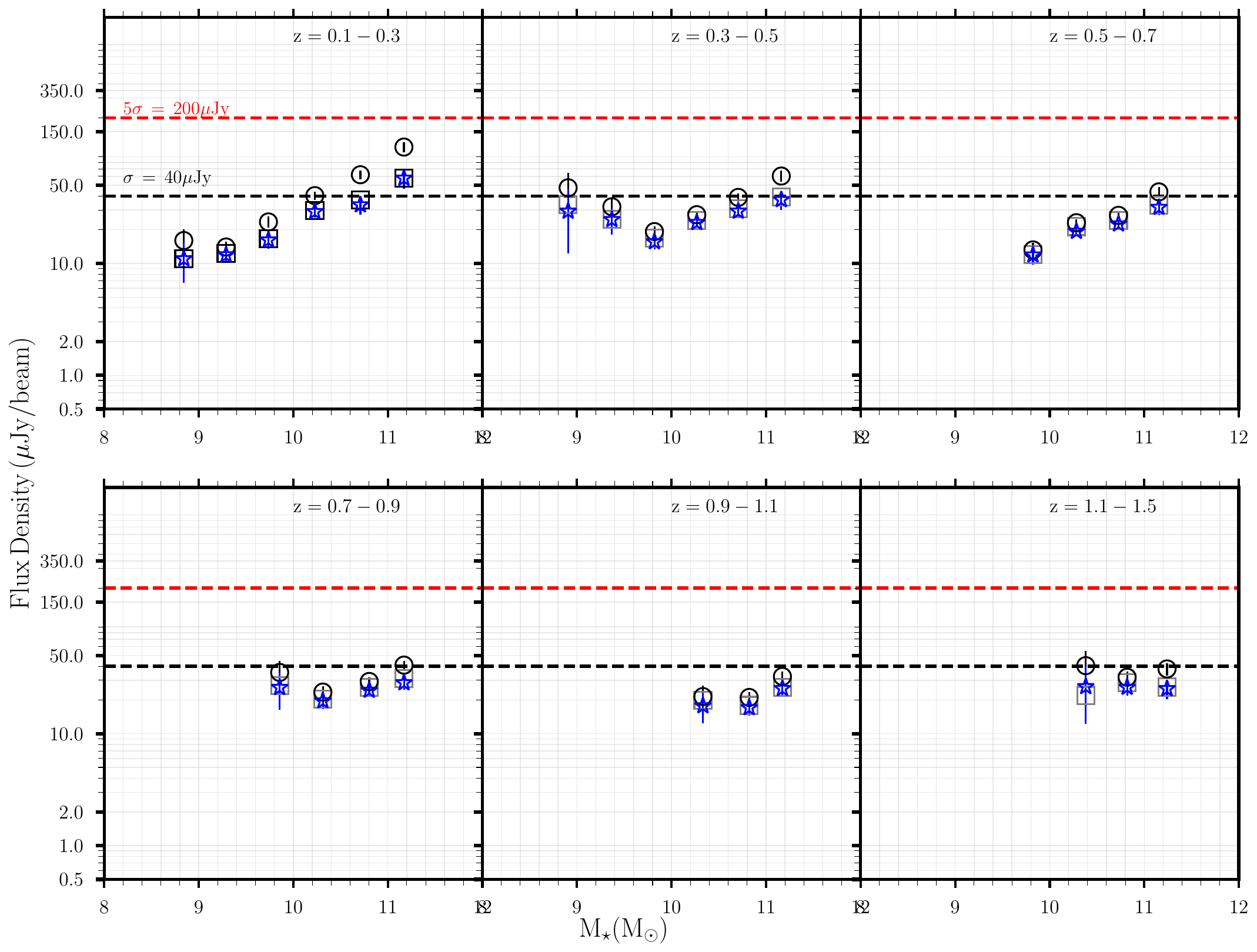}
\\
\includegraphics[width = 0.45\textwidth]{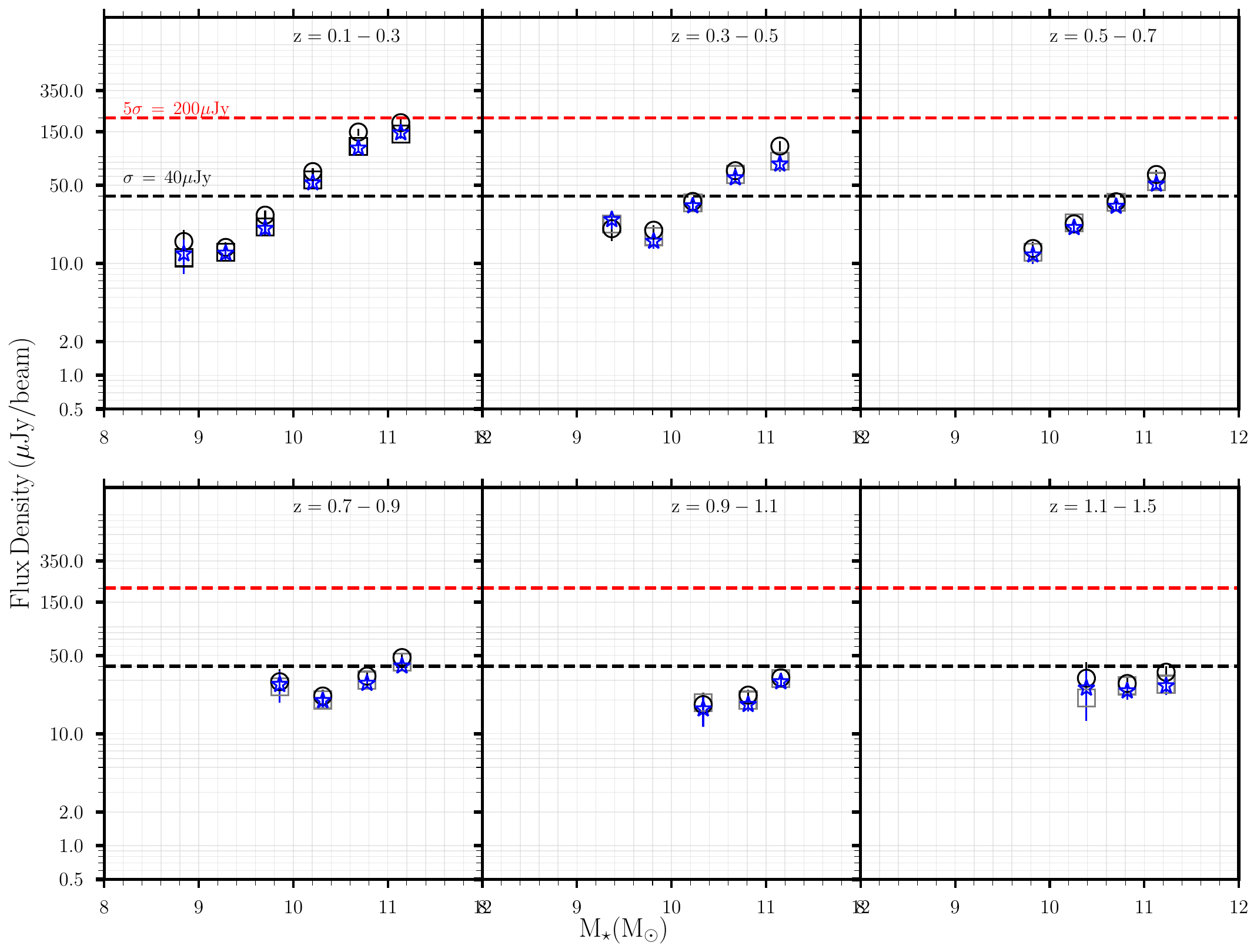}
    \caption{ The 610 MHz radio flux density for each source in the redshift and stellar mass bin, measured as in subsection~\ref{stackmethod.sec} for all galaxies (top panel) and SFGs (bottom panel) for the median stacked images which show a clear detection. The dashed horizontal
lines in each panel represents the $\sigma$ ($\sim40\mu$Jy, black) and 5$\sigma$ ($\sim200\mu$Jy, red) lines from the original 610 MHz image. The open black circles and open black squares represent the integrated and peak fluxes extracted by \textsc{PyBDSF}. The blue stars represent the corrected flux (i.e. ratio between the integrated flux and  the value of the central pixel in the image). The error
bars denote the noise level of radio flux density calculated for each image in each redshift and stellar mass bin.
    }
\label{flx_mass_rms.fig}
\end{figure}

\cite{2020MNRAS.497.5383I} study of the wide-area 610 MHz survey of the ELAIS-N1 field with the GMRT, resulted in a flux-limited catalogue in the presence of noise down to $\sim200\mu$Jy. They corrected for  two important effects, namely Eddington
bias \citep{1913MNRAS..73..359E} and incompleteness, that are key to a sample at faint flux densities. Thus it is not currently known what happens to the number counts and the nature of the sources below $\sim200\mu$Jy at 610 MHz. The dashed horizontal
lines in each panel of Figure~\ref{flx_mass_rms.fig} represents the $\sigma$ ($\sim40\mu$Jy, black) and 5$\sigma$ ($\sim200\mu$Jy, red) lines from the original 610 MHz image.  The comparison to the flux densities extracted from the median stacks showing clear detection of a source at the center suggests that we probing regimes being strongly affected by confusion (i.e. below rms sensitivity threshold). The open black circles and open black squares represent the integrated and peak fluxes extracted by \textsc{PyBDSF}. The blue stars represent the corrected flux (i.e. ratio between the integrated flux and  the value of the central pixel in the image). This correction remains small compared to true variations in stacked fluxes, and essentially all trends remained significant and conclusions unaffected whether or not a bias correction was applied.

\begin{figure}
\centering
\includegraphics[width =
0.45\textwidth]{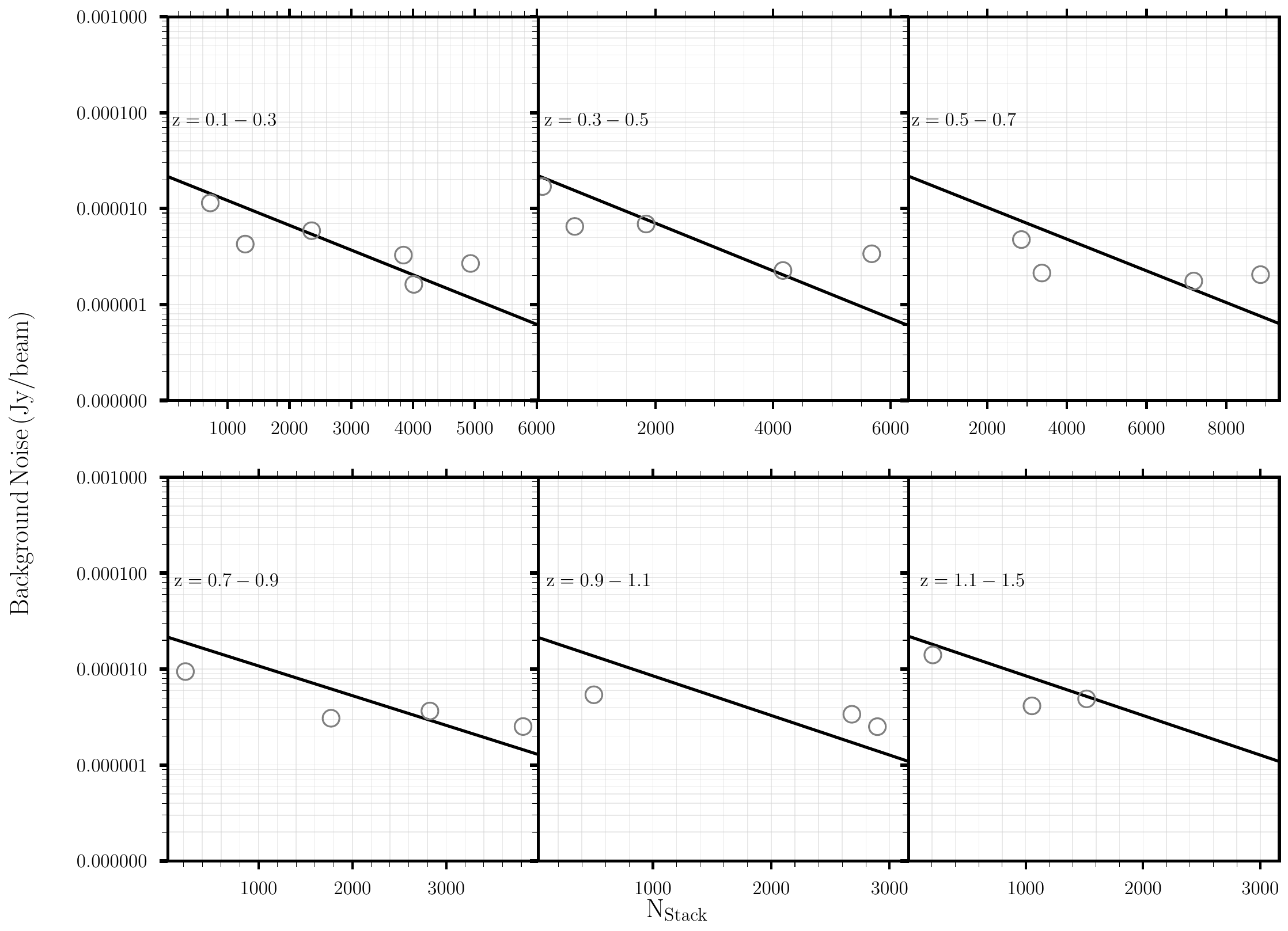}
\\
\includegraphics[width = 0.45\textwidth]{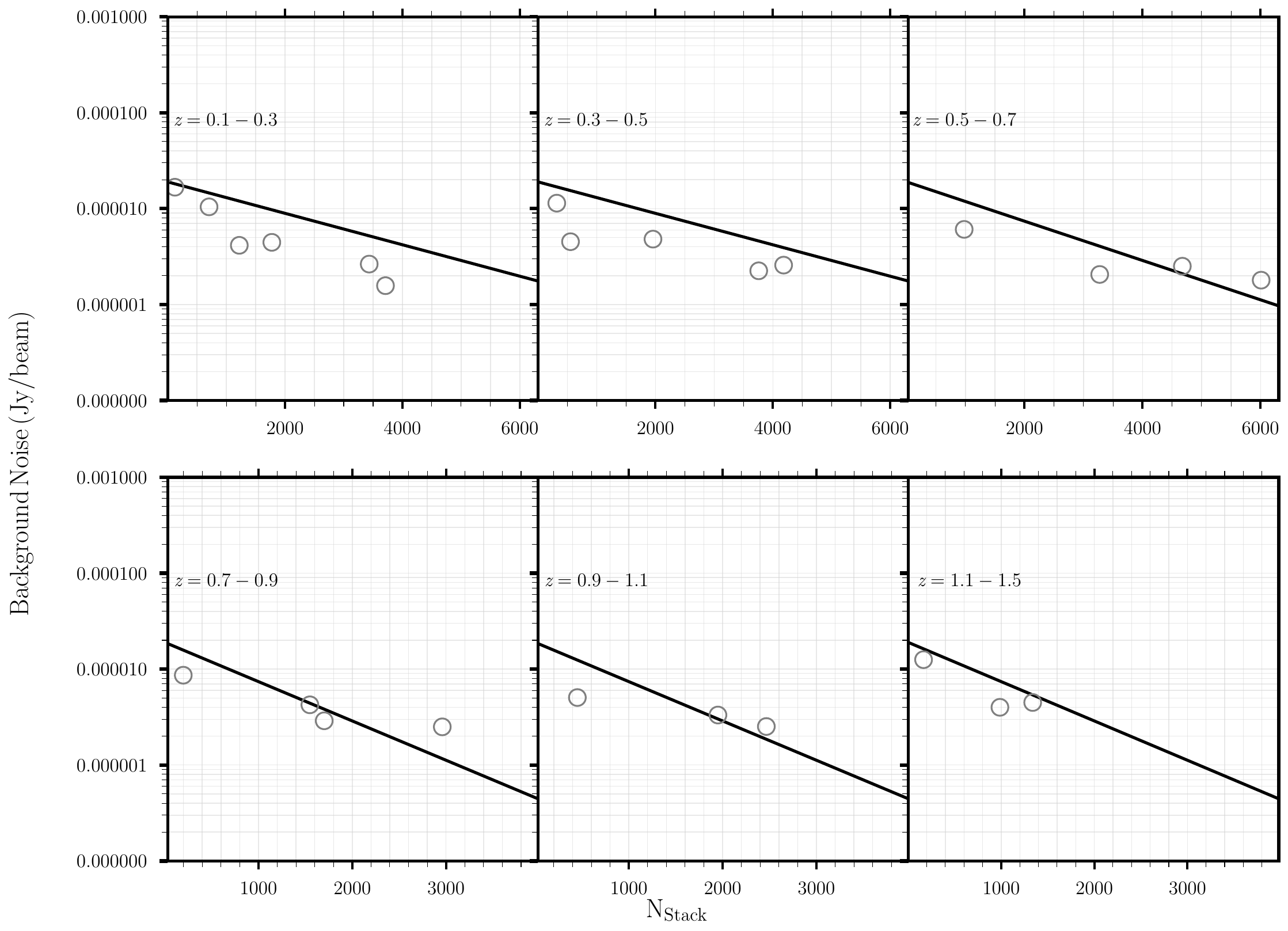}
    \caption{Relationship between the noise in the median
stacked images and the number of sources used in the stacked
sample ($\rm{N_{stack}}$) for all galaxies (top panel) and SFGs (bottom panel) for the median stacked images which show a clear detection. The solid black lines are a noise fit of the data where noise = $\rm{\sigma/\sqrt{\textit{N}}}$, and $\rm{\sigma}$ = $~\sim40\,\mu$Jy. 
}
\label{bkg_rms.fig}
\end{figure}

In our image stacking implementation, we monitor the decrease of the background noise level. This is comparable to the value expected from the typical GMRT rms of $~\sim40\,\mu$Jy divided by $\sqrt{\textit{N}}$ (i.e, a good fit to a $\rm{\sigma/\sqrt{\textit{N}}}$ relationship).
Figure~\ref{bkg_rms.fig} indicates that the noise integrated down as expected following Poissonian statistics. We achieve this by measuring the noise in the stacked postage-stamp images around the detected sources \citep[see][]{2009MNRAS.394....3D,2009MNRAS.394..105G}. 

\section{OTHER SYSTEMATICS}

Studies show that cosmic variance effects are strongest at low redshifts as the effective volume sampled in a redshift bin with $\rm{\bigtriangleup \textit{z}\,=\,0.2}$ increases with redshift. The large area and relatively decreasing redshift range we consider in this work is necessary to minimize the effect of cosmic variance \citep[see][]{2022arXiv220714689O}. We do not account for cosmic variance in our analysis (in line with our previous papers).
\citet{2010MNRAS.407.2131D} and \citet{2011ApJ...731..113M} studies  provide detailed discussions on cosmic variance in the ELAIS-N1 field. 

\begin{figure}
\centering
\centerline{\includegraphics[width=0.32\textwidth]{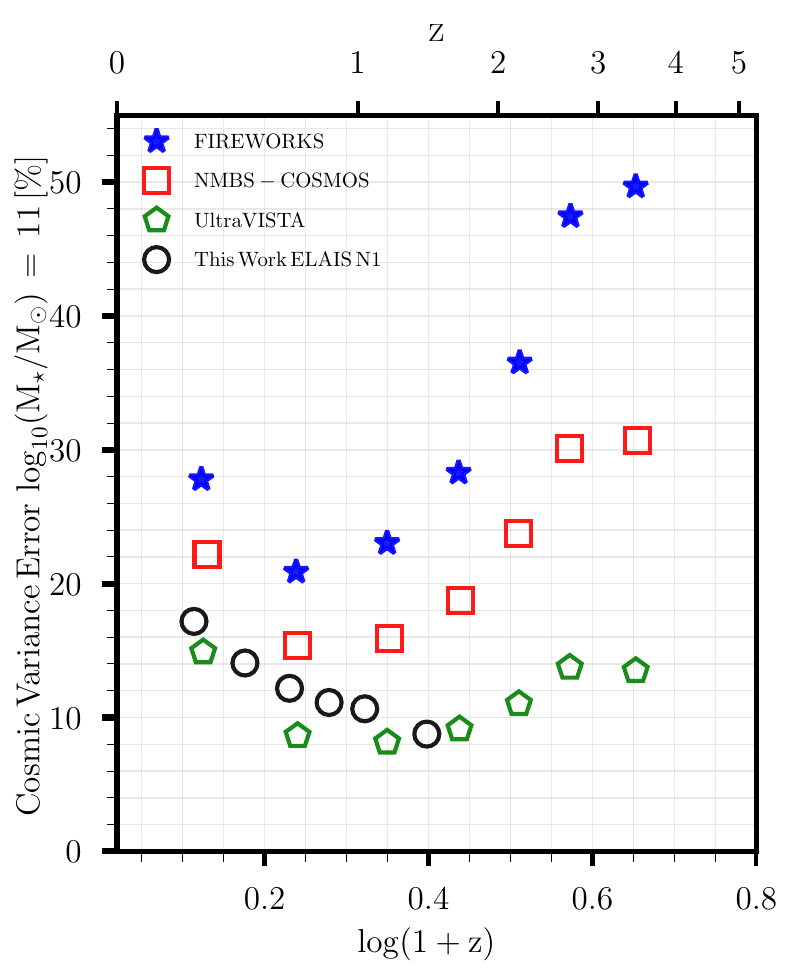}}
\caption{Uncertainty in the number density of galaxies with $\mathrm{\log_{10}(M_{\star}/M_{\odot})\,=\,11}$ due to cosmic variance as a function of redshift calculated using the prescription of \citealt{2011ApJ...731..113M}. The uncertainties in UltraVISTA due to cosmic variance are $\rm{\sim 8\%\,-\,18\%}$ at $\mathrm{\log_{10}(M_{\star}/M_{\odot})\,=\,11}$ over the full redshift range.}
\label{cos_var.fig}
\end{figure}

Cosmic variance
is most pronounced at the high-mass end where galaxies are
more clustered and at low redshift, where the survey volume is
smallest \citep{2013ApJS..206....8M}. In Figure~\ref{cos_var.fig}, we plot the uncertainty
in the abundance of galaxies with $\mathrm{\log_{10}(M_{\star}/M_{\odot})\,=\,11}$ due
to cosmic variance as a function of redshift. Also plotted in Figure ~\ref{cos_var.fig} are the cosmic variance uncertainties from other NIR surveys such as FIREWORKS \citep{2008ApJ...689..653W}, NMBS \citep{2011ApJ...735...86W} and UltraVISTA \citep{2012A&A...544A.156M}. 
In \citet{2021MNRAS.500.4685O}, we estimated that the expected cosmic variance over our relatively large survey area and in our relatively large redshift bins is at the level of 5-10$\%$ and thus will not affect our results disproportionately. Over our full redshift range in this study, the uncertainty from cosmic variance  as a function of redshift calculated using the
prescription of \citet{2011ApJ...731..113M} is $\rm{\sim 8\% - 18\%}$ at $\mathrm{\log_{10}(M_{\star}/M_{\odot})\,=\,11}$.
However, in the interest of simplicity in line with our previous works, we do not consider likely fluctuations from cosmic variance.


\bsp	
\label{lastpage}
\end{document}